\documentclass[a4paper,12pt]{article}
\usepackage{jheppub} 
\usepackage{tcilatex}
\usepackage{amsfonts}
\usepackage{amssymb}
\usepackage{multicol}
\usepackage{graphicx}
\usepackage{float}
\usepackage{caption}
\usepackage{xcolor}
\usepackage[utf8]{inputenc}
\usepackage{amsmath}
\title{Algebraic Realisation of the Zamolodchikov Metric in Narain Theories}
\author[1,2,3]{E.H Saidi}
\author[1,3]{R. Sammani}

\affiliation[1]{ LPHE-MS, Science Faculty, Mohammed V University in Rabat, Morocco.}
\affiliation[2]{Hassan II Academy of Science and Technology, Kingdom of Morocco.}
\affiliation[3]{Centre of Physics and Mathematics, CPM- Morocco.}
\emailAdd{e.saidi@um5r.ac.ma}
\emailAdd{rajae\_sammani@um5.ac.ma}

\abstract{ We revisit Narain conformal field
theories from an algebraic perspective based on finite dimensional Lie algebras $\mathbf{g}$ and
representations $\mathcal{R}_{\mathbf{g}}$, and show how the root and weight lattices can encode the momenta and subsequently the partition functions of Narain theories. \textrm{In this framework, we construct} a realisation of the Zamolodchikov metric of the
moduli space $\mathcal{M}_{\mathbf{g}}$ in terms of Lie\textrm{\ algebraic}
data \textrm{namely} the Cartan matrix K$_{\mathbf{g}}$ and its inverse K$_{%
\mathbf{g}}^{-1}$. Properties regarding the \textrm{ensemble} averaging of
these CFTs and their holographic dual are also derived. \textrm{%
Additionally, we discuss possible generalisations to} NCFTs having
dis-symmetric central charges $(\mathrm{c}_{L},\mathrm{c}_{R})=(\mathrm{s},%
\mathrm{r})$ with $s>r$ \ \textrm{and highlight further features of the
partition function} Z$_{\mathbf{g}}^{(r,r)}$.}


\begin{document}
\notoc
\maketitle
\flushbottom
\newpage
\tableofcontents
\section{Introduction}
\label{sec:intro}

Quantum gravity studies \textrm{revealed} surprising links between disparate
areas of quantum matter. One such connection suggests interrelations between
ensemble \textrm{averaging} \cite{1A},\textrm{\ code-based ensembles \cite%
{2AB,Ange}}, holography and Swampland program \cite{swp1, swp2, Raja} with
regard to the no global symmetry conjecture \cite{NG, chark} in quantum
gravity \cite{1AA}. What makes this particularly fascinating is the concept
of ensemble averaging, firstly discovered for 2D Jackiw-Teitelboim (JT)
gravity \cite{JT} then extended to 3D \cite{1A, 1b, RPP}. It has been shown
that pure gravity in 2D is dual to random matrix theory \cite{JT1}-\cite{JT4}
therefore stretching the usual one-to-one AdS/CFT correspondence to an AdS$%
_{2}$\ dual to an ensemble average instead of one traditional CFT$_{\mathrm{1%
}}$. Adopting an up-down perspective and viewing JT gravity as the
dimensional reduction of 3D gravity \cite{red1, red2}, one might wonder
whether this applies to AdS$_{3}$\ \cite{AT}-\cite{W1} as well.

Indeed three dimensional Chern-Simons (CS) U(1) gravity is dual to an
average over Narain conformal field theories (NCFT$_{2}$) on compact torii
\cite{1A, RPP, Narain, Narain2}. The average is computed over Narain moduli
space parameterised with even integer lattices with a natural distance
measure given by the Zamolodchikov metric \cite{Zchi}. The density of states
is provided by the Siegel-Weil theorem \cite{SW1}-\cite{SW3}.\textrm{\ }%
Using spectral decomposition, the partition function uncovers two pieces:
one is moduli-dependent and another moduli-independent corresponding to the
average \cite{1AB}-\cite{SD3}.

From the bulk perspective, the average is equivalent to the sum over three
dimensional topologies. Since the theory is only defined perturbatively as
it is only checked on torus boundaries, the sum is carried out over
handlebodies \cite{1b, Y, H}. Moreover, adopting diagonal boundary
conditions, the U(1) gravity can be reinterpreted as the asymptotic symmetry
of an AdS$_{\mathrm{3}}$\ gravity coupled to U(1) gauge fields \cite{PT}-%
\cite{Soft2} and the expected Brown-Henneaux boundary gravitons of ordinary
3d gravity \cite{BH} emerge as composites of the asymptotic symmetry.
Furthermore, by promoting the sl$(2,\mathbb{R})$ algebra of AdS$_{\mathrm{3}}
$\ to higher dimensional gauge symmetries, one can again recuperate the U(1)
algebra of the bulk theory dual to the ensemble average from higher spin
gravity \cite{PT, Emb1, Raja2, Raja3,Raja4}.

Generalisations of this theory can be drawn in several ways: $\mathbf{(a)}$
by adding flavours to the partition function \cite{Y}. This follows from
considering fugacities, chemical potentials, conjugate to the U(1) charges. $%
\mathbf{(b)}$ By constructing orbifolds of Narain CFTs \cite{MEER}. The
process of orbifolding consists of architecting new conformal field theories
from established ones by gauging discrete symmetries of the original theory.
There are several incentives for considering orbifolds of Narain CFT but the
most compelling one in the context of ensemble averaging is that it allows
to gauge the global ensemble symmetries \cite{MEER} which might serve in the
restoration of a healthy quantum gravity theory forbidding the presence of
global symmetries. And $\mathbf{(c)}$ by extending Narain lattices to
general non-unimodular even or odd integer lattices \cite{1AA, RPP, MEER}.

In the present investigation, we shall discuss both unorbifolded standard
and generalised Narain CFT and offer a systematic classification based on
finite dimensional Lie algebras $\mathbf{g}$\ and representations\textrm{\ }$%
\mathcal{R}_{\mathbf{g}}$\textrm{. } Building upon this, we give explicit realisations of the Zamolodchikov metric of the moduli space of these CFTs and show that it can be written in terms of the Cartan matrix and its inverse. We first consider Narain conformal field
theories (NCFT$_{2}$) on compact torii \cite{1A, 1AA} and their gravity dual
with symmetry $U(1)^{\mathrm{c}_{L}}\times U(1)^{\mathrm{c}_{R}}$ \cite{PT,
PT2}. These CFT$_{2}$s are given by strings compactified on $\mathbb{T}^{r}$
with radii $\left\{ R_{1},...,R_{r}\right\} $ and underlying symmetry $%
U(1)_{L}^{\mathrm{r}}\times U(1)_{R}^{\mathrm{r}}.$ There are two natural
parametrisations of these CFTs given by two real quadratic forms: $\mathcal{H%
}^{\left( r,r\right) }$ and $\mathfrak{Q}^{\left( r,r\right) },$ each
unveils an interesting structure\textrm{:}

$\left( \mathbf{i}\right) $ A local quadratic form $\mathcal{H}^{\left(
r,r\right) }$ which is a real function of variables $\left( \mathbf{x}%
\right) $ coordinating the moduli space $\mathcal{M}_{\left( r,r\right) }$
of the CFT ($\nabla _{\mathbf{x}}\mathcal{H}^{\left( r,r\right) }\neq 0$).
This is a positive function that can be imagined in terms of scaling
dimensions $\mathcal{H}=h+\bar{h}$ of generic primary field operators $\Phi
_{h,\bar{h}}$ with conformal weight $(h,\bar{h})=(p_{L}^{2}/2,p_{R}^{2}/2)$
and quantized momenta $p_{L/R}$. This $\mathcal{H}^{\left( r,r\right) }$
expands like $\sum_{\text{\textsc{a}}=1}^{2r}$\textsc{u}$^{\text{\textsc{a}}}%
\mathbf{H}_{\text{\textsc{ab}}}$\textsc{v}$^{\text{\textsc{b}}}$ with
characteristic matrix $\mathbf{H}_{\text{\textsc{ab}}}\left( \mathbf{x}%
\right) $ capturing the local target space data of the two dimensional CFTs
and subsequently averaged data on effective field properties in the 3D bulk
dual.

$\left( \mathbf{ii}\right) $ A global real quadratic form $\mathfrak{Q}%
^{\left( r,r\right) }$ that is independent of the moduli ($\nabla _{\mathbf{x%
}}\mathfrak{Q}^{\left( r,r\right) }=0$); thus capturing "topological" data
of the bulk dual of the boundary CFT. This $\mathfrak{Q}^{\left( r,r\right)
} $ can be thought of in terms of generic conformal spins $h-\bar{h}$; and
quite similarly to $\mathcal{H}^{\left( r,r\right) }$, it has the typical
expansion \cite{RPP}
\begin{equation}
\mathfrak{Q}^{\left( r,r\right) }\left( \text{\textsc{u},\textsc{v}}\right)
=\sum_{\text{\textsc{a}}=1}^{2r}\text{\textsc{u}}^{\text{\textsc{a}}}\mathbf{%
Q}_{\text{\textsc{ab}}}\text{\textsc{v}}^{\text{\textsc{b}}}\qquad ,\qquad
\mathfrak{Q}^{\left( r,r\right) }\left( \text{\textsc{u},\textsc{v}}\right)
\in \mathbb{Z}  \label{01}
\end{equation}%
with discrete variables \textsc{u}$^{\text{\textsc{a}}}=\left(
n^{i},w_{a}\right) $ having $r+r$ integral components and characteristic
matrix $\mathbf{Q}_{\text{\textsc{ab}}}$. Here, the integer vector $%
\boldsymbol{n}:=\left( n_{1},...,n_{r}\right) $ labels the Kaluza-Klein (KK)
excitations of the compactified string while the integral $\boldsymbol{w}%
:=\left( w_{1},...,w_{r}\right) $ indexes the winding modes. The global $%
\mathfrak{Q}^{\left( r,r\right) }$ is then characterised by the $2r\times 2r$
square matrix $\mathbf{Q}_{\text{\textsc{ab}}}$ splitting in 4 bloc
sub-matrices as
\begin{equation}
\mathbf{Q}_{\text{\textsc{ab}}}=\left(
\begin{array}{cc}
Q_{ij} & q_{i}^{\text{ }b} \\
\tilde{q}_{\text{ }j}^{a} & \tilde{Q}^{ab}%
\end{array}%
\right)
\end{equation}%
It also allows to define
an even integer lattice $\Lambda _{Q^{\left( r,r\right) }}$ with quadratic
form $\mathfrak{Q}^{\left( r,r\right) }\left( \text{\textsc{u},\textsc{u}}%
\right) $ $\in 2\mathbb{Z}$ as in eq(\ref{01}).

Integrating over the moduli $\left( \mathbf{x}\right) $ defines an
ensemble-averaged free CFT which is conjectured to be holographically dual
to an exotic three dimensional gravity with partition function $Z_{\text{%
\textsc{bulk}}}=Z_{\text{\textsc{cft}}}.$ As we have stated, this gravity\
is believed to be given by a generalised Chern-Simons theory with $U\left(
1\right) ^{r}\times \tilde{U}\left( 1\right) ^{r}$ gauge symmetry; its field
action $\mathcal{S}_{{\small CS}}$ is expressed in terms of the 1-form CS
potentials $\mathcal{A}^{\text{\textsc{a}}}=\mathcal{A}_{\mu }^{\text{%
\textsc{a}}}d$\textsc{x}$^{\mu }$ and their 2-form field strength curvatures
$\mathcal{F}^{\text{\textsc{a}}}=d\mathcal{A}^{\text{\textsc{a}}};$ it reads
approximately as follows \cite{1A,RPP}
\begin{equation}
\mathcal{S}_{{\small CS}}=\frac{1}{8\pi }\int_{\text{\textsc{Y}}_{3}}%
\mathfrak{Q}_{{\small CS}}^{\left( r,r\right) }\qquad ,\qquad \mathfrak{Q}_{%
{\small CS}}^{\left( r,r\right) }=\sum_{\text{\textsc{a}}=1}^{2r}\mathcal{A}%
^{\text{\textsc{a}}}\mathbf{Q}_{\text{\textsc{ab}}}\mathcal{F}^{\text{%
\textsc{b}}}  \label{02}
\end{equation}%
with real threefold \textsc{Y}$_{3}$ being a handlebody whose boundary is a
compact 2D surface; here a 2-torus $S_{t}^{1}\times S_{\sigma }^{1}$ world
sheet. In this topological field action, the global coupling $\mathbf{Q}_{%
\text{\textsc{ab}}}$ is the same matrix as in eq(\ref{01}) and the quadratic
monomials $\mathcal{A}^{\text{\textsc{a}}}\mathcal{F}^{\text{\textsc{b}}}$
refers to the 3-form $\mathcal{A}^{\text{\textsc{a}}}\wedge \mathcal{F}^{%
\text{\textsc{b}}}$.

In this paper, we give a classification of a family of two dimensional Narain conformal
field theories with central charges $\mathrm{c}_{L}=\mathrm{c}_{R}=r$ and
their ensemble average in terms of finite dimensional Lie algebras $\mathbf{g%
}$. The various classes of these conformal field theories are labeled like%
\begin{equation}
\text{NCFT}_{2}^{\mathbf{g}}\qquad ,\qquad \mathbf{g}=\mathtt{A_{r},}\text{ }%
\mathtt{B_{r},}\text{ }\mathtt{C_{r},}\text{ }\mathtt{D_{r},}\text{ }\mathtt{%
E_{6,7,8},}\text{ }\mathtt{F}_{4}\mathtt{,}\text{ }\mathtt{G}_{2}  \label{03}
\end{equation}%
with $\mathbf{g}$ including both simply and non simply laced finite Lie
algebras. Moreover, we investigate the genus-one partition function of these NCFT$%
_{2}^{\mathbf{g}}$s termed below as Z$_{\mathbf{g}}^{(r,r)}\left[ \tau ,\bar{%
\tau};\mathbf{x}\right] .$ It factorises like $\left\vert \eta \left( \tau
\right) \right\vert ^{-2r}\Theta _{\mathbf{g}}^{(r,r)}\left( \tau ,\bar{\tau}%
,\mathbf{x}\right) $ with the usual Dedekin $\eta \left( \tau \right) $ and
the Siegel-Narain theta function $\Theta _{\mathbf{g}}^{(r,r)}$ shown in
this study to be given by the following \textbf{g}-dependent formula%
\begin{equation}
\Theta _{\mathbf{g}}^{(r,r)}\left( \tau _{1},\tau _{2},\mathbf{x}\right)
=\dsum\limits_{\mathbf{\beta },\mathbf{\chi }}e^{-\pi \tau _{2}\left\langle
\mathbf{\beta }^{\vee }.\mathbf{\beta }\right\rangle }e^{2\pi i\tau
_{1}\left\langle \mathbf{\beta }^{\vee }.\mathbf{\chi }\right\rangle
}e^{-\pi \tau _{2}\left\langle \mathbf{\chi }^{\vee }.\mathbf{\chi }%
\right\rangle }  \label{04}
\end{equation}%
where the vectors $\mathbf{\beta }=\sum_{i=1}^{r}n^{i}\mathbf{\beta }_{i}$
and $\mathbf{\chi }$ $=\sum_{j=1}^{r}w_{j}\mathbf{\chi }^{j}$ and their
inner products are defined in the core of the paper [see for instance (\ref%
{at}) and (\ref{ex})]; they capture data on the Lie algebra \textbf{g }and
representations $\mathcal{R}_{\mathbf{g}}$ and have an interpretation in
terms of intersecting 2-cycles in the moduli space of the CFT. Moreover, we
investigate the ensemble averaging $\mathcal{Z}_{\mathbf{g}}^{(r,r)}\left(
\tau ,\bar{\tau}\right) $ over the moduli space which is given by $<Z_{%
\mathbf{g}}^{(r,r)}\left[ \tau ,\bar{\tau};\mathbf{x}\right] >_{\mathcal{M}}$%
; this averaging, denoted shortly as $\mathcal{Z}_{\mathbf{g}}^{(r,r)}\left(
\tau ,\bar{\tau}\right) ,$ kills the dependence on the moduli $\mathbf{x}$;
but manifests itself by some constants c$_{\mathbf{g}}$ bearing the imprint
of $\mathcal{H}_{\mathbf{g}}^{\left( r,r\right) }$\textbf{\ }like $\mathcal{Z%
}_{\mathbf{g}}^{(r,r)}\left( \tau ,\bar{\tau}\right) =\left\vert \eta \left(
\tau \right) \right\vert ^{-2{\small r}}F_{\mathbf{g}}^{(r,r)}$ with $F_{%
\mathbf{g}}^{(r,r)}$ equals to \TEXTsymbol{<}$\Theta _{\mathbf{g}%
}^{(r,r)}\left( \tau _{1},\tau _{2},\mathbf{x}\right) $\TEXTsymbol{>}$_{%
\mathcal{M}}$. In this regard, we show in\textrm{\ appendix C }that the
footprint c$_{\mathbf{g}}=1/vol(\mathcal{M}_{\mathbf{g}}^{({\small D,D)}})$
which is a function of $K_{\mathbf{g}}$ and $K_{\mathbf{g}}^{-1}$ [see eqs(%
\ref{22}-\ref{23})]. Furthermore, we use the obtained results from the
classification (\ref{03}) to give explicit realisations of the Zamolodchikov
metric ds$_{\mathbf{g}}^{2}$ of the moduli space of these CFTs namely
\begin{equation}
\mathcal{M}_{\mathbf{g}}^{(r,r)}=\mathcal{O}(r,r;\mathbb{Z})\backslash
\mathcal{O}(r,r;\mathbb{R})/[\mathcal{O}(r;\mathbb{R})\times \mathcal{O}(r;%
\mathbb{R})]
\end{equation}%
In particular, we show that this metric can be presented like ds$_{\mathbf{g}%
}^{2}=\sum_{i,k=1}^{r}\left( d\mu _{\mathbf{g}}\right) _{i}^{k}\left( d\mu _{%
\mathbf{g}}\right) _{k}^{i}$ with $d\mu _{\mathbf{g}}$ realised in terms of
the Cartan matrix K$_{\mathbf{g}}$ and K$_{\mathbf{g}}^{-1}$ in addition to
the coordinates $\mathbf{x}$ of $\mathcal{M}_{\mathbf{g}}^{(r,r)}$. As well,
we comment on the extension of our proposal to the so-called generalised
NCFTs whose moduli space $\mathcal{O}(s,r;\mathbb{Z})\backslash \mathcal{O}%
(s,r;\mathbb{R})/[\mathcal{O}(s;\mathbb{R})\times \mathcal{O}(r;\mathbb{R})]$
with left central charge $\mathrm{c}_{L}$ greater than the right $\mathrm{c}%
_{R}$ ($s>r$).

The structure of the paper is as follows: \textrm{In \autoref{sec:2}}, we describe
the set up of our proposal for the NCFT$^{\mathbf{su}_{2}}$ and give its
basic properties including the partition function $Z_{\mathbf{su}%
_{2}}^{(1,1)}\left( \tau _{1},\tau _{2},x\right) $ and its ensemble
averaging. \textrm{In \autoref{sec:3}}, we develop the proposal for NCFT$^{\mathbf{%
g}}$ for rank r=2 Lie algebras which are given by su(3), so(5), sp(4) and G$%
_{2}$ with a particular focus on the simply laced case for explicit
calculations. \textrm{In \autoref{sec:4}}, we provide results for NCFT$^{\mathbf{g}%
}$ classified by finite dimensional Lie algebras \textbf{g}. As
illustrations, we describe leading members of the classes of Lie algebras.
\textrm{In \autoref{sec:5}}, we conclude our study by summarising our main \textrm{%
findings} and comment on the so-called generalised Narain-CFTs. Then, we give three appendices: \textrm{\autoref{sec:appA}} concerns Narain
CFTs and gravitational dual. \textrm{\autoref{sec:appB}} regards Root $\Lambda _{r}^{%
\mathbf{g}}$ and weight $\Lambda _{w}^{\mathbf{g}}$ lattices of Lie algebras
$\mathbf{g}$ and their discriminant $\Lambda _{w}^{\mathbf{g}}/\Lambda _{r}^{%
\mathbf{g}}$. \textrm{\autoref{sec:appC}} deals with ensemble averaging, giving
details on the derivation of results presented in the main text and
exhibiting the footprint of $\mathbf{g}$ and representations $\mathcal{R}_{%
\mathbf{g}}$ in the averaged partition function $\mathcal{Z}_{\mathbf{g}%
}^{(r,r)}\left( \tau ,\bar{\tau}\right) $.
\section{Narain CFT$_{2}$ and Lie algebras}
\label{sec:2}
In this section, we establish the foundation for the classification of
standard Narain CFT$_{2}$'s by studying the basic su(2) conformal model with
central charge $(\mathrm{c}_{L},\mathrm{c}_{R})=(\mathrm{1},\mathrm{1})$
corresponding to the level $k=\tilde{k}=1$ in
\begin{equation}
\mathrm{c}_{L}\left( k\right) =\frac{3k}{k+2}\qquad ,\qquad \mathrm{c}_{R}(%
\tilde{k})=\frac{3\tilde{k}}{\tilde{k}+2}
\end{equation}%
We leverage this construction later to propose a classification of Narain
sigma-models with target r-torus $\mathbb{T}^{r}$ leading to a CFT$_{2}$
with central charges $(\mathrm{c}_{L},\mathrm{c}_{R})=(\mathrm{r},\mathrm{r}%
) $ as described in (\ref{03}). This classification will be comprehensively
addressed in the forthcoming sections 3 and 4.

\subsection{Standard Narain CFT$_{2}^{\mathtt{su}_{2}}$}

The standard NCFT$_{2}^{\mathtt{su}_{2}}$ is a subset of conformal field
theories; it concerns $\left( \mathbf{i}\right) $ the compactification of a
\textrm{closed string field} $X=(X_{L}+X_{R})/\sqrt{2}$ on circle $\mathbb{S}%
^{1}\simeq U\left( 1\right) $ with radius R (one 2D free boson), and $\left(
\mathbf{ii}\right) $ has quantized left/right momenta $\left(
p_{L},p_{R}\right) $ labeled by two integers n and w as in eq(\ref{plpr}).
The moduli space of this conformal theory is one dimensional;
\begin{equation}
\mathcal{M}_{1}\simeq \mathcal{O}\left( 1,1,\mathbb{R}\right) /\mathcal{O}%
\left( 1,1,\mathbb{Z}\right) \qquad \dim \mathcal{M}_{1}=1
\end{equation}
it is parameterized by $R$ taking values in the range $0<R<\infty ;$ its
metric $ds_{\mathcal{M}_{1}}^{2}$ is given by $\left( R^{-1}dR\right) ^{2}$
\cite{1A}. By using the T-duality property $R\rightarrow 1/R$, the range of
the radius can be reduced to $1<R<\infty $ indicating that $vol(\mathcal{M}%
_{1})$ has infinite volume; $vol(\mathcal{M}_{1})\sim \int_{1}^{\infty
}d\left( \log R\right) $ behaving like $\left. \log R\right\vert
_{1}^{\infty }$. For later use, we will think about the metric $ds_{\mathcal{%
M}_{1}}^{2}$ like
\begin{equation}
ds_{\mathcal{M}_{1}}^{2}=\left( \mathcal{K}^{-1}d\mathcal{K}\right)
^{2}=\left( -\mathcal{K}d\mathcal{K}^{-1}\right) ^{2}  \label{KK1}
\end{equation}%
with $\mathcal{K}=2R^{2}$ to be commented upon the presentation of our
proposal. Notice that under T-duality ($R\rightarrow 1/R$), we have the
mappings%
\begin{equation}
\begin{tabular}{lll}
$\mathbb{S}^{1}$ & $\quad \leftrightarrow \quad $ & $\mathbb{\tilde{S}}^{1}$
\\
$U\left( 1\right) $ & $\quad \leftrightarrow \quad $ & $\tilde{U}\left(
1\right) $%
\end{tabular}%
\end{equation}%
which expand for left and right sectors $\mathbb{S}_{L/R}^{1}$ as follows;
\textrm{see also appendix C for further details,}%
\begin{equation}
\begin{tabular}{lll}
$\mathbb{S}_{L/R}^{1}$ & $\quad \sim \quad $ & $\mathbb{S}^{1}\pm \mathbb{%
\tilde{S}}^{1}$ \\
$U\left( 1\right) _{L/R}$ & $\quad \sim \quad $ & $U\left( 1\right) \pm
\tilde{U}\left( 1\right) $%
\end{tabular}%
\end{equation}

In standard formulation of NCFT$_{2}^{\mathtt{su}_{2}}$, the momentum vector
$\boldsymbol{p}$ of the compactified string field on a circle is a real 2D
vector $\left( \mathfrak{p}^{\text{\textsc{\b{m}}}}\right) \in \mathbb{R}%
^{1,1}$ splitting like $\mathfrak{p}^{\text{\textsc{\b{m}}}}=\left(
p_{L},p_{R}\right) $. Its quantized\ values form a lattice ${\large \Lambda }%
_{\mathcal{M}_{1}}$ contained in $\mathbb{R}^{1,1}$ with pseudo-metric $\eta
_{\text{\textsc{\b{m}\b{n}}}}=diag(+,-).$ This lattice is defined by the
condition on the spin $\left( p_{L}^{2}-p_{R}^{2}\right) /2\in \mathbb{Z}$
that we write as follows%
\begin{equation}
{\large \Lambda }_{\mathcal{M}_{1}}=\left\{ \mathfrak{p}^{\text{\textsc{\b{m}%
}}}\in \mathbb{R}^{1,1}\quad ;\quad \mathfrak{p}^{\text{\textsc{\b{m}}}}\eta
_{\text{\textsc{\b{m}\b{n}}}}\mathfrak{p}^{\text{\textsc{\b{n}}}}\in 2%
\mathbb{Z}\right\}  \label{la}
\end{equation}%
where
\begin{equation}
\mathfrak{Q}_{\mathcal{M}_{1}}:=\mathfrak{p}^{\text{\textsc{\b{m}}}}\eta _{%
\text{\textsc{\b{m}\b{n}}}}\mathfrak{p}^{\text{\textsc{\b{n}}}%
}=p_{L}^{2}-p_{R}^{2}
\end{equation}
The values of the quantized $p_{L}$ and $p_{R}$ are labeled by two integers:
n and w; they are given by \textrm{\cite{RPP,KN,FUR}}
\begin{equation}
\begin{tabular}{lllll}
$p_{L}$ & $=$ & $\frac{n}{2R}+wR$ & $\in $ & $\mathbb{R}$ \\
$p_{R}$ & $=$ & $\frac{n}{2R}-wR$ & $\in $ & $\mathbb{R}$%
\end{tabular}%
\qquad ,\qquad
\begin{tabular}{lll}
$p_{L}$ & $:=$ & $\left( p_{L}\right) _{n,w}$ \\
$p_{R}$ & $:=$ & $\left( p_{R}\right) _{n,w}$%
\end{tabular}
\label{plpr}
\end{equation}%
the integer $n$ describes the Kaluza-Klein (KK) modes of the compactified
string and the parameter $w$ defines the winding number. By setting\textrm{\
}$\mathfrak{p}^{\text{\textsc{\b{m}}}}=E_{\text{.\textsc{a}}}^{\text{\textsc{%
\b{m}}}}m^{\text{\textsc{a}}}$ and $m^{\text{\textsc{a}}}=(n,w),$ we have
\begin{equation}
E_{\text{ \textsc{a}}}^{\text{\textsc{\b{m}}}}=\left(
\begin{array}{cc}
\frac{1}{2R} & R \\
\frac{1}{2R} & -R%
\end{array}%
\right)  \label{EAM}
\end{equation}%
with modulus independent determinant $\det E_{\text{ \textsc{a}}}^{\text{%
\textsc{\b{m}}}}=-1$ and global quadratic form as%
\begin{equation}
\mathfrak{Q}_{\mathcal{M}_{1}}=m^{\text{\textsc{a}}}\mathbf{Q}_{\text{%
\textsc{ab}}}m^{\text{\textsc{b}}}\qquad ,\qquad \mathbf{Q}_{\text{\textsc{ab%
}}}=E_{\text{\textsc{a}}}^{\text{ \textsc{\b{m}}}}\eta _{\text{\textsc{\b{m}%
\b{n}}}}E_{\text{ \textsc{b}}}^{\text{\textsc{\b{n}}}}
\end{equation}%
The matrix $\mathbf{Q}_{\text{\textsc{ab}}}$ is invariant under $\mathcal{O}%
(1,1,\mathbb{R})/\left[ \mathcal{O}(1)\times \mathcal{O}(1)\right] $ and the
integral form $m^{\text{\textsc{a}}}\mathbf{Q}_{\text{\textsc{ab}}}m^{\text{%
\textsc{b}}}$ is unchanging under $\mathcal{O}(1,1,\mathbb{Z}).$

Using eq(\ref{plpr}), we calculate the useful relations%
\begin{equation}
\begin{tabular}{lll}
$p_{L}^{2}$ & $=$ & $\frac{n^{2}}{4R^{2}}+nw+w^{2}R^{2}$ \\
$p_{R}^{2}$ & $=$ & $\frac{n^{2}}{4R^{2}}-nw+w^{2}R^{2}$%
\end{tabular}%
\qquad \Rightarrow \qquad
\begin{tabular}{lll}
$p_{L}^{2}+p_{R}^{2}$ & $=$ & $\frac{n^{2}}{2R^{2}}+2w^{2}R^{2}$ \\
$p_{L}^{2}-p_{R}^{2}$ & $=$ & $2nw$%
\end{tabular}
\label{LP}
\end{equation}%
showing that $\mathfrak{Q}_{\mathcal{M}_{1}}=p_{L}^{2}-p_{R}^{2}$ is given
by the \emph{coupling} of KK modes \{n\} and the winding \{w\}. Moreover, in
terms of $\mathfrak{p}^{\text{\textsc{\b{m}}}}=(p_{L},p_{R}),$ the global
quadratic form $\mathfrak{Q}_{\mathcal{M}_{1}}:=\mathfrak{Q}\left( \mathfrak{%
p},\mathfrak{p}\right) $ can be formulated like
\begin{equation}
\mathfrak{Q}_{\mathcal{M}_{1}}=\mathfrak{p}^{\text{\textsc{\b{m}}}}\eta _{%
\text{\textsc{\b{m}\b{n}}}}\mathfrak{p}^{\text{\textsc{\b{n}}}}=2nw
\label{qf}
\end{equation}%
In addition to $\mathfrak{Q}_{\mathcal{M}_{1}},$ we also have a twin
quadratic form $\mathcal{H}_{\mathcal{M}_{1}}:=\mathcal{H}\left( \mathfrak{p}%
,\mathfrak{p}\right) $ given by%
\begin{equation}
\mathcal{H}_{\mathcal{M}_{1}}:=p_{L}^{2}+p_{R}^{2}\geq 0
\end{equation}%
it is invariant under $\mathcal{O}\left( 1,\mathbb{R}\right) _{L}\times
\mathcal{O}\left( 1,\mathbb{R}\right) _{R}$ with action $p_{L}\rightarrow
\pm p_{L}$ and $p_{R}\rightarrow \pm p_{R}$. These $\mathfrak{Q}_{\mathcal{M}%
_{1}}$ and $\mathcal{H}_{\mathcal{M}_{1}}$ constitute basic ingredients
required for the derivation of the properties of NCFT$_{2}^{\mathbf{su}%
_{2}}; $ in particular its partition function $Z\left( \tau ,\bar{\tau}%
,R\right) $, its average $\left\langle Z\left( \tau ,\bar{\tau},R\right)
\right\rangle _{\mathcal{M}_{1}}$ and its holographic dual Z$_{grav}$.

Equipped with these tools, we proceed to present our proposal for NCFT$_{2}^{%
\mathbf{su}_{2}};$ to be extended later for higher dimensional Lie algebras
beyond su(2).

\subsection{Refined Narain CFT$_{2}^{\mathtt{su}_{2}}$}

In pursuit of classifying Narain CFT$_{2}$'s and their \textrm{%
generalisations} first considered in \textrm{\cite{JT,RPP}}, we give below\
a refined parametrisation of (\ref{plpr}) by involving both the simple root $%
\mathbf{\alpha }$ (root lattice) and the fundamental weight $\mathbf{\lambda
}$ vector (weight lattice) of the su(2) Lie algebra. The Lie group SU(2)
associated with this algebra is imagined here in terms of the 3-sphere
\begin{equation}
SU(2)\simeq \mathbb{S}^{3}\qquad ,\qquad \mathbb{S}^{3}\simeq \mathbb{S}%
^{1}\times \mathbb{S}^{2}\qquad ,\qquad \mathbb{S}^{1}\simeq U(1)
\end{equation}%
Below, the fiber $\mathbb{S}^{1}$ (with radius R) will be regarded as
Narain's circle of string compactification; as for the base $\mathbb{S}^{2},$
it will be used for geometric interpretations.

\subsubsection{NCFT$_{2}^{\mathtt{su}_{2}}$ at the point $R=1/2$}

The refined parametrisation we propose for the NCFT$_{2}^{\mathtt{su}_{2}}$
description can be introduced in two steps, starting from a special point of
$\mathcal{M}_{1}$ as follows:

\begin{description}
\item[$\left( \mathbf{1}\right) $] Consider first the particular case of a
circle $\mathbb{S}^{1}$ with radius $R=1/2$; this choice corresponds to
sitting at a particular point in the moduli space $\left] 0,\infty \right[ $
that can be split like
\begin{equation}
\mathcal{M}_{1}\simeq \left[ 0,1\right] \cup \left] 1,\infty \right[
\end{equation}%
which we designate as $\mathcal{M}_{\mathtt{su}_{2}}.$ By substituting into (%
\ref{plpr}), we get the local values%
\begin{equation}
\begin{tabular}{lll}
$p_{L}$ & $=$ & $n+\frac{1}{2}w$ \\
$p_{R}$ & $=$ & $n-\frac{1}{2}w$%
\end{tabular}%
\qquad \Rightarrow \qquad
\begin{tabular}{lll}
$p_{L}^{2}+p_{R}^{2}$ & $=$ & $2n^{2}+\frac{1}{2}w^{2}$ \\
$p_{L}^{2}-p_{R}^{2}$ & $=$ & $2nw$%
\end{tabular}
\label{eq}
\end{equation}%
with $\mathfrak{p}^{\text{\textsc{\b{m}}}}=E_{\text{.\textsc{a}}}^{\text{%
\textsc{\b{m}}}}m^{\text{\textsc{a}}}$ and Zweibein as follows%
\begin{equation}
E_{\text{ \textsc{a}}}^{\text{\textsc{\b{m}}}}=\left(
\begin{array}{cc}
1 & +\frac{1}{2} \\
1 & -\frac{1}{2}%
\end{array}%
\right) \qquad ,\qquad \mathbf{Q}_{\text{\textsc{ab}}}=E_{\text{\textsc{a}}%
}^{\text{ \textsc{\b{m}}}}\eta _{\text{\textsc{\b{m}\b{n}}}}E_{\text{
\textsc{b}}}^{\text{\textsc{\b{n}}}}=\left(
\begin{array}{cc}
0 & 1 \\
1 & 0%
\end{array}%
\right)  \label{zw}
\end{equation}%
The generic dependence on the modulus R in (\ref{EAM}) will be restored
later on.

\item[$\left( \mathbf{2}\right) $] Think about the above left $p_{L}$ and
the right $p_{R}$ components of momentum (\ref{eq}) in terms of the vectors $%
\boldsymbol{p}_{L}$ and $\boldsymbol{p}_{R}$ valued in the weight lattice of
su(2) and given by the following linear combinations
\begin{equation}
\begin{tabular}{lll}
$\boldsymbol{p}_{L}$ & $=$ & $\frac{1}{\sqrt{2}}\left( n\mathbf{\alpha }+w%
\mathbf{\lambda }\right) $ \\
$\boldsymbol{p}_{R}$ & $=$ & $\frac{1}{\sqrt{2}}\left( n\mathbf{\alpha }-w%
\mathbf{\lambda }\right) $%
\end{tabular}
\label{p}
\end{equation}%
Here, the $\mathbf{\alpha }$\ and the $\mathbf{\lambda }$\ are respectively
the simple root and the fundamental weight of su(2). From (\ref{p}), we
calculate
\begin{equation}
\begin{tabular}{lll}
$\boldsymbol{p}_{L}^{2}$ & $=$ & $\frac{1}{2}\left( n^{2}\mathbf{\alpha }%
^{2}+w^{2}\mathbf{\lambda }^{2}+2nw\mathbf{\alpha .\lambda }\right) $ \\
$\boldsymbol{p}_{R}^{2}$ & $=$ & $\frac{1}{2}\left( n^{2}\mathbf{\alpha }%
^{2}+w^{2}\mathbf{\lambda }^{2}-2nw\mathbf{\alpha .\lambda }\right) $%
\end{tabular}
\label{PP}
\end{equation}%
and then
\begin{equation}
\begin{tabular}{lllll}
$\boldsymbol{p}_{L}^{2}+\boldsymbol{p}_{R}^{2}$ & $=$ & $n^{2}\mathbf{\alpha
}^{2}+w^{2}\mathbf{\lambda }^{2}$ & $:=$ & $\mathcal{H}_{\mathtt{su}_{2}}$
\\
$\boldsymbol{p}_{L}^{2}-\boldsymbol{p}_{R}^{2}$ & $=$ & $2nw\mathbf{\alpha
.\lambda }$ & $:=$ & $\mathfrak{Q}_{\mathtt{su}_{2}}$%
\end{tabular}
\label{PLR}
\end{equation}%
By substituting $\mathbf{\alpha }^{2}=2$ and $\mathbf{\lambda }^{2}=1/2$ as
well as the usual duality $\mathbf{\alpha .\lambda }=1$, we recover
precisely the relations (\ref{eq}).
\end{description}

\paragraph{\textbf{A) Two comments}}

\begin{description}
\item[$\left( \mathbf{i}\right) $] From eq(\ref{PLR}), we observe that the
root/weight duality relation $\mathbf{\alpha .\lambda }=1$ in the su(2)
algebra captures indeed the \emph{coupling} between the KK and the winding
modes given by
\begin{equation}
\mathfrak{Q}_{\mathtt{su}_{2}}=2\left( n\mathbf{\alpha }\right) \mathbf{.}%
\left( w\mathbf{\lambda }\right)
\end{equation}%
For later generalisation, notice that the root/weight duality in finite Lie
algebras reads in general like $\mathbf{\alpha }_{i}^{\vee }\mathbf{.\lambda
}^{j}=\delta _{i}^{j}$ where the simple co-root $\mathbf{\alpha }_{i}^{\vee
} $ is defined as $2\mathbf{\alpha }_{i}/\mathbf{\alpha }_{i}^{2}.$ For
simply laced Lie algebra such as su(r+1), we have $\mathbf{\alpha }%
_{i}^{2}=2 $ implying $\mathbf{\alpha }_{i}^{\vee }=\mathbf{\alpha }_{i}.$
This property also applies for the so(2r) and exceptional E$_{r}$ Lie
algebras but does not hold for non simply laced B$_{r}$, C$_{r}$, F$_{4}$
and G$_{2}$.

\item[$\left( \mathbf{ii}\right) $] In our proposal, the zweibein $E_{\text{
\textsc{a}}}^{\text{\textsc{\b{m}}}}$ given by eq(\ref{zw}) reads formally
as follows%
\begin{equation}
\boldsymbol{E}_{\text{ \textsc{a}}}^{\text{\textsc{\b{m}}}}=\frac{1}{\sqrt{2}%
}\left(
\begin{array}{cc}
\mathbf{\alpha } & \mathbf{\lambda } \\
\mathbf{\alpha } & -\mathbf{\lambda }%
\end{array}%
\right)
\end{equation}%
with%
\begin{equation}
\mathbf{Q}_{\text{\textsc{ab}}}=\boldsymbol{E}_{\text{\textsc{a}}}^{\text{
\textsc{\b{m}}}}\eta _{\text{\textsc{\b{m}\b{n}}}}\boldsymbol{E}_{\text{
\textsc{b}}}^{\text{\textsc{\b{n}}}}
\end{equation}%
invariant under the $\mathcal{O}(1,1)$ action on the zweibein $\boldsymbol{E}%
_{\text{ \textsc{b}}}^{\text{\textsc{\b{n}}}}$ via orthogonal matrix
transformations $\mathcal{O}_{\text{ \textsc{\b{k}}}}^{\text{\textsc{\b{n}}}%
} $ which act as $\mathcal{O}_{\text{ \textsc{\b{k}}}}^{\text{\textsc{\b{n}}}%
}\boldsymbol{E}_{\text{ \textsc{b}}}^{\text{\textsc{\b{k}}}}$ and preserve $%
\eta _{\text{\textsc{\b{m}\b{n}}}};$ that is ($\mathcal{O}^{T})_{\text{%
\textsc{\b{m}}}}^{\text{ \textsc{\b{p}}}}\eta _{\text{\textsc{\b{p}\b{q}}}}%
\mathcal{O}_{\text{ \textsc{\b{n}}}}^{\text{\textsc{\b{q}}}}=\eta _{\text{%
\textsc{\b{m}\b{n}}}}.$
\end{description}

\paragraph{\textbf{B) Quadratic forms and su(2) Lie algebra}\newline
}

Focussing on eq(\ref{PLR}), we learn that $\mathfrak{Q}_{L}^{\mathtt{su}%
_{2}}:=\boldsymbol{p}_{L}^{2}$ and $\mathfrak{Q}_{R}^{\mathtt{su}_{2}}:=%
\boldsymbol{p}_{R}^{2}$ \ are quadratic forms that can be also presented
like $\mathfrak{Q}_{L/R}^{\mathtt{su}_{2}}=\boldsymbol{m}^{T}.\boldsymbol{Q}%
_{L/R}^{\mathtt{su}_{2}}.\boldsymbol{m}$ where $\boldsymbol{m}^{T}=\left(
n,w\right) $; they are fully defined by the 2$\times $2 matrices $%
\boldsymbol{Q}_{L/R}^{\mathtt{su}_{2}}$ given by
\begin{equation}
\boldsymbol{Q}_{L}^{\mathtt{su}_{2}}=\frac{1}{2}\left(
\begin{array}{cc}
\mathbf{\alpha .\alpha } & \mathbf{\alpha .\lambda } \\
\mathbf{\lambda .\alpha } & \mathbf{\lambda .\lambda }%
\end{array}%
\right) \qquad ,\qquad \boldsymbol{Q}_{R}^{\mathtt{su}_{2}}=\frac{1}{2}%
\left(
\begin{array}{cc}
\mathbf{\alpha .\alpha } & -\mathbf{\alpha .\lambda } \\
-\mathbf{\lambda .\alpha } & \mathbf{\lambda .\lambda }%
\end{array}%
\right)
\end{equation}%
By setting
\begin{equation}
\begin{tabular}{lll}
$\mathcal{H}_{\mathtt{su}_{2}}$ & $=$ & $\mathfrak{Q}_{L}^{\mathtt{su}_{2}}+%
\mathfrak{Q}_{R}^{\mathtt{su}_{2}}$ \\
$\mathfrak{Q}_{\mathtt{su}_{2}}$ & $=$ & $\mathfrak{Q}_{L}^{\mathtt{su}_{2}}-%
\mathfrak{Q}_{R}^{\mathtt{su}_{2}}$%
\end{tabular}%
\end{equation}%
and in bold the matrices%
\begin{equation}
\begin{tabular}{lll}
$\mathbf{H}_{\mathtt{su}_{2}}$ & $=$ & $\boldsymbol{Q}_{L}^{\mathtt{su}_{2}}+%
\boldsymbol{Q}_{R}^{\mathtt{su}_{2}}$ \\
$\mathbf{Q}_{\mathtt{su}_{2}}$ & $=$ & $\boldsymbol{Q}_{L}^{\mathtt{su}_{2}}-%
\boldsymbol{Q}_{R}^{\mathtt{su}_{2}}$%
\end{tabular}%
\end{equation}%
we end up with two interesting matrices%
\begin{equation}
\boldsymbol{H}_{\mathtt{su}_{2}}=\left(
\begin{array}{cc}
\mathbf{\alpha }^{2} & \mathbf{0} \\
\mathbf{0} & \mathbf{\lambda }^{2}%
\end{array}%
\right) \qquad ,\qquad \mathbf{Q}_{\mathtt{su}_{2}}=\left(
\begin{array}{cc}
0 & \mathbf{\alpha .\lambda } \\
\mathbf{\lambda .\alpha } & 0%
\end{array}%
\right)  \label{hd}
\end{equation}%
where $\boldsymbol{H}_{\mathtt{su}_{2}}$ is diagonal and $\mathbf{Q}_{%
\mathtt{su}_{2}}$ off diagonal. The form of the matrix $\mathbf{Q}_{\mathtt{%
su}_{2}}$ is particularly insightful: $\left( \mathbf{i}\right) $ it has
vanishing diagonal entries $\mathbf{Q}_{ii}^{\mathtt{su}_{2}}=0$ (even
integer); and as such it captures only the intersection between the cycles $%
\mathbf{\alpha }$ and $\mathbf{\lambda }$. $\left( \mathbf{ii}\right) $
Because $\mathbf{\alpha .\lambda }=1$, it results that
\begin{equation}
\left\vert \det \mathbf{Q}_{\mathtt{su}_{2}}\right\vert =\left\vert -\left(
\mathbf{\alpha .\lambda }\right) ^{2}\right\vert =1
\end{equation}%
\textrm{showing that} $\mathbf{Q}_{\mathtt{su}_{2}}$ \textrm{describes an
even unimodular lattice}. As anticipated by eq(\ref{qf}), we indeed have $%
\mathfrak{Q}_{\mathtt{su}_{2}}=\mathfrak{p}^{\text{\textsc{\b{m}}}}\eta _{%
\text{\textsc{\b{m}\b{n}}}}\mathfrak{p}^{\text{\textsc{\b{n}}}}=2nw$ which
in our proposal reads as follows%
\begin{equation}
\mathfrak{Q}_{\mathtt{su}_{2}}=\boldsymbol{m}^{T}.\boldsymbol{Q}_{\mathtt{su}%
_{2}}.\boldsymbol{m}\qquad ,\qquad \mathfrak{Q}_{\mathtt{su}_{2}}=\mathfrak{Q%
}_{\mathtt{su}_{2}}\left( \boldsymbol{m},\boldsymbol{m}\right)  \label{m}
\end{equation}%
\textrm{indicating that the lattice} (\ref{la}) is isomorphic to ${\large %
\Lambda }_{\mathtt{su}_{2}}$ with metric ${\large g}_{\text{\textsc{ab}}}$
given by $\boldsymbol{Q}_{\text{\textsc{ab}}}^{_{\mathtt{su}_{{\small 2}}}}$
up to an SL(2,$\mathbb{Z}$) transformation. Explicitly, we have
\begin{equation}
{\large \Lambda }_{\mathtt{su}_{2}}=\left\{ \boldsymbol{m}^{\text{\textsc{a}}%
}\in \mathbb{Z}^{2}\quad ;\quad \mathfrak{Q}_{\mathtt{su}_{2}}=\boldsymbol{m}%
^{\text{\textsc{a}}}{\large g}_{\text{\textsc{ab}}}\boldsymbol{m}^{\text{%
\textsc{b}}}\in 2\mathbb{Z}\right\}
\end{equation}%
with%
\begin{equation}
{\large g}_{\text{\textsc{ab}}}=\boldsymbol{Q}_{\text{\textsc{ab}}}^{_{%
\mathtt{su}_{{\small 2}}}}
\end{equation}
In addition to these relationships, and due to the property $\mathbf{\alpha }%
^{2}=2,$ the fundamental weight can be expressed like $\mathbf{\lambda }=%
\frac{1}{2}\mathbf{\alpha }$ implying a length $\mathbf{\lambda }^{2}=1/2;$
which is simply the inverse of $\mathbf{\alpha }^{2}\mathbf{.}$ Substituting
$\mathbf{\alpha }=2\mathbf{\lambda }$ in (\ref{p}), we find%
\begin{equation}
\begin{tabular}{lll}
$\boldsymbol{p}_{L}$ & $=$ & $\frac{1}{\sqrt{2}}\left( 2n+w\right) \mathbf{%
\lambda }$ \\
$\boldsymbol{p}_{R}$ & $=$ & $\frac{1}{\sqrt{2}}\left( 2n-w\right) \mathbf{%
\lambda }$%
\end{tabular}%
\qquad ,\qquad
\begin{tabular}{lll}
$\boldsymbol{p}_{L}.\mathbf{\alpha }$ & $=$ & $\frac{1}{\sqrt{2}}\left(
2n+w\right) $ \\
$\boldsymbol{p}_{R}.\mathbf{\alpha }$ & $=$ & $\frac{1}{\sqrt{2}}\left(
2n-w\right) $%
\end{tabular}%
\end{equation}%
we deduce that $\left( \mathbf{i}\right) $ the $\boldsymbol{p}_{L}$ and the $%
\boldsymbol{p}_{R}$ determine left/right shifts of the KK\ momentum $%
\boldsymbol{p}_{(n)}=2n\mathbf{\lambda }$ by the amounts $\pm w\mathbf{%
\lambda }$ arising from windings; and $\left( \mathbf{ii}\right) $ with
regard to the factorisation $\mathfrak{Q}_{\mathtt{su}_{2}}=\mathfrak{p}^{%
\text{\textsc{\b{m}}}}\eta _{\text{\textsc{\b{m}\b{n}}}}\mathfrak{p}^{\text{%
\textsc{\b{n}}}}$ of eq(\ref{qf}), we have
\begin{equation}
\mathfrak{p}^{\text{\textsc{\b{m}}}}=\left(
\begin{array}{c}
\frac{1}{\sqrt{2}}\boldsymbol{p}_{L}.\mathbf{\alpha } \\
\frac{1}{\sqrt{2}}\boldsymbol{p}_{R}.\mathbf{\alpha }%
\end{array}%
\right) =\left(
\begin{array}{c}
\frac{2}{\sqrt{2}}\boldsymbol{p}_{L}.\mathbf{\lambda } \\
\frac{2}{\sqrt{2}}\boldsymbol{p}_{R}.\mathbf{\lambda }%
\end{array}%
\right)  \label{pm}
\end{equation}

\subsubsection{NCFT$_{2}^{\mathtt{su}_{2}}$ at generic values of radius R}

In the refined proposal (\ref{p}), we set $R=1/2$ and used $\left( \mathbf{a}%
\right) $ the simple root $\mathbf{\alpha }$ of su(2) with length $\mathbf{%
\alpha }^{2}=2;$ and $\left( \mathbf{b}\right) $ the associated fundamental
weight $\mathbf{\lambda }$ with length $\mathbf{\lambda }^{2}=1/2.$ For
generic values of R, imagined in what follows in terms of the positive
variable $x=1/\left( 2R\right) $, we consider the scaled vector $\mathbf{%
\beta }=x\mathbf{\alpha }$ instead of $\mathbf{\alpha ;}$ and the scaled
weight $\mathbf{\chi }=$ $\mathbf{\lambda }/x$ satisfying the duality
relation $\mathbf{\beta }.\mathbf{\chi }=1$ with%
\begin{equation}
\begin{tabular}{lllll}
$\mathbf{\beta }\left( x\right) $ & $=$ & $\frac{1}{(2R)}\mathbf{\alpha }$ &
$\mathbf{:}=$ & $x\mathbf{\alpha }$ \\
$\mathbf{\chi }\left( x\right) $ & $=$ & $\left( 2R\right) \mathbf{\lambda }$
& $\mathbf{:}=$ & $\frac{1}{x}\mathbf{\lambda }$%
\end{tabular}
\label{at}
\end{equation}%
and the lengths
\begin{equation}
\mathbf{\beta }^{2}\left( x\right) =2x^{2}\qquad ,\qquad \mathbf{\chi }%
^{2}\left( x\right) =\frac{1}{2x^{2}}  \label{ta}
\end{equation}%
With these scaled cycles, the proposal (\ref{p}) extends as follows
\begin{equation}
\begin{tabular}{lll}
$\boldsymbol{p}_{L}\left( x\right) $ & $=$ & $\frac{1}{\sqrt{2}}\left( n%
\mathbf{\beta }+w\mathbf{\chi }\right) $ \\
$\boldsymbol{p}_{R}\left( x\right) $ & $=$ & $\frac{1}{\sqrt{2}}\left( n%
\mathbf{\beta }-w\mathbf{\chi }\right) $%
\end{tabular}
\label{ppr}
\end{equation}%
They are function of the modulus $x.$ For this generic case, eqs(\ref{hd})
get mapped into , : $\ $%
\begin{equation}
\begin{tabular}{lllll}
$\boldsymbol{H}_{\mathtt{su}_{2}}\left( x\right) $ & $=$ & $\left(
\begin{array}{cc}
x^{2}\mathbf{\alpha }^{2} & \mathbf{0} \\
\mathbf{0} & \frac{1}{x^{2}}\mathbf{\lambda }^{2}%
\end{array}%
\right) $ & $=$ & $\left(
\begin{array}{cc}
2x^{2} & \mathbf{0} \\
\mathbf{0} & \frac{1}{2x^{2}}%
\end{array}%
\right) $ \\
&  &  &  &  \\
$\mathbf{Q}_{\mathtt{su}_{2}}\left( x\right) $ & $=$ & $\left(
\begin{array}{cc}
0 & \mathbf{\beta .\chi } \\
\mathbf{\chi .\beta } & 0%
\end{array}%
\right) $ & $=$ & $\left(
\begin{array}{cc}
0 & \mathbf{\alpha .\lambda } \\
\mathbf{\lambda .\alpha } & 0%
\end{array}%
\right) $%
\end{tabular}%
\end{equation}
from which we learn that $\left( \mathbf{i}\right) $ the $\boldsymbol{H}_{%
\mathtt{su}_{2}}$ is a function of the modulus x with $\det \boldsymbol{H}_{%
\mathtt{su}_{2}}=1$ and $\mathbf{Q}_{\mathtt{su}_{2}}^{-1}\boldsymbol{H}_{%
\mathtt{su}_{2}}\mathbf{Q}_{\mathtt{su}_{2}}=\boldsymbol{H}_{\mathtt{su}%
_{2}}^{-1}$, while $\left( \mathbf{ii}\right) $ the $\mathbf{Q}_{\mathtt{su}%
_{2}}$ is x- independent and unimodular
\begin{equation}
\left\vert \det \mathbf{Q}_{\mathtt{su}_{2}}\right\vert =\left\vert -\left(
\mathbf{\beta .\chi }\right) ^{2}\right\vert =\left\vert -\left( \mathbf{%
\alpha .\lambda }\right) ^{2}\right\vert =1
\end{equation}

In what follows, we will often perform calculations at the point $x=1$ in
the moduli space of the Narain CFT$_{2}^{(\mathtt{su}_{2})}$. The
implementation of the modulus $x$ is straightforward and will be applied
when needed for interpretations.

\subsection{Results for NCFT$_{2}^{\mathtt{su}_{2}}$}

To summarize the above description within the framework of our proposal (\ref%
{p}) and (\ref{ppr}), we collect below characteristic results on NCFT$_{2}^{%
\mathtt{su}_{2}}$ in the following list of points:

\begin{description}
\item[$\left( \mathbf{1}\right) $] \textbf{Masses as volume of cycles}: The
positive quadratic form $\mathcal{H}_{\mathtt{su}_{2}}=\boldsymbol{p}%
_{L}^{2}+\boldsymbol{p}_{R}^{2}$ is given by the sum of the self
intersections of \newline
$\left( \mathbf{i}\right) $ the SU(2)$_{\mathbf{\beta }}$- cycle $\mathcal{C}%
_{\mathbf{\beta }}$ generated by the scaled simple root $\mathbf{\beta }=x%
\mathbf{\alpha };$ and \newline
$\left( \mathbf{ii}\right) $ the SU(2)$_{\mathbf{\chi }}$- cycle $\mathcal{C}%
_{\mathbf{\chi }}$ spanned by the scaled fundamental weight $\mathbf{\chi }%
=x^{-1}\mathbf{\lambda }.$\newline
We have%
\begin{equation}
\begin{tabular}{lllllll}
$\mathcal{C}_{\mathbf{\beta }}.\mathcal{C}_{\mathbf{\beta }}$ & $=$ & $%
\mathbf{\beta .\beta }$ & $=$ & $x^{2}\mathbf{\alpha }\mathbf{.}\mathbf{%
\alpha }$ & $=$ & $x^{2}\mathcal{C}_{\mathbf{\alpha }}.\mathcal{C}_{\mathbf{%
\alpha }}$ \\
$\mathcal{C}_{\mathbf{\chi }}.\mathcal{C}_{\mathbf{\chi }}$ & $=$ & $\mathbf{%
\chi .\chi }$ & $=$ & $x^{-2}\mathbf{\lambda }\mathbf{.}\mathbf{\lambda }$ &
$=$ & $x^{-2}\mathcal{C}_{\mathbf{\lambda }}.\mathcal{C}_{\mathbf{\lambda }}$%
\end{tabular}%
\end{equation}%
implying%
\begin{equation}
Area\left( \mathcal{C}_{\mathbf{\beta }}\right) =2x^{2}\qquad ,\qquad
Area\left( \mathcal{C}_{\mathbf{\chi }}\right) =\frac{1}{2x^{2}}
\end{equation}%
By thinking of $\boldsymbol{p}_{L/R}^{2}$ in terms of squared masses $%
\mathbf{\mu }_{L/R}^{2},$ the $\mathcal{H}_{su_{2}}$ is, physically
speaking, nothing but the total squared mass $\mathbf{\mu }_{tot}^{2}=%
\mathbf{\mu }_{L}^{2}+\mathbf{\mu }_{R}^{2}$ of the KK and the winding
modes. Substituting $\boldsymbol{p}_{L/R}^{2}$ by their values, we obtain%
\begin{equation}
\mathbf{\mu }_{tot}^{2}=2x^{2}+\frac{1}{2x^{2}}\qquad \Leftrightarrow \qquad
\mathbf{\mu }_{tot}^{2}=\mathcal{C}_{\mathbf{\beta }}^{2}+\mathcal{C}_{%
\mathbf{\chi }}^{2}
\end{equation}

\item[$\left( \mathbf{2}\right) $] \textbf{Gap energy as intersection of
cycles}: The relative quantity $\mathfrak{Q}_{\mathtt{su}_{2}}=\boldsymbol{p}%
_{L}^{2}-\boldsymbol{p}_{R}^{2}$, which is equal to $2\left( n\mathbf{\beta }%
\right) .\left( w\mathbf{\chi }\right) $, is given by the intersection
between the cycle $n\mathcal{C}_{\mathbf{\beta }}$ and the cycle $w\mathcal{C%
}_{\mathbf{\chi }}.$ We have%
\begin{equation}
\mathcal{C}_{\mathbf{\beta }}.\mathcal{C}_{\mathbf{\chi }}=\mathbf{\beta
.\chi }=\mathbf{\alpha .\lambda }=\mathcal{C}_{\mathbf{\alpha }}.\mathcal{C}%
_{\mathbf{\lambda }}=1  \label{CY}
\end{equation}%
Physically, this represents the gap mass $\mathbf{\mu }_{L}^{2}-\mathbf{\mu }%
_{R}^{2}$ between the left and the right moving particle states. We denote
this gap by $\epsilon _{gap}^{\mathtt{su}_{2}}$ following the terminology
used in the physics of topological insulators and semiconductors \textrm{%
\cite{TW,TDS}}; it is given by $\epsilon _{gap}^{\mathtt{su}_{2}}=2n\times w.
$ Notice that a vanishing gap $\epsilon _{gap}^{\mathtt{su}_{2}}=0$
corresponds to the equality
\begin{equation}
\mathbf{\mu }_{L}^{2}=\mathbf{\mu }_{R}^{2}
\end{equation}%
and the constraint $n\times w=0$ which can be satisfied either by the KK
zero mode $n=0$ or the winding zero mode $w=0.$ From the perspective of (\ref%
{CY}), we have two possibilities $\left( \mathbf{i}\right) $ either the
contraction $\mathcal{C}_{\mathbf{\beta }}\rightarrow 0$ with the
decompactification $\mathcal{C}_{\mathbf{\chi }}\rightarrow \infty ;$ or $%
\left( \mathbf{ii}\right) $ the decompactification $\mathcal{C}_{\mathbf{%
\beta }}\rightarrow \infty $ with the contraction $\mathcal{C}_{\mathbf{\chi
}}\rightarrow 0$ \textrm{\cite{CC}}$.$

\item[$\left( \mathbf{3}\right) $] \textbf{Root}\emph{\ }$\Lambda _{\mathrm{r%
}}^{\mathtt{su}_{2}}$\emph{\ }\textbf{and weight}\emph{\ }$\Lambda _{\mathrm{%
w}}^{\mathtt{su}_{2}}$\emph{\ }\textbf{lattices}: By using the canonical
variables $\boldsymbol{p}_{1}=\left( \boldsymbol{p}_{L}+\boldsymbol{p}%
_{R}\right) /\sqrt{2}$ and $\boldsymbol{p}_{2}=\left( \boldsymbol{p}_{L}-%
\boldsymbol{p}_{R}\right) /\sqrt{2}$ in addition to substituting $%
\boldsymbol{p}_{L}$ and $\boldsymbol{p}_{R}$ by their values, we end up with
new momentum expressions that exhibit the dependencies on both the weight
vector $\mathbf{\lambda }$ and the simple root $\mathbf{\alpha },$%
\begin{equation}
\begin{tabular}{lll}
$\boldsymbol{p}_{1}$ & $=$ & $n\mathbf{\alpha }$ \\
$\boldsymbol{p}_{2}$ & $=$ & $w\mathbf{\lambda }$%
\end{tabular}%
\qquad ,\qquad
\begin{tabular}{lll}
$n$ & $=$ & $\boldsymbol{p}_{1}.\mathbf{\lambda }$ \\
$w$ & $=$ & $\boldsymbol{p}_{2}.\mathbf{\alpha }$%
\end{tabular}
\label{p1}
\end{equation}%
indicating in turn that the free fields $X_{L}$ and $X_{R}$ can be similarly
promoted to vectors $\boldsymbol{X}_{L}$ and $\boldsymbol{X}_{R}$ with
components $U$ and $V$ along the $\mathbf{\alpha }$ and $\mathbf{\lambda }$
directions as follows%
\begin{equation}
\begin{tabular}{lll}
$\boldsymbol{X}_{L}$ & $=$ & $\frac{1}{\sqrt{2}}\left( U\mathbf{\alpha }+V%
\mathbf{\lambda }\right) $ \\
$\boldsymbol{X}_{R}$ & $=$ & $\frac{1}{\sqrt{2}}\left( U\mathbf{\alpha }-V%
\mathbf{\lambda }\right) $%
\end{tabular}%
\qquad ,\qquad
\begin{tabular}{lll}
$U\mathbf{\alpha }$ & $=$ & $\frac{1}{\sqrt{2}}\left( \boldsymbol{X}_{L}+%
\boldsymbol{X}_{R}\right) $ \\
$V\mathbf{\lambda }$ & $=$ & $\frac{1}{\sqrt{2}}\left( \boldsymbol{X}_{L}-%
\boldsymbol{X}_{R}\right) $%
\end{tabular}%
\end{equation}%
and metric%
\begin{equation}
\mathcal{G}_{\text{\textsc{ab}}}=\left(
\begin{array}{cc}
2x^{2} & 0 \\
0 & \frac{1}{2x^{2}}%
\end{array}%
\right) =\left(
\begin{array}{cc}
\frac{1}{2R^{2}} & 0 \\
0 & 2R^{2}%
\end{array}%
\right)
\end{equation}%
See \textrm{appendix C for the full derivation of this metric and other
T-duality properties}. Also, they show that $\boldsymbol{p}_{1}$ labels
precisely the set $\Lambda _{\mathrm{r}}^{\mathtt{su}_{2}}$ which is the
su(2) root lattice while $\boldsymbol{p}_{2}$ parameterises $\Lambda _{%
\mathrm{w}}^{\mathtt{su}_{2}}$ defining the weight lattice. By using (\ref%
{pm}), we also have%
\begin{equation}
\mathfrak{p}^{\text{\textsc{\b{m}}}}=\left(
\begin{array}{c}
n+\frac{1}{2}w \\
n-\frac{1}{2}w%
\end{array}%
\right)  \label{mp}
\end{equation}%
with even integer $\mathfrak{Q}_{\mathtt{su}_{2}}=\mathfrak{p}^{\text{%
\textsc{\b{m}}}}\eta _{\text{\textsc{\b{m}\b{n}}}}\mathfrak{p}^{\text{%
\textsc{\b{n}}}}$. From the relations (\ref{p1}) with $\mathbf{\alpha }=2%
\mathbf{\lambda }$, we learn that the weight\ lattice $\Lambda _{\mathrm{w}%
}^{\mathtt{su}_{2}},$ with sites $\boldsymbol{p}_{2}=w\mathbf{\lambda ,}$
can be decomposed into two isomorphic sublattices as
\begin{equation}
\boldsymbol{p}_{2}^{(even)}=\left( 2w^{\prime }\right) \mathbf{\lambda }%
\qquad ,\qquad \boldsymbol{p}_{2}^{(odd)}=\left( 2w^{\prime }-1\right)
\mathbf{\lambda }
\end{equation}%
with $w^{\prime }$ integer. These sublattices are also isomorphic to the
root lattice $\Lambda _{\mathrm{r}}^{\mathtt{su}_{2}}$. In consequence, the
discriminant $D_{\mathtt{su}_{2}}=\Lambda _{\mathrm{w}}^{\mathtt{su}%
_{2}}/\Lambda _{\mathrm{r}}^{\mathtt{su}_{2}}$ is isomorphic to the group $%
\mathbb{Z}_{2}.$

\item[$\left( \mathbf{4}\right) $] $\mathfrak{Q}_{\mathtt{su}_{2}}$ \textbf{%
and} $\mathcal{H}_{\mathtt{su}_{2}}$ \textbf{as slopes in }$\mathbb{H}_{+}$:
Considering the matrices $\boldsymbol{Q}_{L/R}$\ given by the relations (\ref%
{hd}) and setting $\left( m_{1},m_{2}\right) =\left( n,w\right) $, we have
the useful quantities: \newline
$\left( \mathbf{i}\right) $ the associated quadratic forms%
\begin{equation}
\begin{tabular}{lll}
$\mathfrak{Q}_{L}$ & $=$ & $\boldsymbol{m}^{T}.\boldsymbol{Q}_{L}.%
\boldsymbol{m}$ \\
$\mathfrak{Q}_{R}$ & $=$ & $\boldsymbol{m}^{T}.\boldsymbol{Q}_{R}.%
\boldsymbol{m}$%
\end{tabular}%
\qquad ,\qquad
\begin{tabular}{lll}
$\mathcal{H}_{\mathtt{su}_{2}}$ & $=$ & $\boldsymbol{m}^{T}.\boldsymbol{H}%
_{su_{2}}.\boldsymbol{m}$ \\
$\mathfrak{Q}_{\mathtt{su}_{2}}$ & $=$ & $\boldsymbol{m}^{T}.\boldsymbol{Q}%
_{su_{2}}.\boldsymbol{m}$%
\end{tabular}
\label{LR}
\end{equation}%
$\left( \mathbf{ii}\right) $ the characteristic matrix $\boldsymbol{A}_{%
\mathtt{su}_{2}}$ and the associated quadratic form $\boldsymbol{\Psi }_{%
\mathtt{su}_{2}}$ respectively given by the linear combination
\begin{equation}
\begin{tabular}{lll}
$\boldsymbol{A}_{\mathtt{su}_{2}}$ & $=$ & $i\pi \left( \tau \boldsymbol{Q}%
_{L}-\bar{\tau}\boldsymbol{Q}_{R}\right) $ \\
& $=$ & $i\pi \tau _{1}\boldsymbol{Q}_{\mathtt{su}_{2}}-\pi \tau _{2}%
\boldsymbol{H}_{\mathtt{su}_{2}}$%
\end{tabular}%
\qquad ,\qquad \boldsymbol{A}_{\mathtt{su}_{2}}=\left(
\begin{array}{cc}
-2\pi \tau _{2}x^{2} & i\pi \tau _{1} \\
i\pi \tau _{1} & \frac{-\pi \tau _{2}}{2x^{2}}%
\end{array}%
\right)
\end{equation}%
and%
\begin{equation}
\begin{tabular}{lll}
$\boldsymbol{\Psi }_{n,w}^{\mathbf{su}_{2}}\left( x\right) $ & $=$ & $%
\boldsymbol{m}^{T}.\boldsymbol{A}_{\mathtt{su}_{2}}.\boldsymbol{m}$ \\
& $=$ & $i\pi \tau _{1}\mathfrak{Q}_{\mathtt{su}_{2}}-\pi \tau _{2}\mathcal{H%
}_{\mathtt{su}_{2}}$%
\end{tabular}
\label{AS}
\end{equation}%
where $\tau =\tau _{1}+i\tau _{2}$ is the complex modulus of the 2-torus ($%
\tau _{2}>0$). In these relations, we have $\det \boldsymbol{A}_{\mathtt{su}%
_{2}}=$ $\pi ^{2}\left( \tau \bar{\tau}\right) $; it is independent on the
modulus $x$. \newline
We also have the exponential $\varrho _{n,w}^{\mathbf{su}_{2}}\left(
x\right) =\exp \boldsymbol{\Psi }_{n,w}^{\mathbf{su}_{2}}\left( x\right) $
reading as%
\begin{equation}
\varrho _{n,w}^{\mathbf{su}_{2}}\left( x\right) =e^{i\pi \tau _{1}2nw}\times
\exp \left[ -\pi \tau _{2}\left( 2x^{2}n^{2}+\frac{1}{2x^{2}}w^{2}\right) %
\right]  \label{SA}
\end{equation}%
with $\left\vert \varrho _{n,w}^{\mathbf{su}_{2}}\left( x\right) \right\vert
\geq 0.$ The exponential of $\boldsymbol{\Psi }_{n,w}^{\mathbf{su}%
_{2}}\left( x\right) $ is important for the calculation of the partition
function $Z_{\mathrm{su}_{2}}$ of Narain CFT$_{2}^{\mathbf{su}_{2}}$; it
shows that the $\mathfrak{Q}_{\mathtt{su}_{2}}$ and $\mathcal{H}_{\mathtt{su}%
_{2}}$ are respectively the "slope" in the $\tau _{1}$- and the $\tau _{2}$
directions of the 2-torus upper half plane fibration $\mathbb{H}%
_{+}=H_{+}\otimes Mat\left( 2\right) $,
\begin{equation}
\partial _{\tau _{1}}\boldsymbol{\Psi }_{n,w}^{\mathbf{su}_{2}}=i\pi
\mathfrak{Q}_{\mathtt{su}_{2}}\qquad ,\qquad \partial _{\tau _{2}}%
\boldsymbol{\Psi }_{n,w}^{\mathbf{su}_{2}}=-\pi \mathcal{H}_{\mathtt{su}_{2}}
\end{equation}%
Notice that for large values of $\tau _{2}$, we have for non zero KK and
winding modes ($nw\neq 0$), the remarkable property $\lim_{\tau
_{2}\rightarrow \infty }\varrho _{n,w}^{\mathbf{su}_{2}}=0;$ so we have
\begin{equation}
\lim_{\tau _{2}\rightarrow \infty }\varrho _{n,w}^{\mathbf{su}_{2}}=\delta
_{nw,0}
\end{equation}

\item[$\left( \mathbf{5}\right) $] \textbf{Siegel-Narain} $\Theta _{\mathrm{%
su}_{2}}$ and\ \textbf{Partition function }$Z_{\mathrm{su}_{2}}$: The
genus-one partition function of the Narain CFT$_{2}^{\mathbf{su}_{2}}$ is
defined in terms of the Dedekin eta function $\eta \left( \tau \right)
=q^{1/24}\dprod\nolimits_{n=1}^{\infty }\left( 1-q^{n}\right) $ with $%
q=e^{2\pi i\tau }$\ as follows
\begin{equation}
Z_{\mathrm{su}_{2}}\left[ \tau ,\bar{\tau};R\right] =\frac{1}{\left\vert
\eta \left( \tau \right) \right\vert ^{2}}\Theta _{\mathrm{su}_{2}}\left[
\tau ,\bar{\tau};R\right]  \label{PF}
\end{equation}%
with Siegel-Narain theta function $\Theta _{\mathrm{su}_{2}}=%
\sum_{p_{L},p_{R}}q^{\frac{1}{2}p_{L}^{2}}\bar{q}^{\frac{1}{2}p_{R}^{2}}$
with $\left( p_{L},p_{R}\right) $ belonging to ${\large \Lambda }_{\mathtt{su%
}_{2}}.$ This function $\Theta _{\mathrm{su}_{2}}$ reads explicitly as
follows
\begin{eqnarray}
\Theta _{\mathrm{su}_{2}}\left[ \tau ,\bar{\tau};R\right] &=&\dsum\limits_{%
\boldsymbol{m}\in \mathbb{Z}^{2}}q^{\frac{1}{2}\mathfrak{Q}_{L}\left(
\boldsymbol{m}\right) }\bar{q}^{\frac{1}{2}\mathfrak{Q}_{R}\left(
\boldsymbol{m}\right) }  \notag \\
&=&\dsum\limits_{\boldsymbol{m}\in \mathbb{Z}^{2}}e^{i\pi \tau _{1}\mathfrak{%
Q}_{\mathtt{su}_{2}}\left( \boldsymbol{m}\right) }\times e^{-\pi \tau _{2}%
\mathcal{H}_{\mathtt{su}_{2}}\left( \boldsymbol{m}\right) }
\end{eqnarray}%
it is known to obey the differential equation \textrm{\cite{1A}}%
\begin{equation}
\left[ \tau _{2}^{2}\left( \frac{\partial ^{2}}{\partial \tau _{1}^{2}}+%
\frac{\partial ^{2}}{\partial \tau _{2}^{2}}\right) +\tau _{2}\frac{\partial
}{\partial \tau _{2}}-\frac{1}{4}\left( R\frac{\partial }{\partial R}\right)
^{2}\right] \Theta _{\mathrm{su}_{2}}\left[ \tau ,\bar{\tau};R\right] =0
\label{des}
\end{equation}%
Using the partial Laplacians $\Delta _{H}=-\tau _{2}^{2}\left( \partial
_{\tau _{1}}^{2}+\partial _{\tau _{2}}^{2}\right) $ of the half upper plane H%
$_{+}$ and the $\Delta _{\mathcal{M}_{1}}=-\frac{1}{4}\left( R\partial
_{R}\right) ^{2}$ of the Narain moduli space, the above equation can be
concisely written like%
\begin{equation}
\left( \Delta _{H}-\tau _{2}\frac{\partial }{\partial \tau _{2}}-\Delta _{%
\mathcal{M}_{1}}\right) \Theta _{\mathrm{su}_{2}}\left[ \tau ,\bar{\tau};R%
\right] =0  \label{lap}
\end{equation}%
By substituting $p_{L}^{2}$ and $p_{L}^{2}$ with their expressions in terms
of the matrices $\boldsymbol{Q}_{L}$ and $\boldsymbol{Q}_{R}$ (\ref{LR}), we
can put the partition function (\ref{PF}) into the following form
\begin{equation}
Z_{\mathrm{su}_{2}}\left[ \tau ,\bar{\tau};R\right] =\frac{1}{\left\vert
\eta \left( \tau \right) \right\vert ^{2}}\dsum\limits_{\boldsymbol{m}\in
\mathbb{Z}^{2}}e^{\boldsymbol{\Psi }_{\mathtt{su}_{2}}\left( \boldsymbol{m}%
\right) }
\end{equation}%
with $\boldsymbol{\Psi }_{\mathtt{su}_{2}}\left( \boldsymbol{m}\right) $ as
in (\ref{AS}). For large values of $\tau _{2}$, we have the boundary value
\begin{equation}
\lim_{\tau _{2}\rightarrow \infty }\Theta _{\mathrm{su}_{2}}\left[ \tau ,%
\bar{\tau};R\right] =1
\end{equation}%
given by the KK and winding zero modes.

\item[$\left( \mathbf{6}\right) $] \textbf{Averaged partition function }$%
\left\langle Z_{\mathrm{su}_{2}}\right\rangle _{\mathcal{M}_{su_{2}}}$: From
eq(\ref{ta}), we can express the moduli space metric (\ref{KK1}) in terms of
the self intersection $\mathbf{\beta }^{2}\left( x\right) =2x^{2}$ as follows%
\begin{equation}
ds_{su_{2}}^{2}=\left( \mathcal{K}_{su_{2}}^{-1}d\mathcal{K}_{su_{2}}\right)
^{2}=\left( \frac{2dx}{x}\right) ^{2}  \label{KK2}
\end{equation}%
where we set $\mathcal{K}_{su_{2}}=\mathbf{\beta }^{2}$ which is the self
intersection of the cycle $\mathcal{C}_{\mathbf{\beta }}$; and similarly $%
\mathcal{K}_{su_{2}}^{-1}=\mathbf{\chi }^{2}.$ Using this metric, one seeks
to compute the average of the Siegel-Narain theta function given by
\begin{equation}
F_{\mathrm{su}_{2}}\left( \tau ,\bar{\tau}\right)
=\dint\nolimits_{1}^{\infty }\frac{dR}{2R}\Theta _{\mathrm{su}_{2}}\left[
\tau ,\bar{\tau};R\right]  \label{fsu}
\end{equation}%
However, because $\mathcal{M}_{\mathrm{su}_{2}}$ has infinite measure, this
average diverges \textrm{because }$\Theta _{\mathrm{su}_{2}}$\textrm{\
behaves like R \cite{1A}; other properties of }$F_{\mathrm{su}_{2}}\left(
\tau ,\bar{\tau}\right) $\textrm{\ are derived in appendix C. }By using (\ref%
{des}), \textrm{o}ne can still extract some formal information on $F_{%
\mathrm{su}_{2}}\left( \tau ,\bar{\tau}\right) .$ Integrating by part (\ref%
{fsu}), one ends up with a constraint relation on $F_{\mathrm{su}_{2}}\left(
\tau ,\bar{\tau}\right) $ up to the boundary term $\int_{1}^{\infty
}dR\partial _{R}(R\Theta ^{\mathrm{su}_{2}}).$ By disregarding this boundary
term, one would get the differential equation%
\begin{equation}
\left[ \tau _{2}^{2}\left( \frac{\partial ^{2}}{\partial \tau _{1}^{2}}+%
\frac{\partial ^{2}}{\partial \tau _{2}^{2}}\right) +\tau _{2}\frac{\partial
}{\partial \tau _{2}}\right] F_{\mathrm{su}_{2}}\left( \tau ,\bar{\tau}%
\right) =0
\end{equation}%
whose solution requires knowledge of its behavior for $\tau _{2}\rightarrow
\infty $ which by using (\ref{fsu}), we must have $\lim_{\tau
_{2}\rightarrow \infty }F_{\mathrm{su}_{2}}\left( \tau ,\bar{\tau}\right)
=1. $ Following \cite{1A}, this divergence can be only overcome for higher
dimensional Narain moduli space\textrm{\ }$\mathcal{M}_{d}$ with $d>2;$
\textrm{see appendix A}$\mathrm{.}$

\item[$\left( \mathbf{7}\right) $] \textbf{Modular properties}: Along with
the standard Siegel-Narain theta function $\Theta _{\mathrm{su}_{2}}$, one
can also define a generalised theta function $\Theta _{\boldsymbol{h}}^{%
\mathrm{su}_{{\small 2}}}$ labeled by elements $\boldsymbol{h}$ of the
discriminant group $\mathcal{D}_{\mathtt{su}_{{\small 2}}}=\Lambda _{\mathrm{%
w}}^{\mathtt{su}_{{\small 2}}}/\Lambda _{\mathrm{r}}^{\mathtt{su}_{{\small 2}%
}}\simeq \mathbb{Z}_{2}$ as follows:%
\begin{eqnarray}
\Theta _{\boldsymbol{h}}^{\mathrm{su}_{{\small 2}}}\left[ \tau ,\bar{\tau};R%
\right] &=&\dsum\limits_{\boldsymbol{m}\in \mathbb{Z}^{2}}\exp i\pi \left[
\tau \mathfrak{Q}_{L}^{\mathtt{su}_{{\small 2}}}\left( \boldsymbol{m}+%
\boldsymbol{h}\right) -\bar{\tau}\mathfrak{Q}_{R}^{\mathtt{su}_{{\small 2}%
}}\left( \boldsymbol{m}+\boldsymbol{h}\right) \right]  \notag \\
&=&\dsum\limits_{\boldsymbol{m}\in \mathbb{Z}^{2}}e^{i\pi \tau _{1}\mathfrak{%
Q}_{\mathrm{su}_{{\small 2}}}\left( \boldsymbol{m}+\boldsymbol{h}\right)
}\times e^{-\pi \tau _{2}\mathcal{H}_{\mathrm{su}_{{\small 2}}}\left(
\boldsymbol{m}+\boldsymbol{h}\right) }
\end{eqnarray}%
Under SL(2,$\mathbb{Z}$) transformations generated by $\left( \mathbf{i}%
\right) $ the translation $T:\tau \rightarrow \tau +1$ and $\left( \mathbf{ii%
}\right) $ the inversion $S:\tau \rightarrow -1/\tau $, we have the
following changes \textrm{\cite{RPP}}:%
\begin{equation}
\begin{tabular}{lllll}
$T$ & $:$ & $\Theta _{\boldsymbol{h}}^{\mathrm{su}_{{\small 2}}}\left[ \tau
+1,\bar{\tau}+1;R\right] $ & $=$ & $e^{i\pi \mathfrak{Q}_{\mathtt{su}_{%
{\small 2}}}\left( \boldsymbol{h},\boldsymbol{h}\right) }\Theta _{\mathbf{%
\eta }}^{\mathrm{su}_{{\small 2}}}\left[ \tau ,\bar{\tau};R\right] $ \\
$S$ & $:$ & $\Theta _{\boldsymbol{h}}^{\mathrm{su}_{{\small 2}}}\left[
-1/\tau ,-1/\bar{\tau};R\right] $ & $=$ & $\tau ^{\frac{1}{2}}\bar{\tau}^{%
\frac{1}{2}}\dsum\limits_{\boldsymbol{h}^{\prime }\in \mathcal{D}_{\mathrm{su%
}_{{\small 2}}}}e^{-2i\pi \mathfrak{Q}_{\mathrm{su}_{{\small 2}}}\left(
\boldsymbol{h},\boldsymbol{h}^{\prime }\right) }\Theta _{\boldsymbol{h}%
^{\prime }}^{\mathrm{su}_{{\small 2}}}\left[ \tau ,\bar{\tau};R\right] $%
\end{tabular}%
\end{equation}%
showing that $\Theta _{\boldsymbol{h}}^{\mathrm{su}_{{\small 2}}}$ is
\textrm{only }invariant under $T$ because $\mathfrak{Q}_{\mathtt{su}_{%
{\small 2}}}\left( \boldsymbol{h},\boldsymbol{h}\right) $ is an even
integer. The modular transformations of the partition function $Z_{%
\boldsymbol{h}}^{\mathrm{su}_{{\small 2}}}$ (\ref{PF}) associated with $%
\Theta _{\boldsymbol{h}}^{\mathrm{su}_{{\small 2}}}$ are given by
\begin{equation}
\begin{tabular}{lllll}
$T$ & $:$ & $Z_{\boldsymbol{h}}^{\mathrm{su}_{{\small 2}}}\left[ \tau +1,%
\bar{\tau}+1;R\right] $ & $=$ & $e^{i\pi \mathfrak{Q}_{\mathtt{su}_{{\small 2%
}}}\left( \boldsymbol{h},\boldsymbol{h}\right) }Z_{\boldsymbol{h}}^{\mathrm{%
su}_{{\small 2}}}\left[ \tau ,\bar{\tau};R\right] $ \\
$S$ & $:$ & $Z_{\boldsymbol{h}}^{\mathrm{su}_{{\small 2}}}\left[ -1/\tau ,-1/%
\bar{\tau};R\right] $ & $=$ & $\dsum\limits_{\boldsymbol{h}^{\prime }\in
\mathcal{D}_{\mathrm{su}_{{\small 2}}}}e^{-2i\pi \mathfrak{Q}_{\mathtt{su}%
_{2}}\left( \boldsymbol{h},\boldsymbol{h}^{\prime }\right) }Z_{\boldsymbol{h}%
^{\prime }}^{\mathrm{su}_{{\small 2}}}\left[ \tau ,\bar{\tau};R\right] $%
\end{tabular}%
\end{equation}%
From these transformations, it is clear that the genus-one partition
function $Z_{\boldsymbol{h}}^{\mathrm{su}_{{\small 2}}}$ with $\boldsymbol{h}%
=0$ is modular invariant under $SL(2,\mathbb{Z})$ transformation $\tau
\rightarrow \left( a\tau +b\right) /\left( c\tau +d\right) $ with $ad-bc=1.$
It is important to note for later use that the T-invariance requires an
\emph{even integer} lattice $\Lambda _{d_{L},d_{R}}$ while the invariance S
implies that $\Lambda _{d_{L},d_{R}}$ must be also \emph{self dual}. Notice
that even and self dual lattices exist if and only if $d_{L}-d_{R}=8l$ for
integer $l$. In type II strings, we must have $d_{L}-d_{R}=0$ and for
heterotic string we should have $d_{L}-d_{R}=16$ \cite{KN}.
\end{description}

\section{Rank two Narain CFT$_{2}^{\mathbf{g}}$}
\label{sec:3}
In this section, we initiate our study with NCFT$_{2}^{\mathtt{su}_{3}}$ as
the first member of rank 2 theories which also include G$_{2},$ SO(5) and
SP(4) having the isomorphism $SO(5)\simeq SP(4).$ The general set up for the
rank two unitary NCFT$_{2}^{\mathtt{su}_{3}}$ is quite similar to the
framework outlined in the previous section for NCFT$_{2}^{\mathtt{su}_{2}}$.
The main distinction lies in the presence of two scalar fields $\left(
X_{1},X_{2}\right) $ compactified on the 2-torus $\mathbb{S}^{1}\times
\mathbb{S}^{1}$ generating the Cartan subgroup
\begin{equation}
U(1)\times U(1)\subset SU(3)
\end{equation}%
As in the case of su(2), the string fields are split into left and right
moving components like $X^{a}=(X_{L}^{a}+X_{R}^{a})/\sqrt{2}.$ Their\textrm{%
\ 2D field action }$\mathcal{I}\left[ X_{1},X_{2};\mathcal{G}_{ab},\mathcal{B%
}_{ab}\right] $\textrm{\ generally takes the form }%
\begin{equation}
\mathcal{I}\left[ X;\mathcal{G},\mathcal{B}\right] =\frac{1}{4\pi \alpha
^{\prime }}\dint d^{2}\sigma \dsum\limits_{a,b=1}^{2}\left( \delta ^{\mathrm{%
\alpha \beta }}\partial _{\mathrm{\alpha }}X^{a}\mathcal{G}_{ab}\partial _{%
\mathrm{\beta }}X^{b}+i\varepsilon ^{\mathrm{\alpha \beta }}\partial _{%
\mathrm{\alpha }}X^{a}\mathcal{B}_{ab}\partial _{\mathrm{\beta }}X^{b}\right)
\label{fab}
\end{equation}%
Here, $\mathcal{G}_{ab}$ represents the constant symmetric target space
metric possessing three degrees of freedom, while $\mathcal{B}%
_{ab}=\varepsilon _{ab}\mathcal{B}$ denotes the constant antisymmetric
background field interpreted as a flux. These matrix variables $\left(
\mathcal{G}_{ab},\mathcal{B}_{ab}\right) $ define the coordinates of the
four dimensional moduli space
\begin{equation}
\mathcal{M}_{\mathrm{su}_{3}}=\mathcal{O}(2,2;\mathbb{Z})\backslash \mathcal{%
O}(2,2;\mathbb{R})/\mathcal{O}(2;\mathbb{R})\times \mathcal{O}(2;\mathbb{R})
\label{ms}
\end{equation}%
of the standard Narain CFT$_{2}^{\mathtt{su}_{3}}.$ Distances within this
moduli space are measured by the Zamolodchikov metric \cite{1A}
\begin{equation}
ds_{\mathrm{su}_{3}}^{2}=\mathcal{G}^{ac}\mathcal{G}^{bd}\left( d\mathcal{G}%
_{ab}d\mathcal{G}_{cd}+d\mathcal{B}_{ab}d\mathcal{B}_{cd}\right)  \label{fac}
\end{equation}%
In our proposal, this metric will be realised in terms of the Cartan matrix K%
$_{\mathrm{su}_{3}}$ of su(3) and its inverse K$_{\mathrm{su}_{3}}^{-1}$. A
similar analysis to NCFT$_{2}^{\mathbf{su}_{3}}$ holds also for SO(5) and G$%
_{2}$ where $\mathcal{B}_{ab}$ assumes non zero values.

\subsection{NCFT$_{2}^{su_{3}}$ as an extension of NCFT$_{2}^{su_{2}}$}

The description of the Narain CFT$_{2}^{\mathtt{su}_{3}}$ and its genus-one
partition function $Z_{\mathrm{su}_{3}}$ along the lines of our proposal can
be constructed by starting from eq(\ref{p}) and extending it to the su(3)
symmetry. This su(3) Lie algebra features two simple roots $\left( \mathbf{%
\alpha }_{1},\mathbf{\alpha }_{2}\right) $ and two fundamental weights $%
\left( \mathbf{\lambda }_{1},\mathbf{\lambda }_{2}\right) $ with
intersection matrices%
\begin{equation}
\mathbf{\alpha }_{i}.\mathbf{\alpha }_{j}=K_{ij}\qquad ,\qquad \mathbf{%
\lambda }^{i}.\mathbf{\lambda }^{j}=\tilde{K}^{ij}\qquad ,\qquad \mathbf{%
\alpha }_{i}.\mathbf{\lambda }^{j}=\delta _{i}^{j}
\end{equation}%
reading explicitly as
\begin{equation}
K_{ij}=\left(
\begin{array}{cc}
2 & -1 \\
-1 & 2%
\end{array}%
\right) \qquad ,\qquad \tilde{K}^{ij}=\frac{1}{3}\left(
\begin{array}{cc}
2 & 1 \\
1 & 2%
\end{array}%
\right)  \label{KK}
\end{equation}%
showing in turn that $\mathbf{\alpha }_{i}^{2}=2$ and $\mathbf{\lambda }%
_{i}^{2}=2/3$ as well as $\left\vert K_{ij}\right\vert \geq 1$ and $%
\left\vert \tilde{K}^{ij}\right\vert <1.$ In this NCFT$_{2}^{su_{3}}$ with
central charge $c_{L/R}=2$, the length $\mathbf{\lambda }_{i}^{2}$ differs
from the inverse $1/\mathbf{\alpha }_{i}^{2}$ although it remains confined
to the interval $\left[ 1/2,1\right] ;$ that is $1/2\leq \mathbf{\lambda }%
_{i}^{2}<1.$ Furthermore, the duality relation $\mathbf{\alpha }_{i}.\mathbf{%
\lambda }_{j}=\delta _{ij}$ allows to express the simple roots $\mathbf{%
\alpha }_{i}$ in terms of the fundamental weights $\mathbf{\lambda }_{j}$
and vice versa as follows,%
\begin{equation}
\begin{tabular}{lll}
$\mathbf{\alpha }_{1}$ & $=$ & $2\mathbf{\lambda }_{1}-\mathbf{\lambda }_{2}$
\\
$\mathbf{\alpha }_{2}$ & $=$ & $2\mathbf{\lambda }_{2}-\mathbf{\lambda }_{1}$%
\end{tabular}%
\qquad ,\qquad
\begin{tabular}{lll}
$\mathbf{\lambda }_{1}$ & $=$ & $\frac{2}{3}\mathbf{\alpha }_{1}+\frac{1}{3}%
\mathbf{\alpha }_{2}$ \\
$\mathbf{\lambda }_{2}$ & $=$ & $\frac{1}{3}\mathbf{\alpha }_{1}+\frac{2}{3}%
\mathbf{\alpha }_{2}$%
\end{tabular}
\label{p3}
\end{equation}%
Notice moreover that $\mathbf{\lambda }_{1}$ is the highest weight vector of
the su(3) fundamental representation \textbf{3}; and $\mathbf{\lambda }_{2}$
is the highest weight vector of the conjugate representation \textbf{\={3}}.
The weight vectors $\nu _{i}$ of the states of these representations in the
basis ($\mathbf{\alpha }_{1},\mathbf{\alpha }_{2}$) are given by%
\begin{eqnarray}
\mathbf{3} &:&\nu _{1}=\left( \frac{2}{3}\mathbf{,}\frac{1}{3}\right) ,\quad
\nu _{2}=\left( -\frac{1}{3}\mathbf{,}\frac{1}{3}\right) ,\quad \nu
_{3}=\left( -\frac{1}{3}\mathbf{,-}\frac{2}{3}\right) \\
\mathbf{\bar{3}} &:&\nu _{1}^{\ast }=\left( \frac{1}{3}\mathbf{,}\frac{2}{3}%
\right) ,\quad \nu _{2}^{\ast }=\left( \frac{1}{3}\mathbf{,}-\frac{1}{3}%
\right) ,\quad \nu _{3}^{\ast }=\left( -\frac{2}{3}\mathbf{,}-\frac{1}{3}%
\right)
\end{eqnarray}

\subsubsection{Sitting at the point $x_{i}=1$ in $\mathcal{M}_{\mathbf{su}%
_{3}}$}

Using the two simple roots $\mathbf{\alpha }_{i}$ and the two fundamental
weights $\mathbf{\lambda }_{i}$ of su(3), we can expand the left $%
\boldsymbol{p}_{L}$ and the right $\boldsymbol{p}_{R}$ momentum vectors of
the Narain CFT$_{2}^{\mathtt{su}_{3}}$ for radii $R_{1}=R_{2}=1/2$ as
follows
\begin{equation}
\begin{tabular}{lll}
$\boldsymbol{p}_{L}$ & $=$ & $\frac{1}{\sqrt{2}}\dsum\limits_{i=1}^{2}\left(
n^{i}\mathbf{\alpha }_{i}+w_{i}\mathbf{\lambda }^{i}\right) $ \\
$\boldsymbol{p}_{R}$ & $=$ & $\frac{1}{\sqrt{2}}\dsum\limits_{i=1}^{2}\left(
n^{i}\mathbf{\alpha }_{i}-w_{i}\mathbf{\lambda }^{i}\right) $%
\end{tabular}
\label{p33}
\end{equation}%
with Kaluza-Klein $\left\{ n^{i}\right\} $ modes and winding $\left\{
w_{i}\right\} $ integers. Here, the vielbeins for NCFT$^{\mathbf{su}_{3}}$\
read\textrm{\ }in our proposal as follows%
\begin{equation}
\boldsymbol{E}_{\text{ \textsc{a}}}^{\text{\textsc{\b{m}}}}=\frac{1}{\sqrt{2}%
}\left(
\begin{array}{cc}
\mathbf{\alpha }_{i} & +\mathbf{\lambda }^{j} \\
\mathbf{\alpha }_{i} & -\mathbf{\lambda }^{j}%
\end{array}%
\right)
\end{equation}%
The lengths of these quantum momenta are given by%
\begin{equation}
\begin{tabular}{lll}
$\boldsymbol{p}_{\mathrm{L}}^{2}$ & $=$ & $\frac{1}{2}\dsum%
\limits_{i,j=1}^{2}\left[ n^{i}K_{ij}n^{j}+w_{i}\tilde{K}^{ij}w_{j}+n^{i}%
\left( \mathbf{\alpha }_{i}\mathbf{.\lambda }^{j}\right) w_{j}+w_{i}\left(
\mathbf{\lambda }^{i}\mathbf{.\alpha }_{j}\right) n^{j}\right] $ \\
$\boldsymbol{p}_{\mathrm{R}}^{2}$ & $=$ & $\frac{1}{2}\dsum%
\limits_{i,j=1}^{2}\left[ n^{i}K_{ij}n^{j}+w_{i}\tilde{K}^{ij}w_{j}-n^{i}%
\left( \mathbf{\alpha }_{i}\mathbf{.\lambda }^{j}\right) w_{j}-w_{i}\left(
\mathbf{\lambda }^{i}\mathbf{.\alpha }_{j}\right) n^{j}\right] $%
\end{tabular}%
\end{equation}%
Similarly to the su(2) conformal theory of the previous section, here also
we set $\mathfrak{Q}_{L}^{\mathtt{su}_{3}}:=\boldsymbol{p}_{L}^{2}$ and $%
\mathfrak{Q}_{R}^{\mathtt{su}_{3}}:=\boldsymbol{p}_{R}^{2};$ these are
quadratic forms having the structure
\begin{equation}
\mathfrak{Q}_{L}^{\mathtt{su}_{3}}=\boldsymbol{m}^{T}.\boldsymbol{Q}_{L}^{%
\mathtt{su}_{3}}.\boldsymbol{m}\qquad ,\qquad \mathfrak{Q}_{R}^{\mathtt{su}%
_{3}}=\boldsymbol{m}^{T}.\boldsymbol{Q}_{R}^{\mathtt{su}_{3}}.\boldsymbol{m}
\end{equation}%
where $\boldsymbol{m}^{T}=\left( n^{1},n^{2},w_{1},w_{2}\right) $ is a 4d
integer vector and $\boldsymbol{Q}_{L/R}^{\mathtt{su}_{3}}$ are $4\times 4$
matrices reading in terms of the Cartan matrix $K_{\mathtt{su}_{3}}$ and its
inverse $K_{\mathtt{su}_{3}}^{-1}:=\tilde{K}_{\mathtt{su}_{3}}$ like
\begin{equation}
\boldsymbol{Q}_{L}^{\mathtt{su}_{3}}=\frac{1}{2}\left(
\begin{array}{cc}
K_{ij} & \delta _{i}^{l} \\
\delta _{j}^{k} & \tilde{K}^{kl}%
\end{array}%
\right) \qquad ,\qquad \boldsymbol{Q}_{R}^{\mathtt{su}_{3}}=\frac{1}{2}%
\left(
\begin{array}{cc}
K_{ij} & -\delta _{i}^{l} \\
-\delta _{j}^{k} & \tilde{K}^{kl}%
\end{array}%
\right)
\end{equation}%
By using
\begin{equation}
\begin{tabular}{lll}
$\mathcal{H}_{\mathtt{su}_{3}}$ & $=$ & $\mathfrak{Q}_{L}^{\mathtt{su}_{3}}+%
\mathfrak{Q}_{R}^{\mathtt{su}_{3}}$ \\
$\mathfrak{Q}_{\mathtt{su}_{3}}$ & $=$ & $\mathfrak{Q}_{L}^{\mathtt{su}_{3}}-%
\mathfrak{Q}_{R}^{\mathtt{su}_{3}}$%
\end{tabular}%
\end{equation}%
and denoting in bold letters the associated $4\times 4$ matrices like
\begin{equation}
\begin{tabular}{lll}
$\boldsymbol{H}_{\mathtt{su}_{3}}$ & $=$ & $\boldsymbol{Q}_{L}^{\mathtt{su}%
_{3}}+\boldsymbol{Q}_{R}^{\mathtt{su}_{3}}$ \\
$\boldsymbol{Q}_{\mathtt{su}_{3}}$ & $=$ & $\boldsymbol{Q}_{L}^{\mathtt{su}%
_{3}}-\boldsymbol{Q}_{R}^{\mathtt{su}_{3}}$%
\end{tabular}%
\end{equation}%
we end up with the factorisations%
\begin{equation}
\begin{tabular}{lll}
$\mathcal{H}_{\mathtt{su}_{3}}$ & $=$ & $\boldsymbol{m}^{T}.\boldsymbol{H}_{%
\mathtt{su}_{3}}.\boldsymbol{m}$ \\
$\mathfrak{Q}_{\mathtt{su}_{3}}$ & $=$ & $\boldsymbol{m}^{T}.\boldsymbol{Q}_{%
\mathtt{su}_{3}}.\boldsymbol{m}$%
\end{tabular}%
\end{equation}%
with%
\begin{equation}
\boldsymbol{H}_{\mathtt{su}_{3}}=\left(
\begin{array}{cc}
K_{ij} & 0 \\
0 & \tilde{K}^{kl}%
\end{array}%
\right) \qquad ,\qquad \boldsymbol{Q}_{\mathtt{su}_{3}}=\left(
\begin{array}{cc}
0 & \delta _{i}^{l} \\
\delta _{j}^{k} & 0%
\end{array}%
\right)  \label{hhd}
\end{equation}%
Here also, the matrix $\boldsymbol{H}_{\mathtt{su}_{3}}$is block diagonal
and $\boldsymbol{Q}_{\mathtt{su}_{3}}$ is an off block diagonal unimodular
matrix $\left\vert \det \boldsymbol{Q}_{\mathtt{su}_{3}}\right\vert =1.$ By
substituting $\mathbf{\alpha }_{i}=K_{ij}\mathbf{\lambda }^{j}$ in (\ref{p3}%
), we can put the momenta in the form%
\begin{equation}
\begin{tabular}{lllll}
$\boldsymbol{p}_{L}$ & $=$ & $\frac{1}{\sqrt{2}}\dsum\limits_{i=1}^{2}\left(
K_{ij}n^{j}+w_{i}\right) \mathbf{\lambda }^{i}$ & $\equiv $ & $\frac{1}{%
\sqrt{2}}\dsum\limits_{i=1}^{2}\left( \tilde{n}_{i}+w_{i}\right) \mathbf{%
\lambda }^{i}$ \\
$\boldsymbol{p}_{R}$ & $=$ & $\frac{1}{\sqrt{2}}\dsum\limits_{i=1}^{2}\left(
K_{ij}n^{j}-w_{i}\right) \mathbf{\lambda }^{i}$ & $\equiv $ & $\frac{1}{%
\sqrt{2}}\dsum\limits_{i=1}^{2}\left( \tilde{n}_{i}-w_{i}\right) \mathbf{%
\lambda }^{i}$%
\end{tabular}
\label{p34}
\end{equation}%
with the following features ensues:

\begin{description}
\item[$\left( \mathbf{1}\right) $] the left $\boldsymbol{p}_{L}$ and the
right $\boldsymbol{p}_{R}$ momenta are given by shifts of the KK\ momentum
\begin{equation}
\boldsymbol{p}_{(\mathbf{n})}=\frac{1}{\sqrt{2}}\dsum\limits_{i,j=1}^{2}%
\left( n^{i}K_{ij}\mathbf{\lambda }^{j}\right) \equiv \frac{1}{\sqrt{2}}%
\dsum\limits_{i=1}^{2}\tilde{n}_{j}\mathbf{\lambda }^{j}
\end{equation}%
by the amounts $\pm \sum_{i=1}^{2}\left( w_{i}\mathbf{\lambda }^{i}\right) /%
\sqrt{2}$ induced by non trivial windings.

\item[$\left( \mathbf{2}\right) $] Regarding the factorisation $\mathfrak{Q}%
_{\mathtt{su}_{3}}=\mathfrak{p}^{\text{\textsc{\b{m}}}}\eta _{\text{\textsc{%
\b{m}\b{n}}}}\mathfrak{p}^{\text{\textsc{\b{n}}}}$ with metric $\eta _{\text{%
\textsc{\b{m}\b{n}}}}=diag(+,+,-,-)$, the expressions of the vectors $%
\mathfrak{p}^{\text{\textsc{\b{m}}}}$ in terms of $\boldsymbol{p}_{L}$ and $%
\boldsymbol{p}_{R}$ of the su(3) theory are given by%
\begin{equation}
\mathfrak{p}^{\text{\textsc{\b{m}}}}=\left(
\begin{array}{c}
\frac{1}{\sqrt{2}}\boldsymbol{p}_{L}.\mathbf{\alpha }_{i} \\
\frac{1}{\sqrt{2}}\boldsymbol{p}_{R}.\mathbf{\alpha }_{i}%
\end{array}%
\right) =\left(
\begin{array}{c}
\frac{1}{2}\left( K_{ij}n^{j}+w_{i}\right) \\
\frac{1}{2}\left( K_{ij}n^{j}-w_{i}\right)%
\end{array}%
\right)
\end{equation}

\item[$\left( \mathbf{3}\right) $] The condition $\mathfrak{p}^{\text{%
\textsc{\b{m}}}}\eta _{\text{\textsc{\b{m}\b{n}}}}\mathfrak{p}^{\text{%
\textsc{\b{n}}}}\in 2\mathbb{Z}$ is solved in our proposal like $\mathfrak{Q}%
_{\mathtt{su}_{3}}=\boldsymbol{m}^{T}.\boldsymbol{Q}_{\mathtt{su}_{3}}.%
\boldsymbol{m}$ indicating that the metric ${\large g}_{\text{\textsc{ab}}}$
of the lattice ${\large \Lambda }_{\mathtt{su}_{3}}$ is defined by $%
\boldsymbol{Q}_{\text{\textsc{ab}}}^{\mathtt{su}_{3}};$ that is
\begin{equation}
{\large \Lambda }_{\mathtt{su}_{3}}=\left\{ \boldsymbol{m}\in \mathbb{Z}%
^{4},\quad \mathfrak{Q}_{\mathtt{su}_{3}}=\boldsymbol{m}^{\text{\textsc{a}}}%
{\large g}_{\text{\textsc{ab}}}\boldsymbol{m}^{\text{\textsc{b}}}\quad
|\quad {\large g}_{\text{\textsc{ab}}}=\boldsymbol{Q}_{\text{\textsc{ab}}}^{%
\mathtt{su}_{3}}\right\}
\end{equation}%
where $\boldsymbol{m}=\left( \mathbf{n},\mathbf{w}\right) \in \mathbb{Z}%
^{2}\times \mathbb{Z}^{2}$. For this lattice, we then have the property $%
\mathfrak{Q}_{\mathtt{su}_{3}}=2\mathbf{n}.\mathbf{w}$ which is an even
integer that vanishes for $\mathbf{n}\perp \mathbf{w}.$
\end{description}

\subsubsection{NCFT$_{2}^{su_{3}}$ at generic moduli $\left\{
x_{1},x_{2}\right\} $}

Similarly to the Narain CFT$_{2}^{\mathtt{su}_{2}},$ one can exhibit the
\emph{four} moduli $\left\{ x_{11},x_{12},x_{21},x_{22}\right\} $ of the
su(3) conformal theory including the diagonal $\left\{ x_{1},x_{2}\right\} $
given by the radii $\left( R_{1},R_{2}\right) $ of\ the 2-torus $\mathbb{T}%
^{2}=\mathbb{S}_{1}^{1}\times \mathbb{S}_{2}^{1}$. Instead of the two simple
roots $\left\{ \mathbf{\alpha }_{1},\mathbf{\alpha }_{2}\right\} $ and the
two fundamental weights $\left\{ \mathbf{\lambda }_{1},\mathbf{\lambda }%
_{2}\right\} $, we introduce the "scaled" vectors $\mathbf{\beta }%
_{i}=\sum_{k=1}^{2}x_{i}^{k}\mathbf{\alpha }_{k}$ and "scaled" weights $%
\mathbf{\chi }^{j}=\sum_{l=1}^{2}y_{l}^{j}\mathbf{\lambda }^{l}$ reading
explicitly as
\begin{equation}
\begin{tabular}{lll}
$\mathbf{\beta }_{1}$ & $=$ & $x_{1}^{1}\mathbf{\alpha }_{1}+x_{1}^{2}%
\mathbf{\alpha }_{2}$ \\
$\mathbf{\beta }_{2}$ & $=$ & $x_{2}^{1}\mathbf{\alpha }_{1}+x_{2}^{2}%
\mathbf{\alpha }_{2}$%
\end{tabular}%
\qquad ,\qquad
\begin{tabular}{lll}
$\mathbf{\chi }^{1}$ & $=$ & $y_{1}^{1}\mathbf{\lambda }^{1}+y_{2}^{1}%
\mathbf{\lambda }^{2}$ \\
$\mathbf{\chi }^{2}$ & $=$ & $y_{1}^{2}\mathbf{\lambda }^{1}+y_{2}^{2}%
\mathbf{\lambda }^{2}$%
\end{tabular}
\label{ex}
\end{equation}%
where the coefficients are constrained by the duality relation%
\begin{equation}
\mathbf{\beta }_{i}.\mathbf{\chi }^{j}=\delta _{i}^{j}\qquad \Rightarrow
\qquad \sum_{k=1}^{2}x_{i}^{k}y_{k}^{j}=\delta _{i}^{j}
\end{equation}%
showing that $y_{k}^{i}$ is just the inverse of $x_{i}^{k}$ provided that $%
\det (x_{i}^{k})\neq 0.$ Hence, the four moduli $x_{i}^{k}$ presented in our
proposal parameterise the moduli space $\mathcal{O}(2,2)/\mathcal{O}%
(2)\times \mathcal{O}(2).$ From (\ref{ex}), we also have the "scaled"
intersections%
\begin{equation}
\mathcal{K}_{kl}^{\mathrm{su}_{\mathrm{3}}}=\mathbf{\beta }_{i}.\mathbf{%
\beta }_{j}\qquad ,\qquad \mathcal{\tilde{K}}_{\mathrm{su}_{\mathrm{3}%
}}^{kl}=\mathbf{\chi }^{i}.\mathbf{\chi }^{j}
\end{equation}%
with the expansions%
\begin{equation}
\mathcal{K}_{ij}^{\mathrm{su}_{\mathrm{3}}}=\sum_{k,l=1}^{2}x_{i}^{k}K_{kl}^{%
\mathrm{su}_{\mathrm{3}}}x_{j}^{l}\qquad ,\qquad \mathcal{\tilde{K}}_{%
\mathrm{su}_{\mathrm{3}}}^{ij}=\sum_{k,l=1}^{2}y_{k}^{i}\tilde{K}_{\mathrm{su%
}_{\mathrm{3}}}^{kl}y_{l}^{j}
\end{equation}%
where $K_{kl}^{\mathrm{su}_{\mathrm{3}}}=\mathbf{\alpha }_{i}.\mathbf{\alpha
}_{j}$ and $\tilde{K}_{\mathrm{su}_{\mathrm{3}}}^{kl}=\mathbf{\lambda }^{i}.%
\mathbf{\lambda }^{j}.$ These intersection matrices $\mathcal{K}_{ij}^{%
\mathrm{su}_{\mathrm{3}}}$ and $\mathcal{\tilde{K}}_{\mathrm{su}_{\mathrm{3}%
}}^{ij}$ are functions of $x_{i}^{k}$ and their inverse. For the generic
case, the quantized momenta read as follows%
\begin{equation}
\begin{tabular}{lll}
$\boldsymbol{p}_{L}$ & $=$ & $\frac{1}{\sqrt{2}}\dsum\limits_{i=1}^{2}\left(
n^{i}\mathbf{\beta }_{i}+w_{i}\mathbf{\chi }^{i}\right) $ \\
$\boldsymbol{p}_{R}$ & $=$ & $\frac{1}{\sqrt{2}}\dsum\limits_{i=1}^{2}\left(
n^{i}\mathbf{\beta }_{i}-w_{i}\mathbf{\chi }^{i}\right) $%
\end{tabular}
\label{pr}
\end{equation}%
from which we deduce the vielbeins%
\begin{equation}
\boldsymbol{E}_{\text{ \textsc{a}}}^{\text{\textsc{\b{m}}}}=\frac{1}{\sqrt{2}%
}\left(
\begin{array}{cc}
\mathbf{\beta }_{i} & +\mathbf{\chi }^{j} \\
\mathbf{\beta }_{i} & \mathbf{\chi }^{j}%
\end{array}%
\right)
\end{equation}%
satisfying $\mathbf{Q}_{\text{\textsc{ab}}}=\boldsymbol{E}_{\text{\textsc{a}}%
}^{\text{ \textsc{\b{m}}}}\eta _{\text{\textsc{\b{m}\b{n}}}}\boldsymbol{E}_{%
\text{ \textsc{b}}}^{\text{\textsc{\b{n}}}}$ meaning that they are invariant
under $\mathcal{O}(2,2)/\left[ \mathcal{O}(2)_{L}\times \mathcal{O}(2)_{R}%
\right] .$ The $\mathcal{O}(2,2)$ acts on $\boldsymbol{E}_{\text{\textsc{a}}%
}^{\text{\textsc{\b{m}}}};$ it mixes $\mathbf{\beta }_{i}$ and $\mathbf{\chi
}^{j}$ while preserving the pseudo-metric $\eta _{\text{\textsc{\b{m}\b{n}}}%
}.$ The $\mathcal{O}(2)_{L}$ rotates the two $\mathbf{\beta }_{i}$'s like $%
\mathbf{\beta }_{i}^{\prime }=\left( \mathcal{O}_{L}\right) _{i}^{k}\mathbf{%
\beta }_{k}$, it preserves $\mathbf{\beta }_{i}.\mathbf{\beta }_{j}$. The $%
\mathcal{O}(2)_{R}$ rotates the two $\mathbf{\chi }^{j}$'s; it acts as $%
\mathbf{\chi }^{j\prime }=\left( \mathcal{O}_{R}\right) _{l}^{j}\mathbf{\chi
}^{l}$ and preserves $\mathbf{\chi }^{i}.\mathbf{\chi }^{j}.$ Moreover, the
property $\mathbf{\beta }_{i}.\mathbf{\chi }^{j}=\delta _{i}^{j}$ requires
that $\mathcal{O}_{R}=\mathcal{O}_{L}^{-1}.$

By substituting $\mathbf{\beta }_{i}=\sum_{k=1}^{2}x_{i}^{k}\mathbf{\alpha }%
_{k}$ and $\mathbf{\chi }^{i}=\sum_{l=1}^{2}y_{l}^{j}\mathbf{\lambda }^{l}$,
we get%
\begin{equation}
\begin{tabular}{lll}
$\boldsymbol{p}_{L}$ & $=$ & $\frac{1}{\sqrt{2}}\dsum\limits_{i,k=1}^{2}%
\left( n^{i}x_{i}^{k}K_{kl}+w_{j}y_{l}^{j}\right) \mathbf{\lambda }^{l}$ \\
$\boldsymbol{p}_{R}$ & $=$ & $\frac{1}{\sqrt{2}}\dsum\limits_{i,k=1}^{2}%
\left( n^{i}x_{i}^{k}K_{kl}-w_{j}y_{l}^{j}\right) \mathbf{\lambda }^{l}$%
\end{tabular}
\label{qr}
\end{equation}

In what follows, we often restrict the analysis to the \emph{diagonal} case
where the scaled roots and the scaled weights simplify as,%
\begin{equation}
\begin{tabular}{lll}
$\mathbf{\beta }_{1}$ & $=$ & $x_{1}\mathbf{\alpha }_{1}$ \\
$\mathbf{\beta }_{2}$ & $=$ & $x_{2}\mathbf{\alpha }_{2}$%
\end{tabular}%
\qquad ,\qquad
\begin{tabular}{lll}
$x_{1}$ & $=$ & $\mathbf{\beta }_{1}.\mathbf{\lambda }_{1}$ \\
$x_{2}$ & $=$ & $\mathbf{\beta }_{2}.\mathbf{\lambda }_{2}$%
\end{tabular}
\label{232}
\end{equation}%
and%
\begin{equation}
\begin{tabular}{lll}
$\mathbf{\chi }_{1}$ & $=$ & $\frac{1}{x_{1}}\mathbf{\lambda }_{1}$ \\
$\mathbf{\chi }_{2}$ & $=$ & $\frac{1}{x_{2}}\mathbf{\lambda }_{2}$%
\end{tabular}%
\qquad ,\qquad
\begin{tabular}{lll}
$\mathbf{\chi }_{1}$ & $=$ & $\frac{1}{3x_{1}}\left( 2\mathbf{\alpha }_{1}+%
\mathbf{\alpha }_{2}\right) $ \\
$\mathbf{\chi }_{2}$ & $=$ & $\frac{1}{3x_{2}}\left( \mathbf{\alpha }_{1}+2%
\mathbf{\alpha }_{2}\right) $%
\end{tabular}
\label{233}
\end{equation}%
with positive definite moduli $x_{i}=1/(2R_{i}).$ This permits to fix ideas
and to get an overview; the general relationships follow straightforwardly.

\subsection{Results for diagonal NCFT$_{2}^{\mathtt{su}_{3}}$}

With the above diagonal scaled cycles, we derive the particular intersection
matrices $\mathbf{\beta }_{i}.\mathbf{\beta }_{j}=\mathcal{K}_{ij}^{su_{3}}$
and $\mathbf{\chi }^{i}.\mathbf{\chi }^{j}=\mathcal{\tilde{K}}_{su_{3}}^{ij}$%
\ which explicitly read as follows,
\begin{equation}
\mathcal{K}_{ij}^{su_{3}}=\left(
\begin{array}{cc}
2x_{1}^{2} & -x_{1}x_{2} \\
-x_{1}x_{2} & 2x_{2}^{2}%
\end{array}%
\right) \qquad ,\qquad \mathcal{\tilde{K}}_{su_{3}}^{ij}=\left(
\begin{array}{cc}
\frac{2}{3x_{1}^{2}} & \frac{1}{3x_{1}x_{2}} \\
\frac{1}{3x_{1}x_{2}} & \frac{2}{3x_{2}^{2}}%
\end{array}%
\right)  \label{kkt}
\end{equation}%
From these matrices, we learn that $\mathbf{\beta }_{i}^{2}=2x_{i}^{2}$ and $%
\mathbf{\chi }_{i}^{2}=2/(3x_{i}^{2}).$ We also have the following
expressions for the left and the right momenta%
\begin{equation}
\begin{tabular}{lll}
$\boldsymbol{p}_{L}\left( x_{1},x_{2}\right) $ & $=$ & $\frac{1}{\sqrt{2}}%
\dsum\limits_{i=1}^{2}\left( \mathcal{K}_{ij}n^{j}+w_{i}\right) \mathbf{\chi
}^{i}$ \\
$\boldsymbol{p}_{R}\left( x_{1},x_{2}\right) $ & $=$ & $\frac{1}{\sqrt{2}}%
\dsum\limits_{i=1}^{2}\left( \mathcal{K}_{ij}n^{j}-w_{i}\right) \mathbf{\chi
}^{i}$%
\end{tabular}
\label{ppt}
\end{equation}%
they lead to the matrix
\begin{equation}
\boldsymbol{A}_{\mathtt{su}_{3}}\left( x\right) =\pi \left(
\begin{array}{cc}
-\tau _{2}\mathcal{K}_{ij} & i\tau _{1}\delta _{i}^{\text{ \ }l} \\
i\tau _{1}\delta _{\text{ \ }j}^{k} & -\tau _{2}\mathcal{\tilde{K}}^{kl}%
\end{array}%
\right)
\end{equation}

\subsubsection{Metric of $\mathcal{M}_{\mathbf{su}_{3}}$ and partition
function $Z_{\mathbf{su}_{3}}$}

The realisations (\ref{kkt}-\ref{ppt}) allow the computation of interesting
quantities characterising the (diagonal) NCFT$_{2}^{\mathtt{su}_{3}}$, in
particular the two following:

\begin{description}
\item[$\left( \mathbf{1}\right) $] \textbf{Metric of} $\mathcal{M}_{su_{3}}$%
: the metric of the Narain moduli space generalises (\ref{KK2}) to the rank
2 conformal models. The length $\mathbf{\beta }^{2}=2x^{2}$ in the su(2)
theory extends to the intersection matrix $(\mathcal{K}_{su_{3}})_{ij}=%
\mathbf{\beta }_{i}.\mathbf{\beta }_{j},$ and the 1-form $\mathcal{K}%
_{su_{2}}^{-1}d\mathcal{K}_{su_{2}}$ becomes the generalised 2$\times $2
matrix 1-form $\mathcal{K}_{su_{3}}^{-1}d\mathcal{K}_{su_{3}}$ reading as
follows%
\begin{equation}
d\upsilon _{j}^{i}=\sum_{k=1}^{2}\left( \mathcal{K}_{su_{3}}^{-1}\right)
^{ik}\left( d\mathcal{K}_{su_{3}}\right) _{kj}
\end{equation}%
Using the $\mathcal{K}_{su_{3}}$ and $\mathcal{K}_{su_{3}}^{-1}$ matrices of
(\ref{kkt}), we calculate%
\begin{equation}
d\mathcal{K}_{ij}^{su_{3}}=\left(
\begin{array}{cc}
4x_{1}dx_{1} & -dx_{1}x_{2}-x_{1}dx_{2} \\
-dx_{1}x_{2}-x_{1}dx_{2} & 4x_{2}dx_{2}%
\end{array}%
\right)
\end{equation}%
and then%
\begin{equation}
d\upsilon _{j}^{i}=\left(
\begin{array}{cc}
\frac{7}{3x_{1}} & -\frac{2x_{2}}{3x_{1}^{2}} \\
\frac{2}{3x_{2}} & -\frac{1}{3x_{1}}%
\end{array}%
\right) dx_{1}+\left(
\begin{array}{cc}
-\frac{1}{3x_{2}} & \frac{2}{3x_{1}} \\
-\frac{2x_{1}}{3x_{2}^{2}} & \frac{7}{3x_{2}}%
\end{array}%
\right) dx_{2}
\end{equation}%
In term of this matrix 1-form $d\upsilon _{j}^{i}$, the metric $%
ds_{su_{3}}^{2}$ of the Narain moduli space is given by the trace $%
(d\upsilon _{j}^{i})(d\upsilon _{i}^{j});$ it is expressed like
\begin{equation}
ds_{su_{3}}^{2}=\left( \mathcal{K}_{su_{3}}^{-1}\right) ^{ik}\left( \mathcal{%
K}_{su_{3}}^{-1}\right) ^{jl}\left( d\mathcal{K}_{su_{3}}\right) _{kj}\left(
d\mathcal{K}_{su_{3}}\right) _{li}
\end{equation}%
which explicitly reads as:%
\begin{eqnarray}
ds_{su_{3}}^{2} &=&\left( \frac{50}{9x_{1}^{2}}+\frac{4}{9x_{2}^{2}}+\frac{%
4x_{2}^{2}}{9x_{1}^{4}}\right) dx_{1}^{2}+\left( \frac{50}{9x_{2}^{2}}+\frac{%
4}{9x_{1}^{2}}+\frac{4x_{1}^{2}}{9x_{2}^{4}}\right) dx_{2}^{2}  \notag \\
&&-\left( \frac{28}{9x_{1}x_{2}}+\frac{8x_{1}}{9x_{2}^{3}}+\frac{8x_{2}}{%
9x_{1}^{3}}\right) dx_{1}dx_{2}
\end{eqnarray}

\item[$\left( \mathbf{2}\right) $] \textbf{Partition function of NCFT}$_{2}^{%
\mathtt{su}_{3}}$: the genus-one partition function of the Narain CFT$_{2}^{%
\mathtt{su}_{3}}$ is defined as follows
\begin{eqnarray}
Z_{\mathrm{su}_{3}}\left[ \tau ,\bar{\tau};x\right] &=&\frac{1}{\left\vert
\eta \left( \tau \right) \right\vert ^{4}}\dsum\limits_{\left(
p_{L},p_{R}\right) }q^{\frac{1}{2}p_{L}^{2}}\bar{q}^{\frac{1}{2}p_{R}^{2}}
\notag \\
&=&\frac{1}{\left\vert \eta \left( \tau \right) \right\vert ^{4}}%
\dsum\limits_{(\mathfrak{Q}_{L}^{\mathtt{su}_{3}},\mathfrak{Q}_{R}^{\mathtt{%
su}_{3}})}q^{\frac{1}{2}\mathfrak{Q}_{L}^{\mathtt{su}_{3}}}\bar{q}^{\frac{1}{%
2}\mathfrak{Q}_{R}^{\mathtt{su}_{3}}}
\end{eqnarray}%
Or equivalently, by substituting $\mathfrak{Q}_{L}^{\mathtt{su}_{3}}$ and $%
\mathfrak{Q}_{R}^{\mathtt{su}_{3}}$:
\begin{equation}
Z_{\mathrm{su}_{3}}\left[ \tau ,\bar{\tau};x\right] =\frac{1}{\left\vert
\eta \left( \tau \right) \right\vert ^{4}}\dsum\limits_{\boldsymbol{m}\in
\mathbb{Z}^{2}}\exp \left( \boldsymbol{m}^{T}.\boldsymbol{A}_{\mathrm{su}%
_{3}}.\boldsymbol{m}\right)
\end{equation}%
with $\boldsymbol{A}_{\mathrm{su}_{3}}=i\pi \left( \tau \boldsymbol{Q}_{L}-%
\bar{\tau}\boldsymbol{Q}_{R}\right) $ a 4$\times $4 matrix reading like%
\begin{equation}
\boldsymbol{A}_{\mathrm{su}_{3}}=i\pi \tau _{1}\boldsymbol{Q}_{\mathrm{su}%
_{3}}-\pi \tau _{2}\boldsymbol{H}_{\mathrm{su}_{3}}
\end{equation}%
where the $\boldsymbol{H}_{\mathrm{su}_{3}}$ and the $\boldsymbol{Q}_{%
\mathrm{su}_{3}}$\ matrices are as in eq(\ref{hhd}) with $\mathcal{K}_{ij}$
and $\mathcal{\tilde{K}}^{kl}$\ given by eq(\ref{kkt}). By replacing, we
have the following moduli dependent matrix%
\begin{equation}
\boldsymbol{A}_{\mathrm{su}_{3}}\left( x\right) =\pi \left(
\begin{array}{cccc}
-2\mathrm{\tau }_{2}x_{1}^{2} & \mathrm{\tau }_{2}x_{1}x_{2} & i\mathrm{\tau
}_{1} & 0 \\
\mathrm{\tau }_{2}x_{1}x_{2} & -2\mathrm{\tau }_{2}x_{2}^{2} & 0 & i\mathrm{%
\tau }_{1} \\
i\mathrm{\tau }_{1} & 0 & \frac{-2\mathrm{\tau }_{2}}{3x_{1}^{2}} & \frac{-%
\mathrm{\tau }_{2}}{3x_{1}x_{2}} \\
0 & i\mathrm{\tau }_{1} & \frac{-\mathrm{\tau }_{2}}{3x_{1}x_{2}} & \frac{-2%
\mathrm{\tau }_{2}}{3x_{2}^{2}}%
\end{array}%
\right)
\end{equation}%
with $x_{i}=1/(2R_{i})$ and moduli-independent determinant $\det \boldsymbol{%
A}_{\mathrm{su}_{3}}=\pi ^{4}\left( \tau ^{2}\bar{\tau}^{2}\right) .$ At the
point $x_{i}=1,$ it reduces to:%
\begin{equation}
\left. \boldsymbol{A}_{\mathrm{su}_{3}}\right\vert _{x_{i}=1}=\left(
\begin{array}{cc}
-\pi \mathrm{\tau }_{2}K_{\mathtt{su}_{3}} & i\pi \mathrm{\tau }_{1}I_{%
{\small 2}} \\
i\pi \mathrm{\tau }_{1}I_{{\small 2}} & -\pi \mathrm{\tau }_{2}\tilde{K}_{%
\mathtt{su}_{3}}%
\end{array}%
\right)
\end{equation}
\end{description}

\subsubsection{Flux moduli for so(5) and G$_{2}$}

Following \cite{MEER}, the generalised relations of compactified momentums
can be also presented in terms of the symmetric metric $\mathcal{G}_{ij}$ of
the moduli space (\ref{fac}) and the antisymmetric flux $\mathcal{B}_{ij}$
as follows%
\begin{equation}
\begin{tabular}{lll}
$p_{L,i}$ & $=$ & $n_{i}+\frac{1}{2}\left( \mathcal{G}_{ij}-\mathcal{B}%
_{ij}\right) w^{j}$ \\
$p_{R,i}$ & $=$ & $n_{i}-\frac{1}{2}\left( \mathcal{G}_{ij}+\mathcal{B}%
_{ij}\right) w^{j}$%
\end{tabular}
\label{pn}
\end{equation}%
Using this realisation, the vielbein $\mathfrak{p}^{\text{\textsc{\b{n}}}%
}=E_{\text{ \textsc{b}}}^{\text{\textsc{\b{n}}}}m^{\text{\textsc{b}}}$
relation permits to rewrite $\mathfrak{p}^{\text{\textsc{\b{m}}}}\eta _{%
\text{\textsc{\b{m}\b{n}}}}\mathfrak{p}^{\text{\textsc{\b{n}}}}$ like the
quadratic form $m^{\text{\textsc{a}}}\mathbf{Q}_{\text{\textsc{ab}}}m^{\text{%
\textsc{b}}}$ having the structure,
\begin{equation}
E_{\text{ \textsc{a}}}^{\text{\textsc{\b{m}}}}=\left(
\begin{array}{cc}
\delta _{ij} & +\frac{1}{2}\left( \mathcal{G}_{il}-\mathcal{B}_{il}\right)
\\
\delta _{kj} & -\frac{1}{2}\left( \mathcal{G}_{kl}+\mathcal{B}_{kl}\right)%
\end{array}%
\right) \qquad ,\qquad \mathbf{Q}_{\text{\textsc{ab}}}=\left(
\begin{array}{cc}
0 & I_{d} \\
I_{d} & 0%
\end{array}%
\right)
\end{equation}%
By turning off the flux $\mathcal{B}_{ij}$ and restricting the metric $%
\mathcal{G}_{ij}$ to $\delta _{ij},$ we get%
\begin{equation}
\begin{tabular}{lll}
$p_{L,i}$ & $=$ & $n_{i}+\frac{1}{2}\delta _{ij}w^{j}$ \\
$p_{R,a}$ & $=$ & $n_{a}-\frac{1}{2}\delta _{ij}w^{j}$%
\end{tabular}
\label{pl}
\end{equation}%
From these relations, we deduce
\begin{equation}
E_{\text{ \textsc{a}}}^{\text{\textsc{\b{m}}}}=\left(
\begin{array}{cc}
\delta _{ij} & +\frac{1}{2}\delta _{ij} \\
\delta _{kj} & -\frac{1}{2}\delta _{ij}%
\end{array}%
\right)
\end{equation}%
while $\mathbf{Q}_{\text{\textsc{ab}}}$ remains unchanged as it is
independent of the moduli.

To compare (\ref{pn}) with our realisation using the matrix moduli $%
x_{i}^{k} $, we consider the parametrisation (\ref{pr}) and perform its
projection along the $\sqrt{2}\mathbf{\chi }^{i}$ directions. Setting
\textsc{p}$_{L/R}^{i}=\sqrt{2}\boldsymbol{p}_{L/R}.\mathbf{\chi }^{i}$, this
yields%
\begin{equation}
\begin{tabular}{lll}
$\text{\textsc{p}}_{L}^{k}$ & $=$ & $n^{k}+w_{i}\mathcal{K}^{ik}$ \\
$\text{\textsc{p}}_{R}^{k}$ & $=$ & $n^{k}-\mathcal{K}^{ki}w_{i}$%
\end{tabular}%
\end{equation}%
By comparing these relationships with (\ref{pl}), we learn that non
vanishing flux $\mathcal{B}_{ij}$ requires non symmetric Cartan matrix $%
\mathcal{K}_{ik}=\mathbf{\alpha }_{i}^{\vee }.\mathbf{\alpha }_{j}$ of rank
2 Lie algebras. This is effectively the case of the NCFT$_{2}^{\mathbf{g}}$s
with underlying rank 2 symmetries such as the orthogonal so(5) and the
exceptional G$_{2}$ with Cartan matrices as follows
\begin{equation}
K_{ij}^{\mathbf{so}_{\mathbf{5}}}=\left(
\begin{array}{cc}
2 & -2 \\
-1 & 2%
\end{array}%
\right) \qquad ,\qquad \tilde{K}_{\mathbf{so}_{\mathbf{5}}}^{ij}=\frac{1}{2}%
\left(
\begin{array}{cc}
2 & 2 \\
1 & 2%
\end{array}%
\right)
\end{equation}%
and%
\begin{equation}
K_{ij}^{\mathbf{G}_{2}}=\left(
\begin{array}{cc}
2 & -3 \\
-1 & 2%
\end{array}%
\right) \qquad ,\qquad \tilde{K}_{\mathbf{G}_{2}}^{ij}=\left(
\begin{array}{cc}
2 & 3 \\
1 & 2%
\end{array}%
\right)
\end{equation}
\section{Higher rank NCFT$_{2}^{\text{\textbf{\b{g}}}}$}
\label{sec:4}
In this section, \textrm{we extend the analysis of the} previous
descriptions to higher rank symmetries. \textrm{Specifically, we study}
Narain sigma-models with target torus $\mathbb{T}^{r}\simeq U\left( 1\right)
^{r}$ for the unitary family \textrm{characterised by the Lie algebra}
\begin{equation}
\mathbf{g}_{r}=su(r+1)\qquad ,\qquad u\left( 1\right) ^{r}\subset \mathbf{g}%
_{r}
\end{equation}%
\textrm{We then derive} results for the other families appearing in eq(\ref%
{03}), classified by finite dimensional Lie algebras $\mathbf{g}_{r}$. For
convenience, our focus will be directed towards simply laced Lie algebras; a
similar description holds also for non simply laced symmetries.

\subsection{Narain CFT$_{2}^{\mathbf{su}_{r+1}}$}

\textrm{In the case of} Narain CFT$_{2}^{\mathrm{su}_{\mathrm{r+1}}}$, the
left $\boldsymbol{p}_{L}$ and the right $\boldsymbol{p}_{R}$ momenta of the
r two dimensional scalar fields $X^{i}=(X_{L}^{i}+X_{R}^{i})/\sqrt{2}$
compactified on the r-dimensional torus $\mathbb{T}^{r}=\otimes _{i=1}^{r}%
\mathbb{S}_{i}^{1}$ are given by the \textrm{following} linear combinations
\begin{equation}
\begin{tabular}{lll}
$\boldsymbol{p}_{L}$ & $=$ & $\frac{1}{\sqrt{2}}\dsum\limits_{i=1}^{r}\left(
n^{i}\mathbf{\beta }_{i}+w_{i}\mathbf{\chi }^{i}\right) $ \\
$\boldsymbol{p}_{R}$ & $=$ & $\frac{1}{\sqrt{2}}\dsum\limits_{i=1}^{r}\left(
n^{i}\mathbf{\beta }_{i}-w_{i}\mathbf{\chi }^{i}\right) $%
\end{tabular}
\label{pp}
\end{equation}%
with the constraint relation $\mathbf{\beta }_{i}.\mathbf{\chi }^{j}=\delta
_{i}^{j}$. Here, the $\mathbf{\beta }_{i}$s and the $\mathbf{\chi }^{i}$'s
are linear combinations of the scaled roots $\sum_{k=1}^{r}x_{i}^{k}\mathbf{%
\alpha }_{k}$ and the scaled weights $\sum_{l=1}^{2}y_{l}^{j}\mathbf{\lambda
}^{l}$. \textrm{The coefficients} are conditioned by the duality relation
\textrm{where }$y_{k}^{i}$\textrm{\ is }the inverse of $x_{i}^{k};$ \textrm{%
this leaves} \textrm{the} $r^{2}$ moduli that parameterise the moduli space
of the NCFT$_{2}^{\mathrm{su}_{\mathrm{r+1}}}$. \textrm{The corresponding}
intersection matrices $\mathcal{K}_{kl}^{\mathrm{su}_{\mathrm{r+1}}}=\mathbf{%
\beta }_{i}.\mathbf{\beta }_{j}$ and $\mathcal{\tilde{K}}_{\mathrm{su}_{%
\mathrm{r+1}}}^{ij}=\mathbf{\chi }^{i}.\mathbf{\chi }^{j}$\ read as\
\begin{equation}
\mathcal{K}_{ij}^{\mathrm{su}_{\mathrm{r+1}}}=%
\sum_{k,l=1}^{r}x_{i}^{k}K_{kl}^{\mathrm{su}_{\mathrm{r+1}}}x_{j}^{l}\qquad
,\qquad \mathcal{\tilde{K}}_{\mathrm{su}_{\mathrm{r+1}}}^{ij}=%
\sum_{k,l=1}^{r}y_{k}^{i}\tilde{K}_{\mathrm{su}_{\mathrm{r+1}}}^{kl}y_{l}^{j}
\label{kg}
\end{equation}%
with $K_{kl}^{\mathrm{su}_{\mathrm{r+1}}}=\mathbf{\alpha }_{i}.\mathbf{%
\alpha }_{j}$ and $\tilde{K}_{\mathrm{su}_{\mathrm{r+1}}}^{kl}=\mathbf{%
\lambda }^{i}.\mathbf{\lambda }^{j};$ they are respectively quadratic in $%
x_{i}^{k}$ and $y_{k}^{i}.$ \textrm{For simplicity}, we restrict the
analysis to the diagonal values with $x_{i}^{k}=x_{i}\delta _{i}^{k}$.

The ingredients in the series (\ref{pp}) are as follows:

\begin{description}
\item[$\left( \mathbf{i}\right) $] The integer vector $\boldsymbol{n}%
^{T}=\left( n^{1},...,n^{r}\right) $ \textrm{describing} the KK modes and
the integer vector $\boldsymbol{w}^{T}=\left( w_{1},...,w_{r}\right) $
\textrm{representing} the winding numbers.

\item[$\left( \mathbf{ii}\right) $] The vectors $\mathbf{\beta }_{i}=x_{i}%
\mathbf{\alpha }_{i}$, with fixed label i (e.g: $\mathbf{\beta }_{1}=x_{1}%
\mathbf{\alpha }_{1}$ and so on), are given by diagonal scalings of the
simple roots of su(r+1); the intersection matrix $\mathcal{K}%
_{ij}^{su_{r+1}}=\mathbf{\beta }_{i}.\mathbf{\beta }_{j}$ reads explicitly
as follows%
\begin{equation}
\left(
\begin{array}{ccccccc}
2x_{1}^{2} & -x_{1}x_{2} & 0 & \cdots & 0 & 0 & 0 \\
-x_{1}x_{2} & 2x_{2}^{2} & -x_{2}x_{3} &  & 0 & 0 & 0 \\
0 & -x_{2}x_{3} & 2x_{3}^{2} &  & 0 & 0 & 0 \\
\vdots &  &  & \ddots &  &  & \vdots \\
0 & 0 & 0 &  & 2x_{r-2}^{2} & -x_{r-2}x_{r-1} & 0 \\
0 & 0 & 0 &  & -x_{r-2}x_{r-1} & 2x_{r-1}^{2} & -x_{r-1}x_{r} \\
0 & 0 & 0 & \cdots & 0 & -x_{r-1}x_{r} & 2x_{r}^{2}%
\end{array}%
\right)  \label{ksur}
\end{equation}%
with moduli $x_{i}=1/(2R_{i})$.
\end{description}

The vectors $\mathbf{\chi }_{i}=\mathbf{\lambda }_{i}/x_{i}$ are scaled
fundamental weights of su(r+1) whose expression follows from the usual
duality property $\mathbf{\beta }_{i}\mathbf{.\chi }^{j}=\mathbf{\alpha }_{i}%
\mathbf{.\lambda }^{j}=\delta _{i}^{j}.$ Their intersections $\mathcal{%
\tilde{K}}_{su_{r+1}}^{ij}=\mathbf{\chi }^{i}.\mathbf{\chi }^{j}$ are given
by the inverse of the matrix $\mathcal{K}_{ij}^{su_{r+1}}$.

\subsubsection{Sitting at $x_{i}=1$ in $\mathcal{M}_{\mathbf{su}_{r+1}}$}

\textrm{For computational convenience}, we $\left( \mathbf{i}\right) $ sit
at $x_{i}=1$ in the moduli space of the NCFT$_{2}^{su_{r+1}}$, $\left(
\mathbf{ii}\right) $ use $\mathbf{\lambda }^{i}.\mathbf{\lambda }^{j}=%
\mathbf{\tilde{K}}_{su_{r+1}}^{ij}$ given by the inverse $\left( \mathbf{K}%
_{ij}^{su_{r+1}}\right) ^{-1}$, and \textrm{then} $\left( \mathbf{iii}%
\right) $ substitute $\mathbf{\alpha }_{i}=\mathbf{K}_{ij}\mathbf{\lambda }%
^{j}.$ \textrm{The resulting left and right momenta are expressed as}
\begin{equation}
\begin{tabular}{lll}
$\boldsymbol{p}_{L}$ & $=$ & $\frac{1}{\sqrt{2}}\dsum\limits_{i=1}^{r}\left(
n^{i}\mathbf{K}_{ij}+w_{j}\right) \mathbf{\lambda }^{j}$ \\
$\boldsymbol{p}_{R}$ & $=$ & $\frac{1}{\sqrt{2}}\dsum\limits_{i=1}^{r}\left(
n^{i}\mathbf{K}_{ij}-w_{j}\right) \mathbf{\lambda }^{j}$%
\end{tabular}
\label{lp}
\end{equation}%
From these expansions, we calculate
\begin{equation}
\begin{tabular}{lll}
$\boldsymbol{p}_{L}^{2}$ & $=$ & $\frac{1}{2}\dsum\limits_{i,j=1}^{r}\left(
n^{i}\mathbf{K}_{ij}n^{j}+w_{i}\mathbf{\tilde{K}}^{ij}w_{j}\right)
+\dsum\limits_{i=1}^{r}n^{i}w_{i}$ \\
$\boldsymbol{p}_{R}^{2}$ & $=$ & $\frac{1}{2}\dsum\limits_{i,j=1}^{r}\left(
n^{i}\mathbf{K}_{ij}n^{j}+w_{i}\mathbf{\tilde{K}}^{ij}w_{j}\right)
-\dsum\limits_{i=1}^{r}n^{i}w_{i}$%
\end{tabular}%
\end{equation}%
with\textrm{\ the difference}%
\begin{equation}
\boldsymbol{p}_{L}^{2}-\boldsymbol{p}_{R}^{2}=2\dsum%
\limits_{i=1}^{r}n^{i}w_{i}\in 2\mathbb{Z}
\end{equation}%
Setting $\mathfrak{Q}_{L}^{su_{r+1}}=\boldsymbol{p}_{L}^{2}$ and $\mathfrak{Q%
}_{R}^{su_{r+1}}=\boldsymbol{p}_{R}^{2};$ these relations \textrm{can be
expressed compactly} as follows%
\begin{eqnarray}
\mathfrak{Q}_{L}^{su_{r+1}} &=&\frac{1}{2}\boldsymbol{n}^{T}.\mathbf{K}.%
\boldsymbol{n}+\frac{1}{2}\boldsymbol{w}^{T}.\mathbf{\tilde{K}}.\boldsymbol{w%
}+\frac{1}{2}\left( \boldsymbol{n}^{T}.\boldsymbol{w}+\boldsymbol{w}^{T}.%
\boldsymbol{n}\right) \\
\mathfrak{Q}_{R}^{su_{r+1}} &=&\frac{1}{2}\boldsymbol{n}^{T}.\mathbf{K}.%
\boldsymbol{n}+\frac{1}{2}\boldsymbol{w}^{T}.\mathbf{\tilde{K}}.\boldsymbol{w%
}-\frac{1}{2}\left( \boldsymbol{n}^{T}.\boldsymbol{w}+\boldsymbol{w}^{T}.%
\boldsymbol{n}\right)
\end{eqnarray}%
with matrices $\boldsymbol{Q}_{L}^{su_{r+1}}$ and $\boldsymbol{Q}%
_{R}^{su_{r+1}}$\ given by
\begin{equation}
\boldsymbol{Q}_{L}^{su_{r+1}}=\frac{1}{2}\left(
\begin{array}{cc}
\mathbf{K}_{ij} & \mathbf{\alpha }_{i}\mathbf{.\lambda }^{l} \\
\mathbf{\lambda }^{k}\mathbf{.\alpha }_{j} & \mathbf{\tilde{K}}^{kl}%
\end{array}%
\right) \qquad ,\qquad \boldsymbol{Q}_{R}^{su_{r+1}}=\frac{1}{2}\left(
\begin{array}{cc}
\mathbf{K}_{ij} & -\mathbf{\alpha }_{i}\mathbf{.\lambda }^{l} \\
-\mathbf{\lambda }^{k}\mathbf{.\alpha }_{j} & \mathbf{\tilde{K}}^{kl}%
\end{array}%
\right)
\end{equation}%
\textrm{Putting them into the following definitions}
\begin{equation}
\begin{tabular}{lll}
$\boldsymbol{H}_{su_{r+1}}$ & $=$ & $\boldsymbol{Q}_{L}^{su_{r+1}}+%
\boldsymbol{Q}_{R}^{su_{r+1}}$ \\
$\boldsymbol{Q}_{su_{r+1}}$ & $=$ & $\boldsymbol{Q}_{L}^{su_{r+1}}-%
\boldsymbol{Q}_{R}^{su_{r+1}}$%
\end{tabular}%
\end{equation}%
we get%
\begin{equation}
\boldsymbol{H}_{su_{r+1}}=\left(
\begin{array}{cc}
\mathbf{K}_{ij} & \mathbf{0} \\
\mathbf{0} & \mathbf{\tilde{K}}^{kl}%
\end{array}%
\right) \qquad ,\qquad \boldsymbol{Q}_{su_{r+1}}=\left(
\begin{array}{cc}
0 & \mathbf{\alpha }_{i}\mathbf{.\lambda }^{l} \\
\mathbf{\lambda }^{k}\mathbf{.\alpha }_{j} & 0%
\end{array}%
\right)  \label{hdd}
\end{equation}%
Using $\mathbf{\alpha }_{i}\mathbf{.\lambda }^{l}=\delta _{i}^{l}\mathbf{,}$
we have\footnote{%
\ Given a $2r\times 2r$ matrix M with bloc form $\left( A,B,C,D\right) $
such that AC=CA, we have
\begin{equation*}
\det M=\det \left(
\begin{array}{cc}
A & B \\
C & D%
\end{array}%
\right) =\det \left( AD-CB\right)
\end{equation*}%
}%
\begin{equation}
\det \boldsymbol{Q}_{su_{r+1}}=\det \left( -I_{r}\right) =\left( -\right)
^{r}
\end{equation}%
Physically, the quadratic form $\mathcal{H}_{\mathbf{su}_{r+1}}=\boldsymbol{m%
}^{T}.\boldsymbol{H}_{su_{r+1}}.\boldsymbol{m},$ \textrm{equal to} $%
\boldsymbol{p}_{L}^{2}+\boldsymbol{p}_{R}^{2},$ \textrm{represents} the
total squared mass of the KK and winding modes%
\begin{equation}
\mathcal{H}_{su_{r+1}}=\dsum\limits_{i,j=1}^{r}n^{i}\mathbf{K}%
_{ij}n^{j}+\dsum\limits_{i,j=1}^{r}w_{i}\mathbf{\tilde{K}}^{ij}w_{j}
\end{equation}%
As for the quadratic form $\mathfrak{Q}_{\mathbf{su}_{r+1}}=\boldsymbol{m}%
^{T}.\boldsymbol{Q}_{su_{r+1}}.\boldsymbol{m},$ \textrm{defined by} $%
\boldsymbol{p}_{L}^{2}-\boldsymbol{p}_{R}^{2}$, \textrm{it gives} the gap
mass between left and right moving \textrm{modes} namely%
\begin{equation}
\mathfrak{Q}_{su_{r+1}}=2\dsum\limits_{i=1}^{r}n^{i}w_{i}
\end{equation}%
\textrm{Considering} $\boldsymbol{p}_{1}=\left( \boldsymbol{p}_{L}+%
\boldsymbol{p}_{R}\right) /\sqrt{2}$ and $\boldsymbol{p}_{2}=\left(
\boldsymbol{p}_{L}-\boldsymbol{p}_{R}\right) /\sqrt{2},$ \textrm{we obtain}%
\begin{equation}
\boldsymbol{p}_{1}=\dsum\limits_{i=1}^{r}n^{i}\mathbf{\alpha }_{i}\qquad
,\qquad \boldsymbol{p}_{2}=\dsum\limits_{i=1}^{r}w_{i}\mathbf{\lambda }^{i}
\end{equation}%
indicating that $\boldsymbol{p}_{1}$ labels the root lattice of su(r+1)
while $\boldsymbol{p}_{2}$ parameterises its weight lattice. \textrm{%
Furthermore,} we have
\begin{equation}
\mathcal{H}_{su_{r+1}}=\boldsymbol{p}_{1}^{2}+\boldsymbol{p}_{2}^{2}=%
\boldsymbol{n}^{T}.\mathbf{K}.\boldsymbol{n}+\boldsymbol{w}^{T}.\mathbf{%
\tilde{K}}.\boldsymbol{w}
\end{equation}%
and%
\begin{equation}
\mathfrak{Q}_{su_{r+1}}=2\boldsymbol{p}_{1}.\boldsymbol{p}%
_{2}=2\dsum\limits_{i=1}^{r}n^{i}w_{i}
\end{equation}%
This intersection can be factorised like $\mathfrak{Q}_{\mathbf{su}_{r+1}}=%
\mathfrak{p}^{\text{\textsc{\b{m}}}}\eta _{\text{\textsc{\b{m}\b{n}}}}%
\mathfrak{p}^{\text{\textsc{\b{n}}}}$ with \textrm{diagonal }pseudo-metric $%
\eta _{\text{\textsc{\b{m}\b{n}}}}$ given by $diag({\small [+]}^{r},{\small %
[-]}^{r})$ and
\begin{equation}
\mathfrak{p}^{\text{\textsc{\b{m}}}}=\left(
\begin{array}{c}
\frac{1}{\sqrt{2}}\left( K_{ij}n^{j}+\delta _{i}^{l}w_{l}\right) \\
\frac{1}{\sqrt{2}}\left( K_{kj}n^{j}-\delta _{k}^{l}w_{l}\right)%
\end{array}%
\right)  \label{ppm}
\end{equation}%
For the su(r+1) symmetry, the condition $\mathfrak{p}^{\text{\textsc{\b{m}}}%
}\eta _{\text{\textsc{\b{m}\b{n}}}}\mathfrak{p}^{\text{\textsc{\b{n}}}}\in 2%
\mathbb{Z}$ \textrm{is\ satisfied by} the lattice
\begin{equation}
{\large \Lambda }_{su_{r+1}}=\left\{ \boldsymbol{m}\in \mathbb{Z}^{4},\quad
\mathfrak{Q}_{su_{r+1}}=\boldsymbol{m}^{\text{\textsc{a}}}{\large g}_{\text{%
\textsc{ab}}}\boldsymbol{m}^{\text{\textsc{b}}}\quad |\quad {\large g}_{%
\text{\textsc{ab}}}=(\boldsymbol{Q}_{su_{r+1}})_{\text{\textsc{ab}}}\right\}
\end{equation}%
where $\boldsymbol{m}=\left( \mathbf{n},\mathbf{w}\right) \in \mathbb{Z}%
^{r}\times \mathbb{Z}^{r}$ and the uni-modularity property $\left\vert \det
\boldsymbol{Q}_{su_{r+1}}\right\vert =1$\textrm{\ holds}$.$ For this
lattice, we indeed have the property $\mathfrak{Q}_{su_{r+1}}=2\mathbf{n}.%
\mathbf{w}$ which is an even integer.

\subsubsection{Partition function $Z_{\mathbf{su}_{r+1}}$}

The genus-one partition function $Z_{\mathbf{su}_{r+1}}\left[ \tau ,\bar{\tau%
};\mathbf{x}\right] $ of the compactified conformal theory, with left $%
\boldsymbol{p}_{L}$ and right $\boldsymbol{p}_{R}$ momenta (\ref{pp}), is as
follows
\begin{equation}
Z_{\mathrm{su}_{r+1}}\left[ \tau ,\bar{\tau};\mathbf{x}\right] =\frac{1}{%
\left\vert \eta \left( \tau \right) \right\vert ^{2r}}\vartheta
_{su_{r+1}}\left( \tau ,\bar{\tau};\mathbf{x}\right)
\end{equation}%
where $\eta \left( \tau \right) $ is the Dedekin eta function given by $%
q^{1/24}\dprod\nolimits_{n=1}^{\infty }\left( 1-q^{n}\right) $ and where%
\begin{eqnarray}
\vartheta _{\mathrm{su}_{r+1}}\left( \tau ,\bar{\tau};\mathbf{x}\right)
&=&\dsum\limits_{\left( p_{L},p_{R}\right) }q^{\frac{1}{2}p_{L}^{2}}\bar{q}^{%
\frac{1}{2}p_{R}^{2}}  \notag \\
&=&\frac{1}{\left\vert \eta \left( \tau \right) \right\vert ^{2r}}%
\dsum\limits_{(\mathfrak{Q}_{L}^{\mathtt{su}_{r+1}},\mathfrak{Q}_{R}^{%
\mathtt{su}_{r+1}})}q^{\frac{1}{2}\mathfrak{Q}_{L}^{\mathtt{su}_{r+1}}}\bar{q%
}^{\frac{1}{2}\mathfrak{Q}_{R}^{\mathtt{su}_{r+1}}}
\end{eqnarray}%
with $q=e^{2\pi i\tau }$. Substituting $\tau =\tau _{1}+i\tau _{2}$ ($\func{%
Im}\tau _{2}>0$) and $\mathfrak{Q}_{\mathrm{su}_{r+1}}=p_{L}^{2}-p_{R}^{2}$
as well as $\mathcal{H}_{\mathrm{su}_{r+1}}=p_{L}^{2}+p_{R}^{2},$ we end up
with%
\begin{equation}
Z_{\mathbf{su}_{r+1}}\left[ \tau ,\bar{\tau};\mathbf{x}\right] =\frac{1}{%
\left\vert \eta \left( \tau \right) \right\vert ^{2r}}\dsum\limits_{%
\boldsymbol{m}\in \mathbb{Z}^{2r}}\exp \left( \boldsymbol{m}^{T}.\boldsymbol{%
A}_{su_{r+1}}.\boldsymbol{m}\right)  \label{pa1}
\end{equation}%
where the $2r\times 2r$ matrix $\boldsymbol{A}_{\mathbf{su}_{r+1}}$ is given
by $i\pi \left( \tau \boldsymbol{Q}_{L}^{\mathrm{su}_{r+1}}-\bar{\tau}%
\boldsymbol{Q}_{R}^{\mathrm{su}_{r+1}}\right) $. It also splits in terms of $%
\boldsymbol{Q}_{su_{r+1}}$ and $\boldsymbol{H}_{\mathbf{su}_{r+1}}$ as
follows%
\begin{equation}
\boldsymbol{A}_{\mathbf{su}_{r+1}}=i\pi \tau _{1}\boldsymbol{Q}%
_{su_{r+1}}-\pi \tau _{2}\boldsymbol{H}_{su_{r+1}}  \label{pa2}
\end{equation}%
reading explicitly like%
\begin{equation}
\left. \boldsymbol{A}_{\mathrm{su}_{3}}\right\vert _{x_{i}=1}=\left(
\begin{array}{cc}
-\pi \mathrm{\tau }_{2}\mathcal{K}_{\mathtt{su}_{r+1}} & i\pi \mathrm{\tau }%
_{1}I_{{\small r}} \\
i\pi \mathrm{\tau }_{1}I_{{\small r}} & -\pi \mathrm{\tau }_{2}\mathcal{%
\tilde{K}}_{\mathtt{su}_{r+1}}%
\end{array}%
\right)
\end{equation}%
with $\det \boldsymbol{A}_{\mathrm{su}_{3}}$ equal to $\pi ^{2r}\left( \tau
\bar{\tau}\right) ^{r}$ which is independent of the moduli $x_{i}.$ From
this \textrm{formulation}, \textrm{we deduce that the key} properties of
Narain CFT$^{\mathrm{su}_{r+1}}$s \textrm{are encapsulated in} the\ two
square matrices $\boldsymbol{Q}_{su_{r+1}}$ and $\boldsymbol{H}_{su_{r+1}}$.
\textrm{Therefore, we proceed to determine} the values of these two basic
ingredients \textrm{in the following subsection.}

\subsection{Classification of Narain CFT$_{2}^{\text{\textbf{\b{g}}}}$}

From \textrm{the previous description of} the family NCFT$_{2}^{\mathrm{su}%
_{r+1}}$, we learnt that \textrm{a significant} subset of Narain CFTs can be
classified by finite dimensional Lie algebras as NCFT$_{2}^{\text{\textbf{\b{%
g}}}}$. In this classification, the NCFT$_{2}^{\mathrm{su}_{r+1}}$ addressed
\textrm{earlier} constitutes the first class of Narain conformal field
theories. \textrm{The remaining} CFTs, based on the Lie algebras
\begin{equation}
\text{\textbf{g}}_{r}=B_{r},\quad C_{r},\quad D_{r},\quad F_{4},\quad
G_{2},\quad E_{6,7,8}  \label{g}
\end{equation}%
\textrm{are described in a manner analogous to the class} $A_{r}=\mathrm{su}%
_{r+1}$. For all these NCFT$_{2}^{\text{\textbf{\b{g}}}}$'s, the left $%
\boldsymbol{p}_{L}$ and right $\boldsymbol{p}_{R}$ momenta expand \textrm{in
an equivalent way to} (\ref{pp}) and (\ref{lp}). At $x_{i}=1$ in the Narain
moduli space, we have
\begin{equation}
\begin{tabular}{lll}
$\boldsymbol{p}_{L}$ & $=$ & $\frac{1}{\sqrt{2}}\dsum\limits_{i=1}^{r}\left(
n^{i}\mathbf{K}_{ij}^{(g)}+w_{j}\right) \mathbf{\lambda }^{j}$ \\
$\boldsymbol{p}_{R}$ & $=$ & $\frac{1}{\sqrt{2}}\dsum\limits_{i=1}^{r}\left(
n^{i}\mathbf{K}_{ij}^{(g)}-w_{j}\right) \mathbf{\lambda }^{j}$%
\end{tabular}
\label{K}
\end{equation}%
where $n^{i},$ $w_{j}$\ are integers. Moreover, the $\mathbf{K}_{ij}^{(g)}$
is the intersection matrix $\mathbf{\alpha }_{i}.\mathbf{\alpha }_{j}$ of
the cycles $\mathbf{\alpha }_{i}$ and $\mathbf{\lambda }^{j}$'s are their
duals; i.e: $\mathbf{\alpha }_{i}.\mathbf{\lambda }^{j}=\delta _{i}^{j}.$
\textrm{It is also possible to use the following scaled quantities:} $\left(
\mathbf{i}\right) $ the scaled $\mathbf{\beta }_{i}=\sum_{k}x_{i}^{k}\mathbf{%
\alpha }_{k}$ \textrm{where} $\mathbf{\alpha }_{k}$ \textrm{are} the simple
roots of the Lie algebras (\ref{g}), and $\left( \mathbf{ii}\right) $ the
scaled $\mathbf{\chi }^{i}=\sum_{l}y_{l}^{i}\mathbf{\lambda }^{l}$ with the
constraint $\sum_{k}x_{i}^{k}y_{k}^{j}=\delta _{i}^{j}$ (duality relation).
For the diagonal realisation, we have $\mathbf{\beta }_{i}=x_{i}\mathbf{%
\alpha }_{i}$ and $\mathbf{\chi }^{i}=\frac{1}{x_{i}}\mathbf{\lambda }^{i}.$

\begin{figure}[tbph]
\begin{center}
\includegraphics[width=5cm]{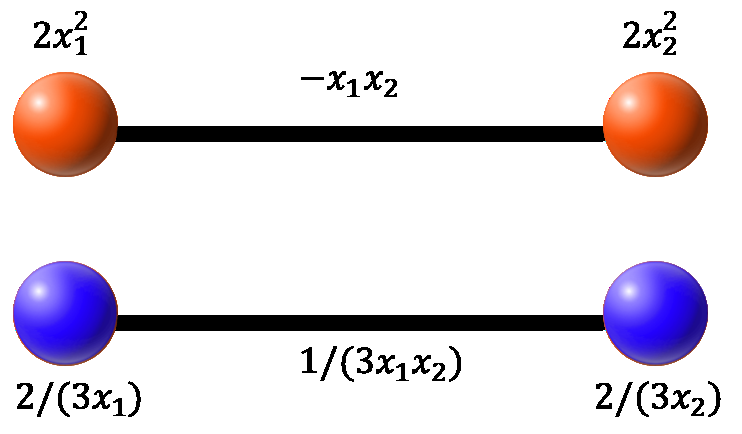}
\end{center}
\par
\vspace{-0.5cm}
\caption{Graphic representations of the intersection matrices $\mathbf{K}^{%
\mathrm{su}_{3}}$ and $\mathbf{\tilde{K}}^{\mathrm{su}_{3}}$. The two nodes
of $\mathbf{K}^{\mathrm{su}_{3}}$ and $\mathbf{\tilde{K}}^{\mathrm{su}_{3}}$
represent the two circles used in the compactification. The $%
x_{i}=1/(2R_{i}) $ gives the real moduli of the 2-torus and its dual. For $%
x_{i}=1,$ the $\mathbf{K}^{\mathrm{su}_{3}}$ \textrm{leads to} the Dynkin
diagram of su(3).}
\label{su3}
\end{figure}

In this generic situation, the Cartan matrix $\mathbf{K}_{ij}^{(g)}$\ in (%
\ref{K}) should be replaced by the scaled $\mathcal{K}_{ij}^{(g)}=\mathbf{%
\beta }_{i}.\mathbf{\beta }_{j}$ and the $\mathbf{\tilde{K}}_{(g)}^{ij}$ by
the scaled $\mathcal{\tilde{K}}_{(g)}^{ij}=\mathbf{\chi }^{i}.\mathbf{\chi }%
^{j}$. \textrm{To clarify the procedure,} we provide an illustration for the
rank 2 algebras. We have
\begin{equation}
\begin{tabular}{lll}
$\mathbf{\beta }_{1}$ & $=$ & $x_{1}\mathbf{\alpha }_{1}$ \\
$\mathbf{\beta }_{2}$ & $=$ & $x_{2}\mathbf{\alpha }_{2}$%
\end{tabular}%
\qquad ,\qquad
\begin{tabular}{lll}
$\mathbf{\chi }^{1}$ & $=$ & $\frac{1}{x_{1}}\mathbf{\lambda }^{1}$ \\
$\mathbf{\chi }^{2}$ & $=$ & $\frac{1}{x_{2}}\mathbf{\lambda }^{2}$%
\end{tabular}
\label{bx}
\end{equation}%
where $x_{i}=1/\left( 2R_{i}\right) .$ Here, $\mathbf{\alpha }_{1},\mathbf{%
\alpha }_{2}$ are the simple roots and $\mathbf{\lambda }^{1},\mathbf{%
\lambda }^{2}$ are the fundamental weights. For the su(3) example, the
intersection matrices $\mathcal{K}^{\mathrm{su}_{3}}$ and $\mathcal{\tilde{K}%
}^{\mathrm{su}_{3}}$\ are given by eq(\ref{kkt}); they can be represented by
the x$_{i}$- dependent\textrm{\ diagram in Figure} \textbf{\ref{su3}}.
Similar relations can be also written down for the other rank 2 Lie algebras
$so(5)\simeq sp(4)$ and G$_{2}$.

\textrm{Next, we give }results for the Narain CFT$^{\mathbf{\text{\b{g}}}}$%
's classified by the simply laced ADE Lie algebras:

\subsubsection{Linear su$\left( r+1\right) $ class: NCFT$_{2}^{\mathbf{su}%
_{r+1}}$}

\begin{figure}[tbph]
\begin{center}
\includegraphics[width=10cm]{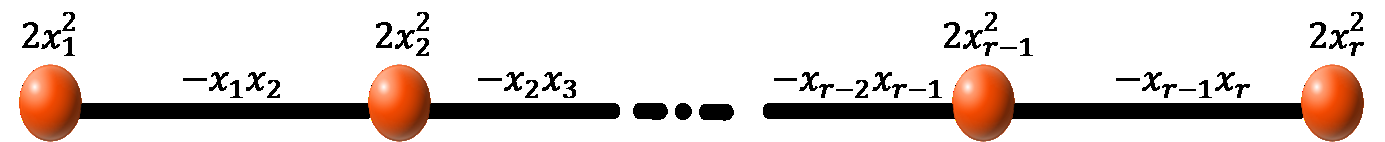}
\end{center}
\par
\vspace{-0.5cm}
\caption{Graphic representation of the intersection matrix $\mathbf{K}%
_{su_{r+1}}$. For the cases of su(4) and su(5), they are explicitly given by
eqs(\protect\ref{40}-\protect\ref{41}).}
\label{surk}
\end{figure}

The class NCFT$^{\mathbf{su}_{r+1}}$ is characterised by the intersection
matrix $\mathcal{K}_{\mathbf{su}_{r+1}}$ (\ref{ksur}) and its inverse $%
\mathcal{\tilde{K}}_{\mathbf{su}_{r+1}}$ \textrm{as depicted by the
following graphical representation in} \textbf{Figure \ref{surk}}. \ \newline

For illustration, \textrm{we consider examples }of rank 3 Narain CFT$%
_{2}^{su_{4}}$ and the rank 4 CFT$_{2}^{su_{5}}:$

\begin{itemize}
\item \textbf{Rank 3 Narain CFT}$_{2}^{su_{4}}$: \newline
For this rank 3 unitary CFT, the intersection matrix $\mathcal{K}_{su_{4}}$,
between the cycles $\mathbf{\beta }_{i}$ as in (\ref{bx}), and the $\mathcal{%
\tilde{K}}_{su_{4}}^{ij}$, between its dual cycles $\mathbf{\chi }^{i}$, are
given by:%
\begin{equation}
\mathcal{K}_{su_{4}}=\left(
\begin{array}{ccc}
2x_{1}^{2} & -x_{1}x_{2} & 0 \\
-x_{1}x_{2} & 2x_{2}^{2} & -x_{2}x_{3} \\
0 & -x_{2}x_{3} & 2x_{3}^{2}%
\end{array}%
\right)  \label{40}
\end{equation}%
and
\begin{equation}
\mathcal{\tilde{K}}_{su_{4}}=\left(
\begin{array}{ccc}
\frac{3}{4x_{1}^{2}} & \frac{1}{2x_{1}x_{2}} & \frac{1}{4x_{1}x_{3}} \\
\frac{1}{2x_{1}x_{2}} & \frac{1}{x_{2}^{2}} & \frac{1}{2x_{2}x_{3}} \\
\frac{1}{4x_{1}x_{3}} & \frac{1}{2x_{2}x_{3}} & \frac{3}{4x_{3}^{2}}%
\end{array}%
\right)
\end{equation}%
We also have $\det \mathcal{K}_{su_{4}}=4x_{1}^{2}x_{2}^{2}x_{3}^{2}.$%
\newline
Here, the condition $\mathfrak{p}^{\text{\textsc{\b{m}}}}\eta _{\text{%
\textsc{\b{m}\b{n}}}}\mathfrak{p}^{\text{\textsc{\b{n}}}}\in 2\mathbb{Z}$
is\ solved by the lattice
\begin{equation}
{\large \Lambda }_{\mathtt{su}_{4}}=\left\{ \boldsymbol{m}\in \mathbb{Z}%
^{6},\quad \mathfrak{Q}_{\mathtt{su}_{4}}=\boldsymbol{m}^{\text{\textsc{a}}}%
{\large g}_{\text{\textsc{ab}}}\boldsymbol{m}^{\text{\textsc{b}}}\quad
|\quad {\large g}_{\text{\textsc{ab}}}=\boldsymbol{Q}_{\text{\textsc{ab}}}^{%
\mathtt{su}_{4}}\right\}
\end{equation}%
with%
\begin{equation}
\boldsymbol{Q}_{\text{\textsc{ab}}}^{\mathtt{su}_{4}}=\left(
\begin{array}{cc}
0 & I_{3} \\
I_{3} & 0%
\end{array}%
\right) \qquad ,\qquad \det \boldsymbol{Q}_{\text{\textsc{ab}}}^{\mathtt{su}%
_{4}}=-1
\end{equation}

\item \textbf{Rank 4 Narain CFT}$_{2}^{su_{5}}$: \newline
For this rank 4 Narain conformal theory, we have%
\begin{equation}
\mathcal{K}_{su_{5}}=\left(
\begin{array}{cccc}
2x_{1}^{2} & -x_{1}x_{2} & 0 & 0 \\
-x_{1}x_{2} & 2x_{2}^{2} & -x_{2}x_{3} & 0 \\
0 & -x_{2}x_{3} & 2x_{3}^{2} & -x_{3}x_{4} \\
0 & 0 & -x_{3}x_{4} & 2x_{4}^{2}%
\end{array}%
\right)  \label{41}
\end{equation}%
with $\det \mathcal{K}_{su_{5}}=5x_{1}^{2}x_{2}^{2}x_{3}^{2}x_{4}^{2}$.
\end{itemize}

The associated $\mathcal{\tilde{K}}_{su_{5}}$ is given by
\begin{equation}
\mathcal{\tilde{K}}_{su_{5}}=\left(
\begin{array}{cccc}
\frac{4}{5x_{1}^{2}} & \frac{3}{5x_{1}x_{2}} & \frac{2}{5x_{1}x_{3}} & \frac{%
1}{5x_{1}x_{4}} \\
\frac{3}{5x_{1}x_{2}} & \frac{6}{5x_{2}^{2}} & \frac{4}{5x_{2}x_{3}} & \frac{%
2}{5x_{2}x_{4}} \\
\frac{2}{5x_{1}x_{3}} & \frac{4}{5x_{2}x_{3}} & \frac{6}{5x_{3}^{2}} & \frac{%
3}{5x_{3}x_{4}} \\
\frac{1}{5x_{1}x_{4}} & \frac{2}{5x_{2}x_{4}} & \frac{3}{5x_{3}x_{4}} &
\frac{4}{5x_{4}^{2}}%
\end{array}%
\right)  \label{42}
\end{equation}%
The condition $\mathfrak{p}^{\text{\textsc{\b{m}}}}\eta _{\text{\textsc{\b{m}%
\b{n}}}}\mathfrak{p}^{\text{\textsc{\b{n}}}}\in 2\mathbb{Z}$ is\ solved by
the lattice
\begin{equation}
{\large \Lambda }_{\mathtt{su}_{5}}=\left\{ \boldsymbol{m}\in \mathbb{Z}%
^{8},\quad \mathfrak{Q}_{\mathtt{su}_{5}}=\boldsymbol{m}^{\text{\textsc{a}}}%
{\large g}_{\text{\textsc{ab}}}\boldsymbol{m}^{\text{\textsc{b}}}\quad
|\quad {\large g}_{\text{\textsc{ab}}}=\boldsymbol{Q}_{\text{\textsc{ab}}}^{%
\mathtt{su}_{5}}\right\}
\end{equation}%
with%
\begin{equation}
\boldsymbol{Q}_{\text{\textsc{ab}}}^{\mathtt{su}_{5}}=\left(
\begin{array}{cc}
0 & I_{4} \\
I_{4} & 0%
\end{array}%
\right) \qquad ,\qquad \det \boldsymbol{Q}_{\text{\textsc{ab}}}^{\mathtt{su}%
_{5}}=+1
\end{equation}

From equations (\ref{40}-\ref{41}), one also learn that for the higher rank
su(r+1) series, the \textrm{determinant is of the form}%
\begin{equation}
\det \mathcal{K}_{su_{r+1}}=(r+1)\dprod\limits_{i=1}^{r}x_{i}^{2}
\end{equation}%
which by setting $x_{i}=1$, \textrm{one recovers the usual value} $\det
K_{su_{r+1}}=r+1$. For the general relationships of (\ref{kg}) namely $%
\mathcal{K}_{ij}^{\mathrm{su}_{\mathrm{r+1}}}=%
\sum_{k,l=1}^{r}x_{i}^{k}K_{kl}^{\mathrm{su}_{\mathrm{r+1}}}x_{j}^{l}$, we
have%
\begin{equation}
\det \mathcal{K}_{su_{r+1}}=\det K_{\mathrm{su}_{\mathrm{r+1}}}\left( \det
x_{i}^{k}\right) ^{2}=(r+1)\left( \det x_{i}^{k}\right) ^{2}
\end{equation}%
\textrm{It is also worth mentioning} that the genus-one partition function
of these NCFT$_{2}^{su_{r+1}}$ is given by eqs(\ref{pa1}-\ref{pa2}).\newline
The condition $\mathfrak{p}^{\text{\textsc{\b{m}}}}\eta _{\text{\textsc{\b{m}%
\b{n}}}}\mathfrak{p}^{\text{\textsc{\b{n}}}}\in 2\mathbb{Z}$ is\ solved by
the lattice
\begin{equation}
{\large \Lambda }_{\mathtt{su}_{r+1}}=\left\{ \boldsymbol{m}\in \mathbb{Z}%
^{2r},\quad \mathfrak{Q}_{\mathtt{su}_{r+1}}=\boldsymbol{m}^{\text{\textsc{a}%
}}{\large g}_{\text{\textsc{ab}}}\boldsymbol{m}^{\text{\textsc{b}}}\quad
|\quad {\large g}_{\text{\textsc{ab}}}=\boldsymbol{Q}_{\text{\textsc{ab}}}^{%
\mathtt{su}_{r+1}}\right\}
\end{equation}%
with%
\begin{equation}
\boldsymbol{Q}_{\text{\textsc{ab}}}^{\mathtt{su}_{r+1}}=\left(
\begin{array}{cc}
0 & I_{r} \\
I_{r} & 0%
\end{array}%
\right) \qquad ,\qquad \det \boldsymbol{Q}_{\text{\textsc{ab}}}^{\mathtt{su}%
_{r+1}}=\left( -\right) ^{r}
\end{equation}

\subsubsection{Orthogonal so(2r) class: NCFT$_{2}^{\mathbf{so}_{2r}}$}

\begin{figure}[tbph]
\begin{center}
\includegraphics[width=12cm]{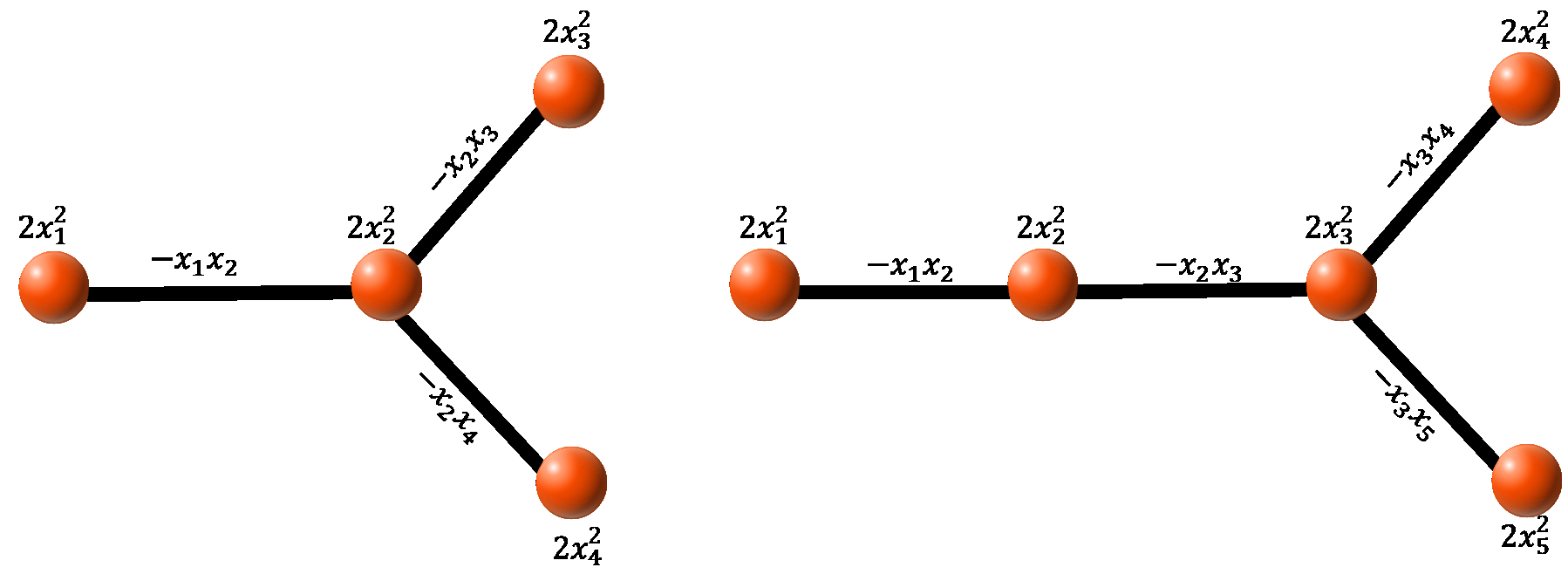}
\end{center}
\par
\vspace{-0.5cm}
\caption{Graphic representations of the intersection matrices $\mathbf{K}^{%
\mathrm{so}_{8}}$ and $\mathbf{K}^{\mathrm{so}_{10}}$.}
\label{so}
\end{figure}

We illustrate the orthogonal class of NCFT$_{2}^{\mathbf{so}_{2r}}$s \textrm{%
by studying the first two} members of the family namely\ the rank 4 NCFT$%
_{2}^{\mathbf{so}_{8}}$ and the rank 5 NCFT$_{2}^{\mathbf{so}_{10}}$
represented by the graphics \textbf{\ref{so}}.

\begin{itemize}
\item \textbf{Rank }4\textbf{\ Narain CFT}$_{2}^{so_{8}}$: \newline
For this rank 4 orthogonal Narain CFT$_{2}^{so_{8}}$, the intersection
matrix $\mathcal{K}_{ij}^{so_{8}}=\mathbf{\beta }_{i}.\mathbf{\beta }_{j},$
and its dual $\mathcal{\tilde{K}}_{so_{8}}^{ij}=\mathbf{\chi }^{i}.\mathbf{%
\chi }^{j}$ are given by%
\begin{equation}
\mathcal{K}_{ij}^{so_{8}}=\left(
\begin{array}{cccc}
2x_{1}^{2} & -x_{1}x_{2} & 0 & 0 \\
-x_{1}x_{2} & 2x_{2}^{2} & -x_{2}x_{3} & -x_{2}x_{4} \\
0 & -x_{2}x_{3} & 2x_{3}^{2} & 0 \\
0 & -x_{2}x_{4} & 0 & 2x_{4}^{2}%
\end{array}%
\right)
\end{equation}%
with determinant $4x_{1}^{2}x_{2}^{2}x_{3}^{2}x_{4}^{2}$; and%
\begin{equation}
\mathcal{\tilde{K}}_{so_{8}}^{ij}=\left(
\begin{array}{cccc}
\frac{1}{x_{1}^{2}} & \frac{1}{x_{1}x_{2}} & \frac{1}{2x_{1}x_{3}} & \frac{1%
}{2x_{1}x_{4}} \\
\frac{1}{x_{1}x_{2}} & \frac{2}{x_{2}^{2}} & \frac{1}{x_{2}x_{3}} & \frac{1}{%
x_{2}x_{4}} \\
\frac{1}{2x_{1}x_{3}} & \frac{1}{x_{2}x_{3}} & \frac{1}{x_{3}^{2}} & \frac{1%
}{2x_{3}x_{4}} \\
\frac{1}{2x_{1}x_{4}} & \frac{1}{x_{2}x_{4}} & \frac{1}{2x_{3}x_{4}} & \frac{%
1}{x_{4}^{2}}%
\end{array}%
\right)
\end{equation}%
\textrm{Using these matrices, the} left and the right momenta \textrm{follow}
from (\ref{K}) and they allow to calculate the $\mathfrak{Q}_{\mathrm{so}%
_{8}}$ and $\mathcal{H}_{\mathrm{so}_{8}}$ as well as the associated
relations such as the genus-one partition function.

\item \textbf{Rank }5\textbf{\ Narain CFT}$_{2}^{so_{10}}$: \newline
For this case, we have%
\begin{equation}
\mathcal{K}_{so_{10}}=\left(
\begin{array}{ccccc}
2x_{1}^{2} & -x_{1}x_{2} & 0 & 0 & 0 \\
-x_{1}x_{2} & 2x_{2}^{2} & -x_{2}x_{3} & 0 & 0 \\
0 & -x_{2}x_{3} & 2x_{3}^{2} & -x_{3}x_{4} & -x_{3}x_{5} \\
0 & 0 & -x_{3}x_{4} & 2x_{4}^{2} & 0 \\
0 & 0 & -x_{3}x_{5} & 0 & 2x_{5}^{2}%
\end{array}%
\right)
\end{equation}%
with determinant $4x_{1}^{2}x_{2}^{2}x_{3}^{2}x_{4}^{2}x_{5}^{2}$; and%
\begin{equation}
\mathcal{\tilde{K}}_{\mathrm{so}_{10}}=\left(
\begin{array}{ccccc}
\frac{1}{x_{1}^{2}} & \frac{1}{x_{1}x_{2}} & \frac{1}{x_{1}x_{3}} & \frac{1}{%
2x_{1}x_{4}} & \frac{1}{2x_{1}x_{5}} \\
\frac{1}{x_{1}x_{2}} & \frac{2}{x_{2}^{2}} & \frac{2}{x_{2}x_{3}} & \frac{1}{%
x_{2}x_{4}} & \frac{1}{x_{2}x_{5}} \\
\frac{1}{x_{1}x_{3}} & \frac{2}{x_{2}x_{3}} & \frac{3}{x_{3}^{2}} & \frac{3}{%
2x_{3}x_{4}} & \frac{3}{2x_{3}x_{5}} \\
\frac{1}{2x_{1}x_{4}} & \frac{1}{x_{2}x_{4}} & \frac{3}{2x_{3}x_{4}} & \frac{%
5}{4x_{4}^{2}} & \frac{3}{4x_{4}x_{5}} \\
\frac{1}{2x_{1}x_{5}} & \frac{1}{x_{2}x_{5}} & \frac{3}{2x_{3}x_{5}} & \frac{%
3}{4x_{4}x_{5}} & \frac{5}{4x_{5}^{2}}%
\end{array}%
\right)
\end{equation}%
\begin{equation*}
\end{equation*}
\end{itemize}

From these two examples, we deduce that for higher rank r of the orthogonal
so(2r) series, \textrm{the determinant is given by}%
\begin{equation}
\det \mathcal{K}_{so_{2r}}=4\dprod\limits_{i=1}^{r}x_{i}^{2}
\end{equation}%
\textrm{Analogously to} the unitary series, the partition function for
orthogonal CFT's reads as
\begin{equation}
Z_{so_{2r}}\left[ \tau ,\bar{\tau};\mathbf{x}\right] =\frac{1}{\left\vert
\eta \left( \tau \right) \right\vert ^{2r}}\dsum\limits_{\boldsymbol{m}\in
\mathbb{Z}^{2}}\exp \left( \boldsymbol{m}^{T}.\boldsymbol{D}_{so_{2r}}.%
\boldsymbol{m}\right)
\end{equation}%
with
\begin{equation}
\boldsymbol{D}_{so_{2r}}=i\pi \left( \tau \boldsymbol{Q}_{L}^{\mathrm{so}%
_{2r}}-\bar{\tau}\boldsymbol{Q}_{R}^{\mathrm{so}_{2r}}\right)
\end{equation}%
reading as follows%
\begin{equation}
\boldsymbol{D}_{so_{2r}}=i\pi \tau _{1}\boldsymbol{Q}_{so_{2r}}-\pi \tau _{2}%
\boldsymbol{H}_{so_{2r}}
\end{equation}%
This analysis extends straightforwardly to the other members of the
orthogonal NCFT$_{2}^{so_{2r}}$ family; \textrm{report to appendix B for
useful properties}.

For these orthogonal CFT's, the condition $\mathfrak{p}^{\text{\textsc{\b{m}}%
}}\eta _{\text{\textsc{\b{m}\b{n}}}}\mathfrak{p}^{\text{\textsc{\b{n}}}}\in 2%
\mathbb{Z}$ is\ solved by the lattice
\begin{equation}
{\large \Lambda }_{so_{2r}}=\left\{ \boldsymbol{m}\in \mathbb{Z}^{2r},\quad
\mathfrak{Q}_{so_{2r}}=\boldsymbol{m}^{\text{\textsc{a}}}{\large g}_{\text{%
\textsc{ab}}}\boldsymbol{m}^{\text{\textsc{b}}}\quad |\quad {\large g}_{%
\text{\textsc{ab}}}=\boldsymbol{Q}_{\text{\textsc{ab}}}^{\mathtt{so}%
_{2r}}\right\}
\end{equation}%
with%
\begin{equation}
\boldsymbol{Q}_{\text{\textsc{ab}}}^{\mathtt{so}_{2r}}=\left(
\begin{array}{cc}
0 & I_{r} \\
I_{r} & 0%
\end{array}%
\right) \qquad ,\qquad \det \boldsymbol{Q}_{\text{\textsc{ab}}}^{\mathtt{so}%
_{2r}}=\left( -\right) ^{r}
\end{equation}%
\textrm{It should be noted that the difference between unitary and
orthogonal Narain CFTs lies in the matrix }$\boldsymbol{H}_{\text{\textsc{ab}%
}}^{\mathtt{so}_{2r}}$\textrm{\ which differs from} $\boldsymbol{H}_{\text{%
\textsc{ab}}}^{su_{r+1}}$.

\subsubsection{Exceptional class: NCFT$_{2}^{\mathbf{e}_{r}}$}

\begin{figure}[tbph]
\begin{center}
\includegraphics[width=10cm]{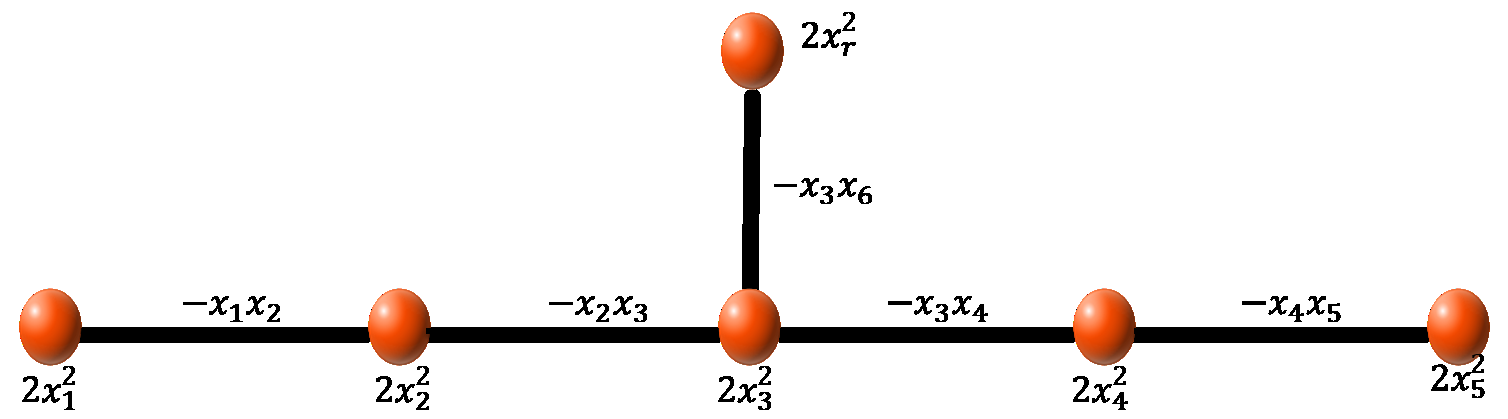}
\end{center}
\par
\vspace{-0.5cm}
\caption{Graphic representation of the intersection matrix $\mathbf{K}^{%
\mathrm{E}_{6}}$.}
\label{e6}
\end{figure}

We illustrate the exceptional class containing three members by giving the
data of the rank 6 conformal theory NCFT$_{2}^{\mathbf{e}_{6}}$ represented
by the graphic \textbf{\ref{e6}}.
For this E$_{6}$ conformal theory, the basic results are captured by the
intersection matrix $\mathcal{K}_{e_{6}}$ of the $\mathbf{\beta }_{i}$%
-cycles;%
\begin{equation}
\mathcal{K}_{e_{6}}=\left(
\begin{array}{cccccc}
2x_{1}^{2} & -x_{1}x_{2} & 0 & 0 & 0 & 0 \\
-x_{1}x_{2} & 2x_{2}^{2} & -x_{2}x_{3} & 0 & 0 & 0 \\
0 & -x_{2}x_{3} & 2x_{3}^{2} & -x_{3}x_{4} & 0 & -x_{3}x_{6} \\
0 & 0 & -x_{3}x_{4} & 2x_{4}^{2} & -x_{4}x_{5} & 0 \\
0 & 0 & 0 & -x_{4}x_{5} & 2x_{5}^{2} & 0 \\
0 & 0 & -x_{3}x_{6} & 0 & 0 & 2x_{6}^{2}%
\end{array}%
\right)
\end{equation}%
with determinant $3x_{1}^{2}x_{2}^{2}x_{3}^{2}x_{4}^{2}x_{5}^{2}x_{6}^{2}$.
Its inverse gives the intersection of the dual cycles; it reads as follows%
\begin{equation}
\mathcal{\tilde{K}}_{e_{6}}=\left(
\begin{array}{cccccc}
\frac{4}{3x_{1}^{2}} & \frac{5}{3x_{1}x_{2}} & \frac{2}{x_{1}x_{3}} & \frac{4%
}{3x_{1}x_{4}} & \frac{2}{3x_{1}x_{5}} & \frac{1}{x_{1}x_{6}} \\
\frac{5}{3x_{1}x_{2}} & \frac{10}{3x_{2}^{2}} & \frac{4}{x_{2}x_{3}} & \frac{%
8}{3x_{2}x_{4}} & \frac{4}{3x_{2}x_{5}} & \frac{2}{x_{2}x_{6}} \\
\frac{2}{x_{1}x_{3}} & \frac{4}{x_{2}x_{3}} & \frac{6}{x_{3}^{2}} & \frac{4}{%
x_{3}x_{4}} & \frac{2}{x_{3}x_{5}} & \frac{3}{x_{3}x_{6}} \\
\frac{4}{3x_{1}x_{4}} & \frac{8}{3x_{2}x_{4}} & \frac{4}{x_{3}x_{4}} & \frac{%
10}{3x_{4}^{2}} & \frac{5}{3x_{4}x_{5}} & \frac{2}{x_{4}x_{6}} \\
\frac{2}{3x_{1}x_{5}} & \frac{4}{3x_{2}x_{5}} & \frac{2}{x_{3}x_{5}} & \frac{%
5}{3x_{4}x_{5}} & \frac{4}{3x_{5}^{2}} & \frac{1}{x_{5}x_{6}} \\
\frac{1}{x_{1}x_{6}} & \frac{2}{x_{2}x_{6}} & \frac{3}{x_{3}x_{6}} & \frac{2%
}{x_{4}x_{6}} & \frac{1}{x_{5}x_{6}} & \frac{2}{x_{6}^{2}}%
\end{array}%
\right)
\end{equation}

For these exceptional CFT's, the condition $\mathfrak{p}^{\text{\textsc{\b{m}%
}}}\eta _{\text{\textsc{\b{m}\b{n}}}}\mathfrak{p}^{\text{\textsc{\b{n}}}}\in
2\mathbb{Z}$ is\ solved by the lattice
\begin{equation}
{\large \Lambda }_{\mathtt{e}_{r}}=\left\{ \boldsymbol{m}\in \mathbb{Z}%
^{2r},\quad \mathfrak{Q}_{\mathtt{e}_{r}}=\boldsymbol{m}^{\text{\textsc{a}}}%
{\large g}_{\text{\textsc{ab}}}\boldsymbol{m}^{\text{\textsc{b}}}\quad
|\quad {\large g}_{\text{\textsc{ab}}}=\boldsymbol{Q}_{\text{\textsc{ab}}}^{%
\mathbf{e}_{r}}\right\}
\end{equation}%
with $r=6,7,8$ and%
\begin{equation}
\boldsymbol{Q}_{\text{\textsc{ab}}}^{\mathbf{e}_{r}}=\left(
\begin{array}{cc}
0 & I_{r} \\
I_{r} & 0%
\end{array}%
\right) \qquad ,\qquad \det \boldsymbol{Q}_{\text{\textsc{ab}}}^{\mathbf{e}%
_{r}}=\left( -\right) ^{r}
\end{equation}%
as well as moduli dependent matrix $\boldsymbol{H}_{\mathbf{e}_{r}}$.

Using the construction of previous section, the genus-one partition function
$Z_{e_{6}}$ for orthogonal Narain CFT$^{e_{6}}$ reads as follows
\begin{equation}
Z_{e_{6}}\left[ \tau ,\bar{\tau};x\right] =\frac{1}{\left\vert \eta \left(
\tau \right) \right\vert ^{12}}\dsum\limits_{\boldsymbol{m}\in \mathbb{Z}%
^{2}}\exp \left( \boldsymbol{m}^{T}.\boldsymbol{E}_{e_{6}}.\boldsymbol{m}%
\right)
\end{equation}%
with
\begin{equation}
\boldsymbol{E}_{e_{6}}=i\pi \left( \tau \boldsymbol{Q}_{L}^{\mathrm{e}_{6}}-%
\bar{\tau}\boldsymbol{Q}_{R}^{\mathrm{e}_{6}}\right)
\end{equation}%
reading also as%
\begin{equation}
\boldsymbol{E}_{e_{6}}=i\pi \tau _{1}\boldsymbol{Q}_{e_{6}}-\pi \tau _{2}%
\boldsymbol{H}_{e_{6}}
\end{equation}%
Similar analysis can be \textrm{replicated} for the two other members of the
exceptional class namely E$_{7}$ and E$_{8}$.

\section{Conclusion and comments}
\label{sec:5}
In this paper, we developed a proposal \textrm{with the aim of classifying a family of}
Narain sigma-models with target torus $\mathbb{T}^{r}$ describing CFT$_{2}$%
's with central charges $(\mathrm{c}_{L},\mathrm{c}_{R})=(\mathrm{r},\mathrm{%
r})$. E\textrm{xploiting this result, we further provided a systematic
classification of the ensemble averaging of these CFTs over their moduli
space. Our findings revealed that this family of Narain CFTs can be indeed realised
in terms of the finite} dimensional Lie algebras with rank r underlying the
level k=1 affine symmetry known to live at the boundary of the 3D bulk
gravity. This allowed us to give explicit realisations of the Zamolodchikov metric  of the moduli space of these CFTs in terms of the associated Cartan matrix and its inverse.

To implement our proposal, we began by \textrm{analysing} the standard NCFT
on a circle $\mathbb{S}^{1}$ with radius R and central charge $\mathrm{c}%
_{L/R}=1$ and showed that \textrm{it can be described using} data on the
su(2) algebra and its representations. In this \textrm{regard}, we found
that this simple NCFT$_{2}^{\mathbf{su}_{2}}$ can be \textrm{concretely
constructed} in terms of \textrm{the} SU(2) symmetry (concisely $\left.
SU(2)_{k}\right\vert _{L/R}$ with level $k=1$) and its representations $%
\mathcal{R}_{\mathbf{su}_{2}}.$ The Kaluza-Klein modes \{n\} of the
compactified string are shown to \textrm{reside} in the root lattice $%
\Lambda _{r}^{\mathbf{su}_{2}}=\mathbb{Z}\mathbf{\alpha }$, generated by the
simple root $\mathbf{\alpha }$ of su(2); while its windings \{w\} are hosted
by the weight lattice $\Lambda _{w}^{\mathbf{su}_{2}}=\mathbb{Z}\mathbf{%
\lambda }$ generated by the fundamental weight $\mathbf{\lambda }$. Using
the fibration property $SU(2)\simeq \mathbb{S}^{2}\times \mathbb{S}^{1}$,
the coordinate variable of the one dimensional moduli space $\mathcal{M}_{%
\mathbf{su}_{2}}$ of the NCFT$_{2}^{\mathbf{su}_{2}}$, describing the size
of the circle $\mathbb{S}^{1}$; has been also \textrm{represented} in terms
of the size of the base 2-cycle $\mathbb{S}^{2}$. As a consequence of this
su(2) based realisation, the Hamiltonian $\mathcal{H}_{\mathbf{su}%
_{2}}=p_{L}^{2}+p_{R}^{2}$ is interpreted \textrm{as a} sum of self
intersection of 2-cycles namely $\mathcal{C}_{\mathbf{\beta }}^{2}+\mathcal{C%
}_{\mathbf{\chi }}^{2}$ which turn \textrm{out} to be equal to $(2x^{2}n^{2}+%
\frac{w^{2}}{2x^{2}})$ with real variable x as in the text. Similarly, the
angular momentum $\mathfrak{Q}_{\mathbf{su}_{2}}=p_{L}^{2}-p_{R}^{2}$ is
\textrm{expressed} by the cycle-intersection $\mathcal{C}_{\mathbf{\beta }}.%
\mathcal{C}_{\mathbf{\chi }}$ and \textrm{is interpreted as the gap} energy
between KK \{n\} and windings \{w\}. Moreover, the genus-one partition
function Z$_{\mathbf{su}_{2}}\left( \tau ,\bar{\tau},R\right) $ and the
underlying Siegel-Narain theta function $\vartheta _{\mathbf{su}_{2}}\left(
\tau ,\bar{\tau},R\right) $ are given by a discrete sum over \textrm{the} $%
U(1)_{L}\times U\left( 1\right) _{R}$ charges belonging to an even integer
lattice $\Lambda _{Q}^{\mathbf{su}_{2}}\subset \Lambda _{w}^{\mathbf{su}%
_{2}} $ characterised by \textrm{the} quadratic form $Q=\sum \boldsymbol{m}^{%
\text{\textsc{a}}}\mathbf{Q}_{\text{\textsc{ab}}}\boldsymbol{m}^{\text{%
\textsc{b}}} $ taking values in 2$\mathbb{Z}$.

\textrm{Building on} this NCFT$_{2}^{\mathbf{su}_{2}}$ \textrm{theory},
\textrm{we proceeded to complete} the classification of Narain theories
\textrm{by incorporating} higher values of $\mathrm{c}_{L/R}$ by \textrm{%
leveraging} the fact that rank r finite Lie algebras $\mathbf{g}$ can be
\textrm{conceptualised as} intersecting building blocks made of r copies of
su(2). Geometrically, this feature corresponds to r intersecting copies of
3-spheres $\mathbb{S}_{i}^{3}$ and, due to the Hofp fibration, \textrm{to
the intersection} of r circles $\mathbb{S}_{i}^{1}$ in the fiber and r
intersecting 2-spheres $\mathbb{S}_{i}^{2}$ in the base. Within this
picture, we obtained a new way to parameterise the quantized left p$_{L}$
and right p$_{L}$ momentums of the NCFT as follows
\begin{equation}
p_{L}=\sum_{i,k=1}^{r}\left( n^{i}x_{i}^{k}\mathbf{\alpha }%
_{k}+w_{i}y_{k}^{i}\mathbf{\lambda }^{k}\right) \quad ,\quad
p_{R}=\sum_{i,k=1}^{r}\left( n^{i}x_{i}^{k}\mathbf{\alpha }%
_{k}-w_{i}y_{k}^{i}\mathbf{\lambda }^{k}\right)
\end{equation}%
where the $\mathbf{\alpha }_{k}$'s are the simple (co-) roots of the Lie
algebra $\mathbf{g}$ and the $\mathbf{\lambda }^{k}$'s its fundamental
weights \textrm{in addition to the integers }$\left\{ n^{i}\right\} $\textrm{%
\ and }$\left\{ w_{i}\right\} $\textrm{\ defining the KK and winding modes}.
The variables $x_{i}^{k}$ and their dual $y_{k}^{i}=\left( x^{-1}\right)
_{k}^{i}$ \textrm{provide} the coordinates of the moduli space $\mathcal{M}_{%
\mathbf{g}}$ (with $\dim \mathcal{M}_{\mathbf{g}}=r^{2}$) of the NCFT$_{2}^{%
\mathbf{g}};$ the constraint relation $x_{i}^{k}y_{k}^{j}=\delta _{j}^{j}$
follows from the root/weight duality relation $\mathbf{\alpha }_{i}^{\vee }.%
\mathbf{\lambda }^{j}=\delta _{j}^{j}.$ By using this coordinate system, we
have also given an explicit realisation of the Zamolodchikov metric of the
moduli space in terms of Cartan matrix $K_{\mathbf{g}}$ of the Lie algebras
and its inverse $K_{\mathbf{g}}^{-1}$. In this regard, we showed that $%
\mathcal{H}_{\mathbf{g}}=p_{L}^{2}+p_{R}^{2}$ is a function of the
coordinate $x_{i}^{k}$ of the moduli space reading like
\begin{equation}
\mathcal{H}_{\mathbf{g}}=\sum n^{i}\left( x_{i}^{k}K_{kl}^{\mathbf{g}%
}x_{j}^{l}\right) n^{j}+\sum w_{i}\left( y_{k}^{i}\tilde{K}_{\mathbf{g}%
}^{kl}y_{l}^{j}\right) w_{j}
\end{equation}%
However, the even integer quadratic form $\mathfrak{Q}_{\mathbf{g}%
}=p_{L}^{2}-p_{R}^{2}$ is independent of $x_{i}^{k};$ it is given by $2%
\mathbf{\vec{n}.\vec{w}}$ and it vanishes for $\mathbf{\vec{n}\bot \vec{w}}$%
.\ The Zamolodchikov metric $ds_{\mathbf{g}}^{2}$ of the moduli space of NCFT%
$_{2}^{\mathbf{g}}$ is given by the trace $(d\upsilon _{\mathbf{g}%
})_{j}^{i}(d\upsilon _{\mathbf{g}})_{i}^{j}$ with matrix 1-form as follows%
\begin{equation}
\left( d\upsilon _{\mathbf{g}}\right) _{j}^{i}=\sum_{k=1}^{r}\left( \mathcal{%
K}_{\mathbf{g}}^{-1}\right) ^{ik}\left( d\mathcal{K}_{\mathbf{g}}\right)
_{kj}
\end{equation}%
As illustrations of our proposal, we \textrm{presented} several examples: we
first considered low dimensional Lie algebras $\mathbf{g}$ with rank $r=1,2$
(namely su$_{2},$ su$_{3},$ so$_{5},$ G$_{2}$); \textrm{see sections 2 and
3. And then, we gave }representatives of \textrm{higher ranked }symmetries
sitting within the families of ADE Lie algebras (su$_{r+1},$ so$_{8},$ so$%
_{10},$ e$_{6}$); \textrm{see section 4}. Along with this description, we
also \textrm{provided} results \textrm{on} the ensemble averaging of these
NCFT$_{2}^{\mathbf{g}}$s by studying the properties of the Siegel-Narain
theta function and averaging of the partition function $\mathcal{Z}_{\mathbf{%
g}}\left[ \tau ,\bar{\tau}\right] $.

To conclude this study, \textrm{we find it relevant to discuss} the r$\times
$s- dimensional moduli space $\mathcal{M}^{(s,r)}=\mathcal{O}(s,r;\mathbb{Z}%
)\backslash \mathcal{O}(s,r;\mathbb{R})/\mathcal{O}(s;\mathbb{R})\times
\mathcal{O}(r;\mathbb{R})$ of the so-called generalised Narain CFTs and
their ensemble average. These are extended conformal field theories having
\textrm{distinct} central charges $\mathrm{c}_{L}=\mathrm{s}$ and $\mathrm{c}%
_{R}=\mathrm{r};$ say $\mathrm{s>r}$. This generalisation labeled as GNCFT$%
^{(s,r)}$ could, roughly speaking, be derived from NCFT$^{(r,r)}$ by
deforming the left $\boldsymbol{p}_{L}^{2}$ by a positive quantity while
fixing $\boldsymbol{p}_{R}^{2}$ as follows
\begin{equation}
\boldsymbol{p}_{L}^{2}\rightarrow \boldsymbol{p}_{L}^{2}+\mathbf{\pi }%
_{L}^{2}+\eta \boldsymbol{p}.\mathbf{\pi }_{L}\qquad ,\qquad \boldsymbol{p}%
_{R}^{2}\rightarrow \boldsymbol{p}_{R}^{2}
\end{equation}%
with some number $\eta $ that we set to zero \textrm{in the rest of this
discussion}. In this shift, the extra term $\mathbf{\pi }_{L}^{2}$ \textrm{%
is required to be a} quadratic form that can be thought of like $\sum_{\text{%
\textsc{\.{a}}},\text{\textsc{\.{b}}=1}}^{\nu }\tilde{m}^{\text{\textsc{\.{a}%
}}}\mathfrak{M}_{\text{\textsc{\.{a}\.{b}}}}\tilde{m}^{\text{\textsc{\.{b}}}%
} $ with $\nu $ integers \{$\tilde{m}^{\text{\textsc{\.{a}}}}$\}$_{1\leq
\text{\textsc{\.{a}}}\leq \nu }$ and matrix $\mathfrak{M}_{\text{\textsc{\.{a%
}\.{b}}}}$ encoding dissymmetric data in the left sector that violate the
left/right correspondence of Narain theories. Under this shift, the
quadratic form $\mathcal{H}_{\mathbf{g}}$ gets modified like $\left(
p_{L}^{2}+p_{R}^{2}\right) +\mathbf{\pi }_{L}^{2}$ and the $\mathfrak{Q}_{%
\mathbf{g}}$ becomes $\mathfrak{\tilde{Q}}=\left( p_{L}^{2}-p_{R}^{2}\right)
+\mathbf{\pi }_{L}^{2}.$ For $\mathfrak{\tilde{Q}}$ to describe a
generalised NCFT, it must be $\left( \mathbf{i}\right) $ moduli independent;
thus requiring $\mathfrak{M}_{\text{\textsc{\.{a}\.{b}}}}$ to be a constant
matrix, and $\left( \mathbf{ii}\right) $ an even integer form requiring $%
\mathbf{\pi }_{L}^{2}\in 2\mathbb{Z}$. Typical examples of the $\left(
2r+\nu \right) \times 2\left( r+\nu \right) $ matrix of the deformed $%
\mathfrak{\tilde{Q}}$ are given by%
\begin{equation}
\mathbf{\tilde{Q}}=\left(
\begin{array}{cc}
\mathbf{Q}_{\text{\textsc{ab}}} & 0 \\
0 & \mathfrak{M}_{\text{\textsc{\.{a}\.{b}}}}%
\end{array}%
\right) \qquad ,\qquad \mathfrak{M}_{\text{\textsc{\.{a}\.{b}}}}=\left(
\begin{array}{cc}
0 & I_{\nu \times \nu } \\
I_{\nu \times \nu } & 0%
\end{array}%
\right)
\end{equation}%
with $\mathfrak{M}_{\text{\textsc{\.{a}\.{b}}}}$ having analogous features
as $\mathbf{Q}_{\text{\textsc{ab}}}$. In this regard, one can write down
several examples of possible $\mathfrak{M}_{\text{\textsc{\.{a}\.{b}}}}$;
for instance the two simple following: $\left( a\right) $ for generic
integer $\nu ,$ and $\left( b\right) $ for even integer $\nu =2\nu ^{\prime
} $ as follows;
\begin{equation}
\left( a\right) :\mathfrak{M}_{\text{\textsc{\.{a}\.{b}}}}=\left(
\begin{array}{ccc}
2m_{1} &  &  \\
& \ddots &  \\
&  & 2m_{\nu }%
\end{array}%
\right) \qquad ,\qquad \left( b\right) :\mathfrak{M}_{\text{\textsc{\.{a}\.{b%
}}}}=\left(
\begin{array}{cc}
0 & I_{\nu ^{\prime }\times \nu ^{\prime }} \\
I_{\nu ^{\prime }\times \nu ^{\prime }} & 0%
\end{array}%
\right)
\end{equation}%
We suspect that these generalised NCFTs might be also classified by Lie
algebras, \textrm{though} we have not yet developed rigourous realisations
of such type of deformations nor their classification using Lie algebras
\textbf{g} \textrm{while insuring the covering of all} moduli space
dimensions; i.e: $\dim \mathcal{M}^{(s,r)}=r\times s.$ Progress in this
direction will be reported in a future occasion.\newpage
\section*{Appendices}
\appendix
We give three appendices: Appendix A discusses the ensemble average of
Narain conformal field theories and their bulk dual in relation with our
classification proposal. Appendix B examines useful properties of the root $%
\Lambda _{r}^{\mathbf{g}}$ and the weight $\Lambda _{w}^{\mathbf{g}}$
lattices of finite dimensional Lie algebras $\mathbf{g}$. In appendix C, we
derive the metric of the moduli space $\mathcal{M}_{\mathbf{g}}$ and give
explicit calculations regarding the averaging of genus-one partition
function Z$_{\mathbf{g}}^{\text{\textsc{ncft}}}$ \textrm{to highlight} the
footprint of $\mathbf{g}$ and representations $\mathcal{R}_{\mathbf{g}}$.

\section{Narain CFTs and gravitational dual}
\label{sec:appA}
The genus-one partition functions Z$_{\mathbf{g}}^{\text{\textsc{cft}}}$ of
the Narain CFTs have been considered in the main text in link with the
classification in terms of $\mathbf{g}$. In this appendix, we give
properties of its averaging over the moduli space,
\begin{equation}
<Z_{\mathbf{g}}^{\text{\textsc{cft}}}\left( \tau ,\bar{\tau};\mathbf{x}%
\right) >_{\mathcal{M}}:=\mathcal{Z}_{\mathbf{g}}^{\text{\textsc{cft}}%
}\left( \tau ,\bar{\tau}\right) ,
\end{equation}%
Here, we \textrm{describe} the two dimensional surface (string world sheet)
in terms of a 2- torus $\Sigma =\mathbb{S}_{\sigma }^{1}\times \mathbb{S}%
_{t_{E}}^{1}$, where $\mathbb{S}_{\sigma }^{1}$ parameterizes space while $%
\mathbb{S}_{t_{E}}^{1}$ \textrm{represents} the Euclidean time.

\subsubsection*{Averaged partition function of Narain CFTs}

The genus-one partition function $Z_{{\small D}}\left( \tau ,\bar{\tau};%
\mathbf{x}\right) $ of the sigma-model with target space $\mathbb{T}^{%
{\small D}}$ (here $D=r$) is a function of $\left( \mathbf{1}\right) $ the
modular parameters $\left( \tau ,\bar{\tau}\right) $ of the 2-torus and $%
\left( \mathbf{2}\right) $ the coordinates $\mathbf{x}=\left(
x_{i}^{k}\right) $ of the moduli space $\mathcal{M}_{{\small D}}.$ \textrm{%
It takes the form}
\begin{equation}
Z_{{\small D}}\left( \tau ,\bar{\tau};\mathbf{x}\right) =\frac{\Theta _{%
{\small D}}\left( \tau ,\bar{\tau};\mathbf{x}\right) }{\left\vert \eta
\left( \tau \right) \right\vert ^{2{\small D}}}
\end{equation}%
where $\Theta _{{\small D}}$ is the Siegel-Narain theta function obeying the
differential equation%
\begin{equation}
\left( \Delta _{{\small H}}-D\tau _{{\small 2}}\frac{\partial }{\partial
\tau _{{\small 2}}}-\Delta _{\mathcal{M}_{{\small D}}}\right) \Theta _{%
{\small D}}\left( \tau ,\bar{\tau};\mathbf{x}\right) =0  \label{dift}
\end{equation}%
and the asymptotic property $\lim_{\tau _{2}\rightarrow \infty }\Theta _{%
{\small D}}=1.$ \textrm{For each NCFT}$_{2}^{\mathbf{g}}$\textrm{\ theory of
our classification, there is an associated partition function expressed like}%
\begin{equation}
Z_{\mathbf{g}}^{({\small D,D)}}\left( \tau ,\bar{\tau};\mathbf{\xi }\right) =%
\frac{\Theta _{\mathbf{g}}^{({\small D,D)}}\left( \tau ,\bar{\tau};\mathbf{x}%
\right) }{\left\vert \eta \left( \tau \right) \right\vert ^{2{\small D}}}
\label{zg}
\end{equation}%
with $\Theta _{\mathbf{g}}^{({\small D,D)}}$ as in eq(\ref{04}) namely%
\begin{equation}
\Theta _{\mathbf{g}}^{({\small D,D)}}\left( \tau _{1},\tau _{2},\mathbf{x}%
\right) =\dsum\limits_{\mathbf{\beta },\mathbf{\chi }}e^{-\pi \tau _{2}\left[
\left\langle \mathbf{\beta }^{\vee }.\mathbf{\beta }\right\rangle
+\left\langle \mathbf{\chi }^{\vee }.\mathbf{\chi }\right\rangle \right]
}e^{2\pi i\tau _{1}\left\langle \mathbf{\beta }^{\vee }.\mathbf{\chi }%
\right\rangle }
\end{equation}%
and where
\begin{equation}
\mathbf{\beta }=\sum_{i=1}^{D}n^{i}\mathbf{\beta }_{i}=%
\sum_{i,k=1}^{D}n^{i}x_{i}^{k}\mathbf{\alpha }_{k},\quad \mathbf{\chi }%
=\sum_{j=1}^{D}w_{j}\mathbf{\chi }^{j}=\sum_{j,l=1}^{D}w_{j}y_{l}^{j}\mathbf{%
\lambda }^{k},\quad \sum_{k=1}^{D}x_{i}^{k}y_{k}^{j}=\delta _{i}^{j}
\end{equation}%
with%
\begin{equation}
\mathcal{K}_{ij}^{\mathbf{g}}=\mathbf{\beta }_{i}^{\vee }.\mathbf{\beta }%
_{j}\qquad ,\qquad \left( \mathcal{K}_{\mathbf{g}}^{-1}\right) ^{ij}=\mathbf{%
\chi }^{\vee i}.\mathbf{\chi }^{j}
\end{equation}%
reading explicitly as ($D=r$)
\begin{equation}
\mathcal{K}_{ij}^{\mathbf{g}}=\sum_{k,l=1}^{r}x_{i}^{k}K_{kl}^{\mathbf{g}%
}x_{j}^{l}\qquad ,\qquad \left( \mathcal{K}_{\mathbf{g}}^{-1}\right)
^{ij}=\sum_{k,l=1}^{r}y_{k}^{i}\tilde{K}_{\mathbf{g}}^{kl}y_{l}^{j}
\label{68}
\end{equation}%
By introducing the change of variables
\begin{equation}
x_{i}^{k}=\dsum\limits_{\text{\textsc{m=1}}}^{r^{2}}\left( \Gamma _{\text{%
\textsc{m}}}\right) _{i}^{k}x^{\text{\textsc{m}}}\qquad ,\qquad
y_{j}^{l}=\dsum\limits_{\text{\textsc{n=1}}}^{r^{2}}\left( \bar{\Gamma}^{%
\text{\textsc{n}}}\right) _{j}^{l}y_{\text{\textsc{n}}}
\end{equation}%
and%
\begin{equation}
\boldsymbol{K}_{\text{\textsc{mn}}}^{\mathbf{g}}=\Gamma _{\text{\textsc{m}}%
}^{T}K^{\mathbf{g}}\Gamma _{\text{\textsc{n}}}\qquad ,\qquad \boldsymbol{K}_{%
\mathbf{g}}^{\text{\textsc{mn}}}=(\bar{\Gamma}^{\text{\textsc{m}}})^{T}%
\tilde{K}_{\mathbf{g}}\bar{\Gamma}^{\text{\textsc{n}}}
\end{equation}%
with $\Gamma _{\text{\textsc{m}}}$ and $\bar{\Gamma}^{\text{\textsc{m}}}$\
as
\begin{equation}
\sum_{k=1}^{r}\left( \Gamma _{\text{\textsc{m}}}\right) _{i}^{k}\left( \bar{%
\Gamma}^{\text{\textsc{n}}}\right) _{k}^{j}=\frac{1}{r^{2}}\delta _{\text{%
\textsc{m}}}^{\text{\textsc{n}}}\delta _{i}^{j},\qquad \dsum\limits_{\text{%
\textsc{m=1}}}^{r^{2}}x^{\text{\textsc{m}}}y_{\text{\textsc{m}}%
}=r^{2},\qquad y_{\text{\textsc{m}}}=\frac{x_{\text{\textsc{m}}}}{\left\Vert
x\right\Vert ^{2}}
\end{equation}%
where we have set $\sum_{\text{\textsc{m=1}}}^{r^{2}}x^{\text{\textsc{m}}}x_{%
\text{\textsc{m}}}=\left\Vert x\right\Vert ^{2}$. By substituting in (\ref%
{68}), we can express the above matrices $\mathcal{K}_{ij}^{\mathbf{g}}$ and
$\left( \mathcal{K}_{\mathbf{g}}^{-1}\right) ^{ij}$ in terms of the new
coordinates \{$x^{\text{\textsc{m}}}$\} like%
\begin{equation}
\mathcal{K}_{ij}^{\mathbf{g}}=\dsum\limits_{\text{\textsc{m,n=1}}}^{r^{2}}(%
\boldsymbol{K}_{\text{\textsc{mn}}}^{\mathbf{g}})_{ij}\left( x^{\text{%
\textsc{m}}}x^{\text{\textsc{n}}}\right) \qquad ,\qquad \left( \mathcal{K}_{%
\mathbf{g}}^{-1}\right) ^{ij}=\dsum\limits_{\text{\textsc{m,n=1}}}^{r^{2}}(%
\boldsymbol{K}_{\mathbf{g}}^{\text{\textsc{mn}}})^{ij}y_{\text{\textsc{m}}%
}y_{\text{\textsc{n}}}
\end{equation}%
Focussing on the Siegel-Narain theta function classified by \textbf{g}, we
can define the average%
\begin{equation}
F_{\mathbf{g}}^{({\small D,D)}}\left( \tau ,\bar{\tau}\right) =\frac{1}{vol(%
\mathcal{M}_{\mathbf{g}}^{({\small D,D)}})}\dint\nolimits_{\mathcal{M}_{%
{\small D}}}d\mathbf{\mu }\Theta _{\mathbf{g}}^{({\small D,D)}}\left( \tau ,%
\bar{\tau};\mathbf{x}\right)  \label{mo}
\end{equation}%
with
\begin{equation}
d\mathbf{\mu }=\sqrt{\det G}\dprod\limits_{i,k=1}^{r}dx_{i}^{k}=\sqrt{\det
\mathcal{G}}\dprod\limits_{\text{\textsc{m=1}}}^{r^{2}}dx^{\text{\textsc{m}}}
\end{equation}%
and metric
\begin{eqnarray}
ds_{\mathbf{g}}^{2} &=&\sum_{i,j,k,l=1}^{r}\left( \mathcal{K}_{\mathbf{g}%
}^{-1}\right) ^{ik}\left( \mathcal{K}_{\mathbf{g}}^{-1}\right) ^{jl}\left( d%
\mathcal{K}_{\mathbf{g}}\right) _{kj}\left( d\mathcal{K}_{\mathbf{g}}\right)
_{li}  \notag \\
&=&\sum_{i,j,k,l=1}^{r}G_{kl}^{ij}\left( dx_{i}^{k}\right) \left(
dx_{j}^{l}\right) \\
&=&\dsum\limits_{\text{\textsc{m,n=1}}}^{r^{2}}\mathcal{G}_{\text{\textsc{mn}%
}}dx^{\text{\textsc{m}}}dx^{\text{\textsc{n}}}  \notag
\end{eqnarray}%
as well as the total volume
\begin{equation}
\int_{\mathcal{M}_{\mathbf{g}}^{({\small D,D)}}}1d\mathbf{\mu }=vol(\mathcal{%
M}_{\mathbf{g}}^{({\small D,D)}})
\end{equation}%
In the large limit of $\func{Im}\tau $, we have the property
\begin{equation}
\lim_{\tau _{2}\rightarrow \infty }F_{\mathbf{g}}^{({\small D,D)}}\left(
\tau ,\bar{\tau}\right) =\frac{1}{vol(\mathcal{M}_{\mathbf{g}}^{({\small D,D)%
}})}  \label{lm}
\end{equation}%
following from the asymptotic $\lim_{\tau _{2}\rightarrow \infty }\Theta _{%
\mathbf{g}}^{({\small D,D)}}=1.$

Using (\ref{mo}), we can also deduce the mean value $<Z_{\mathbf{g}}^{(%
{\small D,D)}}\left( \tau ,\bar{\tau};\mathbf{x}\right) >_{\mathcal{M}_{%
{\small D}}}$ which is given by
\begin{equation}
\mathcal{Z}_{\mathbf{g}}^{({\small D,D)}}\left( \tau ,\bar{\tau}\right)
=\int_{\mathcal{M}_{{\small D}}}d\mathbf{\mu }Z_{\mathbf{g}}^{({\small D,D)}%
}\left( \tau ,\bar{\tau};\mathbf{x}\right)
\end{equation}%
and reading as follows
\begin{equation}
\mathcal{Z}_{\mathbf{g}}^{({\small D,D)}}\left( \tau ,\bar{\tau}\right) =%
\frac{F_{\mathbf{g}}^{({\small D,D)}}\left( \tau ,\bar{\tau}\right) }{%
\left\vert \eta \left( \tau \right) \right\vert ^{2{\small D}}}
\end{equation}%
From eq(\ref{dift}), we see that the above $F_{{\small D}}\left( \tau ,\bar{%
\tau}\right) $ verifies the differential equation%
\begin{equation}
\left( \Delta _{{\small H}}-D\tau _{{\small 2}}\frac{\partial }{\partial
\tau _{{\small 2}}}\right) F_{\mathbf{g}}^{({\small D,D)}}\left( \tau ,\bar{%
\tau}\right) =0  \label{lapl}
\end{equation}%
In this relation, the laplacian $\Delta _{{\small H}}$ on the half plane is
\textrm{defined} by $-\tau _{{\small 2}}^{2}(\partial _{\tau
_{1}}^{2}+\partial _{\tau _{2}}^{2}).$ Moreover, \textrm{due to }$F_{\mathbf{%
g}}^{({\small D,D)}}\left( \tau ,\bar{\tau}\right) /\left\vert \eta \left(
\tau \right) \right\vert ^{2{\small D}}$\textrm{\ being }modular invariant,
the averaged $F_{\mathbf{g}}^{({\small D,D)}}\left( \tau ,\bar{\tau}\right) $
transforms under modular transformations by scaling factors with weights $%
\left( u,v\right) =\left( D/2,D/2\right) $. Indeed, by performing the change
\begin{equation}
\tau \rightarrow \tau ^{\prime }=\frac{a\tau +b}{c\tau +d}
\end{equation}%
with integers $\left( a,b,c,d\right) $ constrained as $ad-bc=1,$ and using
the known property
\begin{equation}
\left\vert \eta \left( \tau ^{\prime }\right) \right\vert =\left\vert c\tau
+d\right\vert ^{1/2}\left\vert \eta \left( \tau \right) \right\vert
\end{equation}%
we have the following scaling%
\begin{equation}
F_{\mathbf{g}}^{({\small D,D)}}\left( \tau ^{\prime },\bar{\tau}^{\prime
}\right) =\left( c\tau +d\right) ^{\frac{D}{2}}\left( c\bar{\tau}+d\right) ^{%
\frac{D}{2}}F_{\mathbf{g}}^{({\small D,D)}}\left( \tau ,\bar{\tau}\right)
\end{equation}%
Furthermore, \textrm{upon imposing} the modular transformation $\tau
_{2}^{\prime }=\tau _{2}/\left\vert c\tau +d\right\vert ^{2},$ it follows
that the function $W_{\mathbf{g}}^{({\small D,D)}}\left( \tau ,\bar{\tau}%
\right) :=\tau _{{\small 2}}^{D/2}F_{\mathbf{g}}^{({\small D,D)}}\left( \tau
,\bar{\tau}\right) $ is modular invariant and behaves for large values of $%
\tau _{2}$ like
\begin{equation}
\frac{1}{vol(\mathcal{M}_{\mathbf{g}}^{({\small D,D)}})}\tau _{2}^{{\small D}%
/2}
\end{equation}%
due to (\ref{lm}). This volume is function of the Cartan matrix $K_{\mathbf{g%
}}.$ By substituting into (\ref{lapl}) and using $\Delta _{{\small H}}(\tau
_{{\small 2}}^{-s}W_{\mathbf{g}}^{({\small D,D)}})=-s(s-1)\tau _{{\small 2}%
}^{-s}W_{\mathbf{g}}^{({\small D,D)}}$ with $s=D/2,$ we end up with the
following differential equation
\begin{equation}
\left[ \Delta _{H}+s(s-1)\right] W_{\mathbf{g}}^{({\small D,D)}}\left( \tau ,%
\bar{\tau}\right) =0\qquad ,\qquad s=D/2
\end{equation}%
showing that $W_{\mathbf{g}}^{({\small D,D)}}$ is an eigen function of $%
\Delta _{H}$ with eigenvalue $-s(s-1).$ Candidates of such eigen functions
are given by the (real analytic) Eisenstein series defined as follows \cite%
{1A, RPP}%
\begin{equation}
E_{s}\left( \tau ,\bar{\tau}\right) =\dsum\limits_{c,d}\frac{\tau _{{\small 2%
}}^{s}}{\left\vert c\tau +d\right\vert ^{2s}}\qquad ,\qquad \tau _{2}=\func{%
Im}\tau  \label{es}
\end{equation}%
with coprime integers $\left( c,d\right) \simeq \left( -c,-d\right) .$ This
series can be also imagined in terms of sum is over all modular images of
the function $\tau _{2}^{s};$ and as such it can defined as follows%
\begin{equation}
E_{s}\left( \tau ,\bar{\tau}\right) =\dsum\limits_{\gamma \in SL_{2}/\Pi }%
\func{Im}\left( \mathrm{\gamma }\tau \right) ^{s}
\end{equation}%
showing that it is modular invariant. In this writing, the $\mathrm{\gamma }$
belongs to the coset group $\mathrm{SL}_{2}/\Pi _{1}$ with generic
transformations as follows%
\begin{eqnarray}
SL(2,\mathbb{Z}) &=&\left\{ \gamma =\left(
\begin{array}{cc}
a & b \\
c & d%
\end{array}%
\right) ,\text{ }ad-bc=1\right\} \\
\Pi (1,\mathbb{Z}) &=&\left\{ \boldsymbol{\pi }_{n}=\left(
\begin{array}{cc}
1 & n \\
0 & 1%
\end{array}%
\right) ,\text{ \qquad }n\in \mathbb{Z}\right\}
\end{eqnarray}%
Recall that under these group transformations, the imaginary part $\tau _{2}=%
\func{Im}\tau $ is invariant under the translations generated by $%
\boldsymbol{\pi }_{1}$ while it is mapped into $\tau _{2}/\left\vert c\tau
+d\right\vert ^{2}$ under \textrm{the S application}. Notice also that the
series (\ref{es}) \textrm{converges for} $\func{Re}s>1$ corresponding
\textrm{in our case} to $D>2$ \textrm{\cite{1A}}$.$

In sum, we have the following features:

\begin{description}
\item[$\left( \mathbf{i}\right) $] The function $\tau _{2}^{s}$ is an
eigenfunction of the Laplacian $\Delta _{H}$ with eigenvalue $-s(s-1)$; the
same is valid for its modular images. \textrm{So, the} $\lim_{\tau
_{2}\rightarrow \infty }$ $E_{s}=\tau _{2}^{s}$ \textrm{correspond} to the
identity element of the $\mathrm{SL}\left( 2,\mathbb{Z}\right) $ group ($%
a=d=1$). As such $E_{{\small D}/2}$ satisfies all of the desired properties
of $W_{\mathbf{g}}^{({\small D,D)}}.$

\item[$\left( \mathbf{ii}\right) $] The functions $E_{{\small D}/2}$ and $W_{%
\mathbf{g}}^{({\small D,D)}}$\ can be \textrm{superimposed}; their
difference
\begin{equation}
W_{\mathbf{g}}^{({\small D,D)}}-c_{\mathbf{g}}E_{{\small D}/2}  \label{22}
\end{equation}%
grows at infinity slowly than $\tau _{2}^{D/2}$; and hence must vanish
\textrm{\cite{1A}}. In this relation, we have
\begin{equation}
c_{\mathbf{g}}=\frac{1}{vol(\mathcal{M}_{\mathbf{g}}^{({\small D,D)}})}
\label{23}
\end{equation}
\end{description}

\subsubsection*{Gravitational dual}

The 3D bulk partition function $Z_{\text{\textsc{bulk}}}$ of the
gravitational dual \textrm{\cite{2A}} of the averaged genus-one partition
function $Z^{\text{\textsc{cft}}}$ of Narain CFT$_{2}$ with target space $%
\mathbb{T}^{d}$ is expressed as a sum over handlebodies Y as follows%
\begin{equation}
Z_{\text{\textsc{bulk}}}=\sum_{\text{handlebadies \textsc{Y}}}Z_{\text{%
\textsc{cs}}}^{\text{\textsc{y}}}
\end{equation}%
Here Y is the thermal AdS$_{3}$ given by $\mathcal{D}\times \mathbb{S}%
_{t}^{1}$ where $\mathcal{D}$ is a disc with boundary $\mathbb{S}_{\sigma
}^{1}$ that we \textrm{concisely} write as
\begin{equation}
Y=\mathcal{D}\times \mathbb{S}_{t}^{1}\qquad ,\qquad \partial Y=\mathbb{S}%
_{\sigma }^{1}\times \mathbb{S}_{t}^{1}
\end{equation}%
and the $Z_{\text{\textsc{cs}}}^{\text{\textsc{y}}}$ is the partition
function of $U(1)^{d}\times U(1)^{d}$ Chern-Simons fields on Y. In this
regard, recall that in 3D Einstein gravity \textrm{possesses no} physical
bulk excitations \textrm{other than the} boundary gravitons first described
by Brown and Henneaux \cite{BH}. The path integral of the Chern-Simons field
on $\mathcal{D}\times \mathbb{S}_{t}^{1}$ simply equals the partition
function of the Brown-Henneaux modes. \textrm{Thus, }the bulk $Z_{\text{%
\textsc{cs}}}^{\text{\textsc{y}}}$ is given by the thermal partition
function of boundary gravitons \textrm{composites of the} D boundary photons
whose partition function is
\begin{equation}
Z_{\text{\textsc{cs}}}^{\text{\textsc{y}}}=\frac{1}{\left\vert \eta \left(
\tau \right) \right\vert ^{2d}}=q^{-\frac{2d}{24}}\dprod\limits_{n=1}^{%
\infty }\frac{1}{\left\vert 1-q^{n}\right\vert ^{2d}}
\end{equation}%
This \textrm{matches} the vacuum character of d copies of the $U(1)\times
U(1)$ current algebra \cite{vac}. This expression can be verified in a
direct bulk computation in $U(1)^{2d}$ Chern-Simons theory, just as in the
gravity case.

To get the full partition function $Z_{\text{\textsc{bulk}}},$ we need to
sum over all handlebodies,%
\begin{equation}
Z_{\text{\textsc{bulk}}}=\sum_{\mathrm{\gamma \in \Pi \backslash SL}_{2}}%
\frac{1}{\left\vert \eta \left( \mathrm{\gamma }\tau \right) \right\vert
^{2d}}
\end{equation}%
By factorising $\left\vert \eta \left( \tau \right) \right\vert ^{-2d}$ like
the product $(\tau _{2}^{-d/2}\left\vert \eta \left( \tau \right)
\right\vert ^{-2d})\times \tau _{2}^{d/2}$ with modular invariant $\tau
_{2}^{-d/2}\left\vert \eta \left( \tau \right) \right\vert ^{-2d}$, and
thinking about the factor $\tau _{2}^{d/2}$ in terms of $E_{d/2},$ we end up
with the following expression of the bulk partition function%
\begin{equation}
Z_{\text{\textsc{bulk}}}^{\mathbf{g}}=\frac{c_{\mathbf{g}}E_{d/2}}{\tau
_{2}^{d/2}\left\vert \eta \left( \tau \right) \right\vert ^{2d}}
\end{equation}%
with $c_{\mathbf{g}}$ as in (\ref{23}).

\section{Root and weight lattices of Lie algebras $\mathbf{g}$}
\label{sec:appB}
In this appendix, we\textrm{\ provide a summary of} useful tools on the root
$\Lambda _{r}$ and the weight $\Lambda _{w}$ lattices of finite dimensional
Lie algebras $\mathbf{g}$. We also \textrm{describe} the congruency classes
of the coset $\Lambda _{w}/\Lambda _{r}$. These tools \textrm{play an
integral role }in the main text.

\subsubsection*{Lattices $\Lambda _{r}$ and $\Lambda _{weight}$}

Root lattice of a finite dimensional Lie algebra $\Lambda _{root}:=\Lambda
_{r}^{\mathbf{g}}$ and its weight $\Lambda _{weight}:=\Lambda _{w}^{\mathbf{g%
}}$ are generated by \textrm{the} simple roots $\mathbf{\alpha }_{i}$ and
fundamental weights $\mathbf{\lambda }^{i}$. Sites \textbf{r}$_{\mathbf{n}}$%
\textbf{\ }in the lattice $\Lambda _{r}^{\mathbf{g}}$ are given by the
integral expansion $\sum n^{i}\mathbf{\alpha }_{i}.$ \textrm{Similarly, sites%
} \textbf{k}$_{\mathbf{w}}$ in the weight lattice are given by $\sum w_{i}%
\mathbf{\lambda }^{i}$. Formally, we have
\begin{equation}
\begin{tabular}{lll}
$\Lambda _{r}^{\mathbf{g}}$ & $=$ & $\mathbb{Z}\mathbf{\alpha }_{1}+...+%
\mathbb{Z}\mathbf{\alpha }_{r}$ \\
$\Lambda _{w}^{\mathbf{g}}$ & $=$ & $\mathbb{Z}\mathbf{\lambda }_{1}+...+%
\mathbb{Z}\mathbf{\lambda }_{r}$%
\end{tabular}%
\end{equation}%
Because of the relationship $\mathbf{\alpha }_{i}=K_{ij}\mathbf{\lambda }%
^{j} $ between simple roots and fundamental weights, we have the inclusion $%
\Lambda _{r}^{\mathbf{g}}\subset \Lambda _{w}^{\mathbf{g}}$. Putting into $%
\mathbf{r}_{\mathbf{n}}=\sum_{i}n^{i}\mathbf{\alpha }_{i}$, we end up with
the following expansion%
\begin{equation}
\mathbf{r}_{\mathbf{n}}=\sum_{i,j=1}^{r}\left( K_{ij}n^{i}\right) \mathbf{%
\lambda }^{j}=\sum_{j=1}^{r}m_{j}\mathbf{\lambda }^{j}
\end{equation}%
with $m_{j}=\sum_{i}n^{i}K_{ij}.$\ Moreover, seen that the number of
congruency classes is equal to the number of distinct elements of the coset $%
\Gamma :=\Lambda _{w}^{\mathbf{g}}/\Lambda _{r}^{\mathbf{g}},$ it follows
that the factor $\Gamma $ is a finite group with representative element $%
\mathbf{\bar{\upsilon}}=\sum \bar{v}^{i}\mathbf{\alpha }_{i}$ that has a
definite expression for each algebra. As such, given a weight $\mathbf{%
\lambda }=\sum w_{i}\mathbf{\lambda }^{i},$ the congruency classes are
defined as follows%
\begin{equation}
\mathbf{\lambda .\bar{\upsilon}}=\sum_{i=1}^{r}w_{i}\bar{v}^{i}\qquad \func{%
mod}\left\vert \Gamma \right\vert
\end{equation}%
In this regards, recall that the order $\left\vert \Gamma \right\vert $ of
the coset group $\Gamma $ is known to be equal to the determinant of the
Cartan matrix $K_{ij},$ that is $\det K=\left\vert \Gamma \right\vert $.

\subsubsection*{Lattices for classical Lie algebras}

For the classical Lie algebras ABCD families,%
\begin{equation}
\mathbf{A}_{r}=su(r+1),\quad \mathbf{B}_{r}=so(2r+1),\quad \mathbf{C}%
_{r}=sp(2r),\quad \mathbf{D}_{r}=so(2r)
\end{equation}%
the following characteristic features \textrm{are noteworthy:}%
\begin{equation}
\begin{tabular}{|c|c|c|c|c|c|}
\hline
Lie algebra & $\mathbf{\bar{\upsilon}}$ & $\Gamma $ & $\left\vert \Gamma
\right\vert $ & $\mathbf{\lambda .\bar{\upsilon}}$ & $\func{mod}$ \\ \hline
$\mathbf{A}_{r}$ & $\left( 1,2,...,r-1,r\right) $ & $\mathbb{Z}_{r+1}$ & $%
r+1 $ & $\mathbf{\lambda }_{1}+2\mathbf{\lambda }_{2}+...r\mathbf{\lambda }%
_{r}$ & $r+1$ \\ \hline
$\mathbf{B}_{r}$ & $\left( 0,...,0,1\right) $ & $\mathbb{Z}_{2}$ & $2$ & $%
\mathbf{\lambda }_{r}$ & $2$ \\ \hline
$\mathbf{C}_{r}$ & $\left( 1,2,...,r-1,r\right) $ & $\mathbb{Z}_{2}$ & $2$ &
$\mathbf{\lambda }_{1}+2\mathbf{\lambda }_{2}+...r\mathbf{\lambda }_{r}$ & $%
2 $ \\ \hline
$\mathbf{D}_{2m}$ & $\left( 0,...,0,1,1\right) $ & $\mathbb{Z}_{4}$ & $4$ & $%
\mathbf{\lambda }_{2m-1}+\mathbf{\lambda }_{2m}$ & $4$ \\ \hline
\end{tabular}%
\end{equation}%
\begin{equation*}
\end{equation*}%
where we have singled out even and odd ranks ($r=2m$ and $r=2m+1$) for the
orthogonal $D_{r}.$ Notice that for $so(4m),$ the two fundamental spinor
representations are each self-conjugate while for $so(2m+2)$ they are
pairwise conjugates.\newline
For fundamental representations $\mathbf{\lambda }_{l},$ we have the
following properties: For $A_{r},$ we have the congruency classes%
\begin{equation}
\begin{tabular}{|c|c|c|c|}
\hline
\textbf{A}$_{r}$ & fund $\mathbf{\lambda }_{l}$ & dim & class \\ \hline
tensor repres & $T_{[i_{1}...i_{l}]}$ & $\frac{\left( r+1\right) !}{l!\left(
r+1-l\right) !}$ & $\overline{l}\in \mathbb{Z}_{r+1}$ \\ \hline
vector $\mathbf{\lambda }_{1}$ & $V_{j}$ & $r+1$ & $\overline{\mathbf{1}}$
\\ \hline
$\mathbf{\lambda }_{r}$ dual of $\mathbf{\lambda }_{1}$ & $\tilde{V}^{j}$ & $%
r+1$ & $\overline{\mathbf{r}}\simeq \overline{\mathbf{1}}^{\ast }$ \\ \hline
$adj=\mathbf{\lambda }_{1}+\mathbf{\lambda }_{r}$ & $T_{i}^{j}$ & $\left(
r+1\right) ^{2}-1$ & $\overline{\mathbf{0}}$ \\ \hline
\end{tabular}%
\end{equation}%
with $\tilde{V}^{j}=\frac{1}{r!}\varepsilon
^{ji_{1}...i_{r}}T_{[i_{1}...i_{r}]}$ and T$_{i}^{j}$ traceless. For
illustration, we give the characteristics of the example of SU(3),%
\begin{equation}
\begin{tabular}{|c|c|c|c|}
\hline
weight & $\mathbf{\lambda }_{1}$ & $\mathbf{\lambda }_{2}$ & $\mathbf{%
\lambda }_{1}+\mathbf{\lambda }_{2}$ \\ \hline
Repres & $3$ & $3^{\ast }$ & $adj=8$ \\ \hline
$\mathbb{Z}_{3}$ class & $\overline{\mathbf{1}}$ & $\overline{\mathbf{2}}$ &
$\overline{\mathbf{3}}\equiv \overline{\mathbf{0}}$ \\ \hline
tensors & $T_{i}$ & $T^{i}$ & $T_{i}^{j}$ \\ \hline
\end{tabular}%
\end{equation}%
where $T^{i}=\frac{1}{2}\varepsilon ^{ijk}T_{\left[ jk\right] }$ and $%
T_{i}^{j}$ is traceless.

For the other Lie algebras, we have:
\begin{equation}
\begin{tabular}{|c|c|c|c|c|c|}
\hline
{\small Lie algebra} & \multicolumn{2}{|c|}{\small Reprentations} & $%
\left\vert \Gamma \right\vert $ & class $\overline{\mathbf{0}}$ & class $%
\overline{\mathbf{1}}$ \\ \hline
$\mathbf{B}_{r}$ & $\left\{ \mathbf{\lambda }_{i}\right\} _{1\leq i\leq r-1}$
& {\small spinor:} $\mathbf{\lambda }_{r}$ & $\mathbb{Z}_{2}$ & $\left\{
\mathbf{\lambda }_{i}\right\} _{1\leq i\leq r-1}$ & $\mathbf{\lambda }_{r}$
\\ \hline
$\mathbf{C}_{r}$ & $\left.
\begin{array}{c}
\left\{ \mathbf{\lambda }_{2i}\right\} \\
\left\{ \mathbf{\lambda }_{2i+1}\right\}%
\end{array}%
\right. $ & {\small adjoint:} $\theta =2\mathbf{\lambda }_{1}$ & $\mathbb{Z}%
_{2}$ & $\left\{ \mathbf{\lambda }_{2i}\right\} $ & $\left\{ \mathbf{\lambda
}_{2i+1}\right\} $ \\ \hline
$\mathbf{D}_{2m}$ & \multicolumn{1}{|l|}{$\left.
\begin{array}{c}
\left\{ \mathbf{\lambda }_{2i+1}\right\} \\
\left\{ \mathbf{\lambda }_{2i}\right\}%
\end{array}%
\right. $} & \multicolumn{1}{|l|}{{\small spinors:} $\left.
\begin{array}{c}
\mathbf{\lambda }_{2m-1} \\
\mathbf{\lambda }_{2m}%
\end{array}%
\right. $} & \multicolumn{1}{|l|}{$\mathbb{Z}_{4}$} & \multicolumn{1}{|l|}{$%
\left\{ \mathbf{\lambda }_{i}\right\} _{1\leq i\leq 2m-2}$} &
\multicolumn{1}{|l|}{$\left.
\begin{array}{c}
\mathbf{\lambda }_{2m-1} \\
\mathbf{\lambda }_{2m}%
\end{array}%
\right. $} \\ \hline
\end{tabular}%
\end{equation}%
\begin{equation*}
\end{equation*}%
The particular family $D_{2m+1}$ with odd rank will be considered later.
Meanwhile, we give the following properties:

\begin{description}
\item[$\left( i\right) $] \textbf{Case B}$_{r}\simeq so(2r+1)$\newline
The $\left( r-1\right) $ fundamental weight $\mathbf{\lambda }_{l}$
representations are labeled by $1\leq j\leq r-1;$ they correspond to \emph{%
rank-l} antisymmetric tensor $T_{\left[ i_{1}...i_{l}\right] }.$ The
completely symmetric traceless tensors $T_{\left( i_{1}...i_{l}\right) }$
give the irreducible representations $l\mathbf{\lambda }_{1}$:%
\begin{equation}
\begin{tabular}{lll}
$T_{\left[ i_{1}...i_{l}\right] }$ & $\qquad \leftrightarrow \qquad $ & $%
\mathbf{\lambda }_{l}$ \\
$T_{\left( i_{1}...i_{l}\right) }$ & $\qquad \leftrightarrow \qquad $ & $l%
\mathbf{\lambda }_{1}$ \\
spinor 2$^{r}$ & $\qquad \leftrightarrow \qquad $ & $\mathbf{\lambda }_{r}$
\\
$T_{\left[ ij\right] }$ & $\qquad \leftrightarrow \qquad $ & $\mathbf{%
\lambda }_{2}=$adj%
\end{tabular}%
\end{equation}

\item[$\left( ii\right) $] \textbf{Case C}$_{r}\simeq sp(2r)$\newline
Unlike $\mathbf{A}_{r},$ $\mathbf{B}_{r}$ and $\mathbf{D}_{r}$, the
antisymmetrized tensors T$_{\left[ i_{1}...i_{l}\right] }$ \textrm{do} not
automatically correspond to irreducible representation of C$_{r}$ \textrm{%
given that a}ny pair of indices can be contracted with the invariant rank-2
antisymmetric metric $\Omega ^{\left[ ij\right] }$. Here, one can disregard
the distinction between upper and lower indices, because the defining and
all other irreducible representations of $sp(2n)$ are self-conjugate.%
\begin{equation}
\begin{tabular}{|c|c|c|}
\hline
tensor & weight & $\dim $ \\ \hline
$V_{i}$ & $\mathbf{\lambda }_{1}$ & $2r$ \\ \hline
$\left.
\begin{array}{c}
T_{\left[ i_{1}i_{2}...i_{l}\right] }\text{ {\small with}} \\
\Omega ^{i_{1}i_{2}}T_{\left[ i_{1}i_{2}...i_{l}\right] }=0%
\end{array}%
\right. $ & $\mathbf{\lambda }_{l}$ & $2\left( r+1-l\right) \frac{\left(
2r+1\right) !}{l!\left( 2r+2-l\right) !}$ \\ \hline
$T_{\left( i_{1}...i_{l}\right) }$ & $l\mathbf{\lambda }_{1}$ & $2rl$ \\
\hline
spinor & $\mathbf{\lambda }_{r}$ & $2^{r}$ \\ \hline
\end{tabular}%
\end{equation}%
\begin{equation*}
\
\end{equation*}

\item[$\left( iii\right) $] \textbf{Case D}$_{r}\simeq so(2r)$\newline
The first $\left( r-2\right) $ fundamental weights $\left\{ \mathbf{\lambda }%
_{l}\right\} _{1\leq l\leq r-2}$ of the D$_{r}$ algebra correspond to the
completely antisymmetric tensors $T_{\left[ i_{1}i_{2}...i_{l}\right] }$
\textrm{while} traceless symmetric tensor $T_{\left(
i_{1}i_{2}...i_{l}\right) }$ of rank $l$ corresponds \textrm{per} usual to
weight $l\mathbf{\lambda }_{1}.$\newline
The two other fundamental weights $\mathbf{\lambda }_{r-1}$ and $\mathbf{%
\lambda }_{r}$ \textrm{define} the two spinor representations with dimension
$2^{r-1}.$\newline
$\bullet $ \textbf{Subfamily D}$_{2m}\simeq so(4m)\newline
$The fundamental representations $\mathbf{\lambda }_{j}$ with $1\leq j\leq
2m-2$ reside in the congruency class $\mathbf{\bar{0}}$. This includes the
adjoint representation $\mathbf{\lambda }_{2}.$ The fundamental spinor
representations $\mathbf{\lambda }_{2m-1}$ and $\mathbf{\lambda }_{2m}$
\textrm{occupy} class $\mathbf{\bar{1}}$. Completely \emph{antisymmetric}
tensors $T_{[i_{1}...i_{l}]}$ of rank $l=2m-1$ and $l=2m$ reside in class $%
\overline{\mathbf{2}}$.\newline
$\bullet $ \textbf{Subfamily }$\mathbf{D}_{2m+1}=so(4m+2)$\newline
For the algebra $so(4m+2),$ the congruence vector $\mathbf{\bar{\upsilon}}%
=\sum \bar{v}^{i}\mathbf{\alpha }_{i}$ is given modulo $4$ by the following
\begin{equation}
\mathbf{\bar{\upsilon}}=\left( 2,4,6,...,4m-2,2m-1,2m+1\right)
\end{equation}%
Completely \emph{antisymmetric} tensors $T_{[i_{1}...i_{l}]}$ of even $l=2j$
(odd $l=2j-1$) rank less than $2m-1$ reside in class $\overline{\mathbf{0}}$
($\overline{\mathbf{2}}$). Completely \emph{symmetric} tensors $%
T_{(i_{1}...i_{l})}$ of any even (odd) rank also \textrm{occupy} class $%
\overline{\mathbf{0}}$ ($\overline{\mathbf{2}}$). The rank $2m$ completely
\emph{antisymmetric} tensor \textrm{are} in class $\overline{\mathbf{0}}$,
while the rank- $2m+1$ \emph{antisymmetric} tensor divides into two
representations, both in class $\overline{\mathbf{2}}$.
\end{description}

The fundamental highest weight representations divide up like%
\begin{equation}
\begin{tabular}{|c|c|}
\hline
fundamental repr. & class \\ \hline
$\lambda _{2},\lambda _{4},\lambda _{6},...,\lambda _{2m-2}$ & $\overline{%
\mathbf{0}}$ \\ \hline
$\lambda _{2m}$ & $\overline{\mathbf{1}}$ \\ \hline
$\lambda _{1},\lambda _{3},\lambda _{2},...,\lambda _{2m-1}$ & $\overline{%
\mathbf{2}}$ \\ \hline
$\lambda _{2m+1}$ & $\overline{\mathbf{3}}$ \\ \hline
\end{tabular}%
\end{equation}

\section{Metric of $\mathcal{M}_{\mathbf{g}}$ and average of $\Theta _{%
\mathbf{g}}$ function}
\label{sec:appC}
In this appendix, we \textrm{provide detailed derivations for the} results
reported in the main text. First, we calculate the metric $\mathcal{G}_{%
\text{\textsc{ab}}}$ of the moduli space. Then, we\textrm{\ evaluate} the
Siegel-Narain theta function $\Theta _{\mathbf{g}}\left[ \tau _{1},\tau _{2};%
\mathbf{x}\right] $ at $\tau _{1}=0$; this permits to identify the value of
the constant c$_{\mathbf{g}}$ and the dependence in $\tau _{2}$\ of the
average $<\Theta _{\mathbf{g}}>_{\mathcal{M}}.$

\subsubsection*{String field in the Cartan basis}

Here we construct the field action S$\left[ X\right] $ of a compactified
string field $X\left( t,\sigma \right) $ on a circle that is manifestly
invariant under T-duality. This symmetry exchanges $\mathbb{S}^{1}$ with
radius $R$ into a dual circle $\mathbb{\tilde{S}}^{1}$ with radius $1/R$;
thus permitting to think about the action of T on $\mathbb{S}^{1}$ in terms
of the duplication $\mathbb{S}^{1}\times \mathbb{\tilde{S}}^{1}$ and the
permutation%
\begin{equation*}
T:\mathbb{S}^{1}\times \mathbb{\tilde{S}}^{1}\qquad \rightarrow \qquad
\mathbb{\tilde{S}}^{1}\times \mathbb{S}^{1}
\end{equation*}%
\textrm{with} matrix representation%
\begin{equation}
T=\left(
\begin{array}{cc}
0 & 1 \\
1 & 0%
\end{array}%
\right) \qquad ,\qquad T^{2}=I_{id}  \label{T}
\end{equation}%
Below, we show that this T-invariant action S$\left[ X\right] $, \textrm{from%
} now on denoted like $S\left[ \Phi ^{\text{\textsc{a}}}\right] ,$ has the
form%
\begin{equation}
S_{\mathbf{su}_{2}}\left[ \Phi ^{\text{\textsc{a}}}\right] \simeq
\dint\nolimits_{\mathbb{R}^{1,1}}dtd\sigma \text{ }\sum_{\text{\textsc{a},%
\textsc{b}=1}}^{2}\mathcal{G}_{\text{\textsc{ab}}}^{\mathbf{su}_{2}}\left(
\partial _{\mathrm{\gamma }}\Phi ^{\text{\textsc{a}}}\right) \left( \partial
^{\mathrm{\gamma }}\Phi ^{\text{\textsc{a}}}\right)
\end{equation}%
with the doublet $\Phi ^{\text{\textsc{a}}}$ given by ($U,V$) \textrm{as
shown below} and the metric $\mathcal{G}_{\text{\textsc{ab}}}$ given by%
\begin{equation}
\mathcal{G}_{\text{\textsc{ab}}}^{\mathbf{su}_{2}}=\left(
\begin{array}{cc}
\frac{1}{2R^{2}} & 0 \\
0 & 2R^{2}%
\end{array}%
\right) =\left(
\begin{array}{cc}
2x^{2} & 0 \\
0 & \frac{1}{2x^{2}}%
\end{array}%
\right)
\end{equation}%
where we have set $x=1/(2R).$ The T-duality \textrm{exchanges} the circles $%
\mathbb{S}^{1}$ and $\mathbb{\tilde{S}}^{1}$; it acts like $T:\sqrt{2}%
R\rightarrow 1/(\sqrt{2}R)$ and \textrm{maps} the field components like $%
U\leftrightarrow V.$

In our setting, \textrm{parameterised by} the su(2) simple root $\mathbf{%
\alpha }$ and its fundamental weight $\mathbf{\lambda }$, the above
T-duality can be imagined in terms of "scaled" simple root $\mathbf{\beta }%
\left( x\right) $ and "scaled" fundamental weight $\mathbf{\chi }\left(
x\right) $. They are given by
\begin{equation}
\begin{tabular}{lll}
$\mathbf{\beta }$ & $=$ & $x\mathbf{\alpha }$ \\
$\mathbf{\chi }$ & $=$ & $\frac{1}{x}\mathbf{\lambda }$%
\end{tabular}%
\qquad ,\qquad
\begin{tabular}{lll}
$\mathbf{\beta }^{2}$ & $=$ & 2$x^{2}$ \\
$\mathbf{\chi }^{2}$ & $=$ & $\frac{1}{2x^{2}}$%
\end{tabular}%
\qquad ,\qquad
\begin{tabular}{lll}
$\mathbf{\beta .\lambda }$ & $=$ & 1 \\
$\mathbf{\beta }^{2}\mathbf{\chi }^{2}$ & $=$ & $1$%
\end{tabular}%
\end{equation}%
with $\mathbf{x}\in \left] 0,\infty \right[ $ being the coordinate of the
moduli space $\mathcal{M}_{\mathbf{su}_{2}}$ that can be decomposed as the
union of two intervals
\begin{equation}
\mathcal{M}_{\mathbf{su}_{2}}=\left] 0,1\right[ \cup \left[ 1,\infty \right[
\end{equation}%
In the basis ($\mathbf{\alpha },\mathbf{\lambda }$), the 1-cycle $\mathbb{S}%
^{1}$ is thought of in terms of $\mathbf{\beta }$; and the dual 1-cycle $%
\mathbb{\tilde{S}}^{1}$ \textrm{is} imagined as $\mathbf{\chi }.$ The
operator representing T is \textrm{defined} by the matrix (\ref{T}) having
two eigenstates $\mathbb{S}_{\pm }^{1}=\mathbb{S}^{1}\pm \mathbb{\tilde{S}}%
^{1}$ with eigenvalues $\pm 1$ as follows%
\begin{equation}
T=\left(
\begin{array}{cc}
+1 & 0 \\
0 & -1%
\end{array}%
\right) \qquad ,\qquad \mathbb{S}_{L}^{1}=\mathbb{S}_{\mathbb{+}}^{1}\qquad
,\qquad \mathbb{S}_{R}^{1}=\mathbb{S}_{\mathbb{-}}^{1}
\end{equation}

To derive \textrm{the} components $\left( U,V\right) $ of the field doublet $%
\Phi ^{\text{\textsc{a}}}$ and its T-duality properties, we start from the
canonical momenta
\begin{equation}
\begin{tabular}{lllll}
$\boldsymbol{p}_{1}$ & $=$ & $\frac{1}{\sqrt{2}}\left( \boldsymbol{p}_{L}+%
\boldsymbol{p}_{R}\right) $ & $=$ & $n\mathbf{\alpha :}=p\mathbf{\alpha }$
\\
$\boldsymbol{p}_{2}$ & $=$ & $\frac{1}{\sqrt{2}}\left( \boldsymbol{p}_{L}-%
\boldsymbol{p}_{R}\right) $ & $=$ & $w\mathbf{\lambda :}=q\mathbf{\lambda }$%
\end{tabular}%
\end{equation}%
where we have \textrm{used} the parametrisation \textrm{(\ref{p}). We }get
the expressions of $\boldsymbol{p}_{1}$ and $\boldsymbol{p}_{2},$%
\begin{equation}
\boldsymbol{p}_{1}=n\mathbf{\alpha :}=p\mathbf{\alpha }\qquad ,\qquad
\boldsymbol{p}_{2}=w\mathbf{\lambda :}=q\mathbf{\lambda }
\end{equation}%
exchanged under the permutation $\mathbf{\alpha }\leftrightarrow \mathbf{%
\lambda }$ and $p\leftrightarrow q.$ Using the action of T-duality on left
and right momenta $\boldsymbol{p}_{\pm }$ namely $T:$ $\boldsymbol{p}%
_{L}\leftrightarrow \boldsymbol{p}_{L}$ and $T:$ $\boldsymbol{p}%
_{R}\leftrightarrow -\boldsymbol{p}_{R}$, we learn that we have:%
\begin{equation}
T:p\rightarrow p\qquad ,\qquad T:q\rightarrow -q
\end{equation}%
As for $p_{L}$ and $p_{R}$, the left $X_{L}(t+\sigma )$ and the right $%
X_{R}\left( t-\sigma \right) $ moving modes can be also promoted to vectors $%
\boldsymbol{X}_{L}$ and $\boldsymbol{X}_{R}$ with components $U(t;\sigma )$
and $V(t;\sigma )$ along the $\mathbf{\beta }$ and $\mathbf{\chi }$
directions as follows%
\begin{equation}
\begin{tabular}{lll}
$\boldsymbol{X}_{L}$ & $=$ & $\frac{1}{\sqrt{2}}\left( U\mathbf{\beta }+V%
\mathbf{\chi }\right) $ \\
$\boldsymbol{X}_{R}$ & $=$ & $\frac{1}{\sqrt{2}}\left( U\mathbf{\beta }-V%
\mathbf{\chi }\right) $%
\end{tabular}%
\end{equation}%
From these relationships, we deduce the canonical fields%
\begin{equation}
\begin{tabular}{lll}
$U\mathbf{\beta }$ & $=$ & $\frac{1}{\sqrt{2}}\left( \boldsymbol{X}_{L}+%
\boldsymbol{X}_{R}\right) $ \\
$V\mathbf{\chi }$ & $=$ & $\frac{1}{\sqrt{2}}\left( \boldsymbol{X}_{L}-%
\boldsymbol{X}_{R}\right) $%
\end{tabular}%
\end{equation}%
and well as their gradients%
\begin{equation}
\begin{tabular}{lll}
$\mathbf{\beta }\left( \partial _{\mathrm{\gamma }}U\right) $ & $=$ & $\frac{%
1}{\sqrt{2}}\partial _{\mathrm{\gamma }}\left( \boldsymbol{X}_{L}+%
\boldsymbol{X}_{R}\right) $ \\
$\mathbf{\chi }\left( \partial _{\mathrm{\gamma }}V\right) $ & $=$ & $\frac{1%
}{\sqrt{2}}\partial _{\mathrm{\gamma }}\left( \boldsymbol{X}_{L}-\boldsymbol{%
X}_{R}\right) $%
\end{tabular}%
\end{equation}%
\textrm{The} corresponding two dimensional field action $S\left[ U,V\right] $
reads as follows:%
\begin{equation}
S_{\mathbf{su}_{2}}\left[ U,V\right] \simeq \dint\nolimits_{\mathbb{R}%
^{1,1}}dtd\sigma \text{ }\left[ \mathbf{\beta }^{2}\left( \partial _{\mathrm{%
\gamma }}U\right) \left( \partial ^{\mathrm{\gamma }}U\right) +\mathbf{\chi }%
^{2}\left( \partial _{\mathrm{\gamma }}V\right) \left( \partial ^{\mathrm{%
\gamma }}V\right) \right]
\end{equation}%
and splits like $S\left[ U\right] +S\left[ V\right] $ with
\begin{eqnarray}
S\left[ U\right] &\simeq &\mathbf{\beta }^{2}\dint\nolimits_{\mathbb{R}%
^{1,1}}dtd\sigma \text{ }\left( \partial _{\mathrm{\gamma }}U\right) \left(
\partial ^{\mathrm{\gamma }}U\right) \\
S\left[ V\right] &\simeq &\mathbf{\chi }^{2}\dint\nolimits_{\mathbb{R}%
^{1,1}}dtd\sigma \left( \partial _{\mathrm{\gamma }}V\right) \left( \partial
^{\mathrm{\gamma }}V\right)
\end{eqnarray}%
indicating in turn \textrm{that} the metric $ds_{\mathbf{su}_{2}}^{2}=\left(
d\mu \right) ^{2}$ of the moduli subspace $\left] 0,\infty \right[ $ is
given by $\left( d\mu \right) ^{2}\simeq \left( dR/R\right) ^{2}.$ By
setting $\Phi ^{\text{\textsc{a}}}=\left( U,V\right) ,$ we can also rewrite
this action as follows%
\begin{equation}
S_{\mathbf{su}_{2}}\left[ U,V\right] \simeq \dint\nolimits_{\mathbb{R}%
^{1,1}}dtd\sigma \text{ }\mathcal{G}_{\text{\textsc{ab}}}\left( \partial _{%
\mathrm{\gamma }}\Phi ^{\text{\textsc{a}}}\right) \left( \partial ^{\mathrm{%
\gamma }}\Phi ^{\text{\textsc{a}}}\right)
\end{equation}%
with%
\begin{equation}
\mathcal{G}_{\text{\textsc{ab}}}^{\mathbf{su}_{2}}=\left(
\begin{array}{cc}
\mathbf{\beta }^{2} & 0 \\
0 & \mathbf{\chi }^{2}%
\end{array}%
\right) =\left(
\begin{array}{cc}
2x^{2} & 0 \\
0 & \frac{1}{2x^{2}}%
\end{array}%
\right)
\end{equation}%
This construction extends straightforwardly to higher dimensional NCFT$^{%
\mathbf{g}}$s with Lie algebra \textbf{g }and metric $\mathcal{G}_{\text{%
\textsc{ab}}}^{\mathbf{g}}$. For these theories, the metric $ds_{\mathbf{g}%
}^{2}$ has the form $\left( \mathbf{\chi ^{2}.}d\mathbf{\beta }^{2}\right)
^{2}.$

\subsubsection*{Computing $\Theta _{\mathbf{su}_{2}}\left[ \protect\tau _{2};%
\mathbf{x}\right] $ and $\Theta _{\mathbf{g}}\left[ \protect\tau _{2};%
\mathbf{x}\right] $}

Here, we study properties of the function $\Theta _{\mathbf{g}}\left[ \tau ,%
\bar{\tau};R\right] $\ and $F_{\mathbf{g}}\left( \tau ,\bar{\tau}\right) $
of eq(\ref{fsu}) by considering first the case $\mathrm{su}_{2}$ and turn
after to generic Lie algebras $\mathbf{g}$.

\paragraph{\textbf{A)} Higher ranks NCFT$_{\mathbf{su}_{2}}^{(1,1)}$\newline
}

Recall that the Siegel-Narain theta function is given by%
\begin{equation}
\Theta _{\mathrm{su}_{2}}\left[ \tau _{1},\tau _{2};x\right]
=\dsum\limits_{n,w}e^{2i\pi \tau _{1}nw}\exp \left[ -\pi \tau _{2}\left(
2x^{2}n^{2}+\frac{1}{2x^{2}}w^{2}\right) \right]  \label{2p}
\end{equation}%
Its "averaged" function $F_{\mathrm{su}_{2}}\left( \tau ,\bar{\tau}\right) $
\textrm{in terms of the variable x can be formulated as:}
\begin{equation}
F_{\mathrm{su}_{2}}\left( \tau _{1},\tau _{2}\right) \sim
\dint\nolimits_{0}^{\infty }\frac{dx}{x}\Theta _{\mathrm{su}_{2}}\left[ \tau
_{1},\tau _{2};x\right]  \label{FT}
\end{equation}%
\textrm{where we set the measure }$dx^{l}/x^{l}=ldx/x$\textrm{\ with }$l=\pm
1,$\textrm{\ and }$x=1/(2R).$\textrm{\ }To compute $F_{\mathrm{su}_{2}}$ and
later $F_{\mathbf{g}}$, we start from eq(\ref{SA}) and \textrm{calculate}
the double sum of $\varrho _{n,w}^{\mathbf{su}_{2}}\left( x;\tau _{1},\tau
_{2}\right) $. \textrm{Given that }$\Theta _{\mathrm{su}_{2}}$ is invariant
by translation in the $\tau _{1}$ direction; i.e: $\Theta _{\mathrm{su}_{2}}%
\left[ \tau _{1}+1\right] =\Theta _{\mathrm{su}_{2}}\left[ \tau _{1}\right]
, $ one \textrm{exploits} this symmetry to set $\tau _{1}=0$. This feature
\textrm{also manifests via} the contribution in $\tau _{1}$ in (\ref{2p})
which is given by the phase factor e$^{2i\pi \tau _{1}nw}=(-)^{2\tau _{1}}$
indicating that we have%
\begin{equation}
\Theta _{\mathrm{su}_{2}}\left[ \tau _{1},\tau _{2};x\right] =(-)^{2\tau
_{1}}\vartheta _{\mathrm{su}_{2}}\left[ \tau _{2};x\right]
\end{equation}%
with
\begin{equation*}
\vartheta _{\mathrm{su}_{2}}\left[ \tau _{2};x\right] =\dsum\limits_{n,w}%
\exp \left[ -\pi \tau _{2}\left( 2x^{2}n^{2}+\frac{1}{2x^{2}}w^{2}\right) %
\right]
\end{equation*}%
In what follows, we set $\tau _{1}$ to zero and \textrm{focus on} computing $%
\vartheta _{\mathrm{su}_{2}}\left[ \tau _{2};x\right] $ and $F_{\mathrm{su}%
_{2}}\left( \tau _{2}\right) .$ In this regard, notice that the above $%
\vartheta _{\mathrm{su}_{2}}\left[ \tau _{2};x\right] $ factorises like
\begin{equation*}
\vartheta _{\mathrm{su}_{2}}\left[ \tau _{2};x\right] =\bar{K}_{\mathrm{su}%
_{2}}\left[ \tau _{2};x\right] \times \bar{W}_{\mathrm{su}_{2}}\left[ \tau
_{2};x\right]
\end{equation*}%
with factors $\bar{K}_{\mathrm{su}_{2}}$ and $\bar{W}_{\mathrm{su}_{2}}$
\textrm{as} given by the series%
\begin{equation}
\begin{tabular}{lll}
$\bar{K}_{\mathrm{su}_{2}}\left[ \tau _{2};x\right] $ & $=$ & $%
\dsum\limits_{n\in \mathbb{Z}}e^{-\left( 2\pi \tau _{2}x^{2}\right) n^{2}}$
\\
$\bar{W}_{\mathrm{su}_{2}}\left[ \tau _{2};x\right] $ & $=$ & $%
\dsum\limits_{w\in \mathbb{Z}}e^{-\frac{\pi \tau _{2}}{2x^{2}}w^{2}}$%
\end{tabular}%
\end{equation}%
converging to%
\begin{eqnarray}
\bar{K}_{\mathrm{su}_{2}}\left[ \tau _{2};x\right] &=&\sqrt{\frac{1}{\tau
_{2}2x^{2}}}=\sqrt{\frac{1}{2\tau _{2}}}\frac{1}{\left\vert x\right\vert } \\
\bar{W}_{\mathrm{su}_{2}}\left[ \tau _{2};x\right] &=&\sqrt{\frac{2x^{2}}{%
\tau _{2}}}=\sqrt{\frac{2}{\tau _{2}}}\left\vert x\right\vert
\end{eqnarray}%
From these \textrm{formulas}, we deduce the following features:

\begin{itemize}
\item The $\vartheta _{\mathrm{su}_{2}}\left[ \tau _{2};x\right] $ converges
to the value%
\begin{equation}
\vartheta _{\mathrm{su}_{2}}\left[ \tau _{2};x\right] =\bar{K}\left(
x\right) \bar{W}\left( x\right) =\frac{1}{\tau _{2}}
\end{equation}%
manifestly invariant under T-transformation. It is independent of the
modulus $x$. Substituting into (\ref{FT}), we find that $F_{\mathrm{su}%
_{2}}\left( \tau _{2}\right) $ diverge%
\begin{equation}
F_{\mathrm{su}_{2}}\left( \tau _{2}\right) =\frac{1}{\tau _{2}}%
\dint\nolimits_{0}^{\infty }\left( \log x\right) dx=\infty
\end{equation}%
Notice \textrm{also that}%
\begin{eqnarray}
\kappa _{\mathrm{su}_{2}}\left( \tau _{2}\right)
&:&=\dint\nolimits_{1}^{\infty }\frac{dx}{x}\bar{K}_{\mathrm{su}_{2}}\left[
\tau _{2};x\right]  \notag \\
&=&\sqrt{\frac{1}{2\tau _{2}}}\dint\nolimits_{1}^{\infty }\frac{dx}{x^{2}}=%
\sqrt{\frac{1}{2\tau _{2}}}
\end{eqnarray}%
and%
\begin{eqnarray}
\omega _{\mathrm{su}_{2}}\left( \tau _{2}\right)
&:&=\dint\nolimits_{1}^{\infty }\frac{dx}{x}W_{\mathrm{su}_{2}}\left[ \tau
_{2};x\right]  \notag \\
&=&\sqrt{\frac{2}{\tau _{2}}}\dint\nolimits_{1}^{\infty }dx=\infty
\end{eqnarray}

\item The functions $\bar{K}\left( x\right) $ and $\bar{W}\left( x\right) $
read in terms of the scaled root $\mathbf{\beta }=x\mathbf{\alpha }$ and the
scaled weight $\mathbf{\chi }=\mathbf{\lambda }/x$ as follows
\begin{equation*}
\bar{K}_{\mathrm{su}_{2}}\left[ \tau _{2};x\right] =\sqrt{\frac{\mathbf{\chi
}^{2}}{\tau _{2}}}\qquad ,\qquad \bar{W}_{\mathrm{su}_{2}}\left[ \tau _{2};x%
\right] =\sqrt{\frac{\mathbf{\beta }^{2}}{\tau _{2}}}
\end{equation*}%
\textrm{exchangeable under} T-duality.
\end{itemize}

\paragraph{B) Higher ranks NCFT$_{\mathbf{g}}^{(r,r)}$\newline
}

From the above su(2) \textrm{set up}, \textrm{we can draw the generalisation
towards} NCFT$_{\mathrm{g}}^{(r,r)}$s labeled by higher dimensional Lie
algebras. We have%
\begin{equation}
\Theta _{\mathbf{g}}\left[ \tau _{2};x\right] =\dsum\limits_{n,w}\exp \left[
-\pi \tau _{2}\left( n^{i}\mathcal{K}_{ij}^{\mathbf{g}}n^{i}+w_{i}\mathcal{%
\tilde{K}}_{\mathbf{g}}^{ij}w_{j}\right) \right]
\end{equation}%
with%
\begin{equation}
\mathcal{K}_{ij}^{\mathbf{g}}=x_{i}^{k}K_{kl}^{\mathbf{g}}x_{j}^{l}\qquad
,\qquad \mathcal{\tilde{K}}_{\mathbf{g}}^{ij}=y_{k}^{i}K_{\mathbf{g}%
}^{kl}y_{l}^{j}  \label{KG}
\end{equation}%
It factorises $\Theta _{\mathbf{g}}\left[ \tau _{2};\mathbf{x}\right] =\bar{K%
}_{\mathbf{g}}\left[ \tau _{2};\mathbf{x}\right] \times \bar{W}_{\mathbf{g}}%
\left[ \tau _{2};\mathbf{x}\right] $ with factors $\bar{K}_{\mathbf{g}}$ and
$\bar{W}_{\mathbf{g}}$ given by the series%
\begin{equation}
\begin{tabular}{lll}
$\bar{K}_{\mathbf{g}}\left[ \tau _{2};\mathbf{x}\right] $ & $=$ & $%
\dsum\limits_{n\in \mathbb{Z}}\exp \left[ -n^{i}\left( \pi \tau _{2}\mathcal{%
K}_{ij}^{\mathbf{g}}\right) n^{i}\right] $ \\
$\bar{W}_{\mathbf{g}}\left[ \tau _{2};\mathbf{x}\right] $ & $=$ & $%
\dsum\limits_{w\in \mathbb{Z}}\exp \left[ -w_{i}(\pi \tau _{2}\mathcal{%
\tilde{K}}_{\mathbf{g}}^{ij})w_{j}\right] $%
\end{tabular}%
\end{equation}%
converging to%
\begin{equation}
\bar{K}_{\mathbf{g}}\left[ \tau _{2};\mathbf{x}\right] =\sqrt{\frac{\pi ^{r}%
}{\det \left( \pi \tau _{2}\mathcal{K}_{ij}^{\mathbf{g}}\right) }}\qquad
,\qquad \bar{W}_{\mathbf{g}}\left[ \tau _{2};\mathbf{x}\right] =\sqrt{\frac{%
\pi ^{r}}{\det \left( \pi \tau _{2}\mathcal{\tilde{K}}_{\mathbf{g}%
}^{ij}\right) }}
\end{equation}%
By using the relationships (\ref{KG}), we also have%
\begin{eqnarray*}
\bar{K}_{\mathbf{g}}\left[ \tau _{2};\mathbf{x}\right] &=&\frac{1}{\tau
_{2}^{r/2}\left( \det K_{ij}^{\mathbf{g}}\right) ^{\frac{1}{2}}}\times \frac{%
1}{\left\vert \det x_{i}^{k}\right\vert } \\
\bar{W}_{\mathbf{g}}\left[ \tau _{2};\mathbf{x}\right] &=&\frac{\left( \det
K_{ij}^{\mathbf{g}}\right) ^{\frac{1}{2}}}{\tau _{2}^{r/2}}\times \left\vert
\det x_{i}^{k}\right\vert
\end{eqnarray*}

From these values, we obtain the following results:

\begin{itemize}
\item The $\Theta _{\mathbf{g}}\left[ \tau _{2};\mathbf{x}\right] $
converges to the value%
\begin{equation}
\Theta _{\mathbf{g}}\left[ \tau _{2};\mathbf{x}\right] =\bar{K}_{\mathbf{g}}%
\left[ \tau _{2};\mathbf{x}\right] \bar{W}_{\mathbf{g}}\left[ \tau _{2};%
\mathbf{x}\right] =\frac{1}{\tau _{2}^{r}}
\end{equation}%
manifestly invariant under T-duality. It is independent of the modulus $%
x_{i}^{k}$. Substituting into (\ref{FT}), we find that $F_{\mathbf{g}}\left[
\tau _{2}\right] $ diverges%
\begin{equation}
F_{\mathbf{g}}\left[ \tau _{2}\right] =\dint\nolimits_{0}^{\infty }\sqrt{%
\det \mathcal{G}_{\text{\textsc{ab}}}}\dprod\limits_{i,k}dx
\end{equation}

\item The functions $\bar{K}_{\mathbf{g}}\left[ \tau _{2};\mathbf{x}\right] $
and $\bar{W}_{\mathbf{g}}\left[ \tau _{2};\mathbf{x}\right] $ read in terms
of the scaled root $\mathbf{\beta }_{i}=\sum x_{i}^{k}\mathbf{\alpha }_{k}$
and the scaled weight $\mathbf{\chi }^{j}=\sum w_{l}^{j}\mathbf{\lambda }%
^{l} $ as follows%
\begin{eqnarray}
\bar{K}_{\mathbf{g}}\left[ \tau _{2};\mathbf{x}\right] &=&\frac{1}{\tau
_{2}^{r/2}(\mathbf{\beta }_{i}.\mathbf{\beta }_{j})^{\frac{1}{2}}}\  \\
\bar{W}_{\mathbf{g}}\left[ \tau _{2};\mathbf{x}\right] &=&\frac{\left(
\mathbf{\chi }^{i}.\mathbf{\chi }^{j}\right) ^{\frac{1}{2}}}{\tau _{2}^{r/2}}%
\
\end{eqnarray}
\end{itemize}


\newpage

\begin{thebibliography}{99}
\bibitem{1A} Maloney, A., Witten, E. Averaging over Narain moduli space. J.
High Energ. Phys. 2020, 187 (2020). https://doi.org/10.1007/JHEP10(2020)187,
arXiv:2006.04855v2 [hep-th]

\bibitem{2AB} Ofer Aharony, Anatoly Dymarsky, Alfred D. Shapere, Holographic
description of Narain CFTs and their code-based ensembles, JHEP 2024, 343
(2024), arXiv:2310.06012v1 [hep-th].

\bibitem{Ange} Angelinos, N., Chakraborty, D. \& Dymarsky, A. Optimal Narain
CFTs from codes. J. High Energ. Phys. 2022, 118 (2022).
https://doi.org/10.1007/JHEP11(2022)118

\bibitem{swp1} van Beest, M., Calder\'{o}n-Infante, J., Mirfendereski, D.,
\& Valenzuela, I. (2022). Lectures on the swampland program in string
compactifications. Physics Reports, 989, 1-50.

\bibitem{swp2} Palti, E. (2019). The swampland: introduction and review.
Fortschritte der Physik, 67(6), 1900037.

\bibitem{Raja} Sammani, R., Boujakhrout, Y., Saidi, E. H., Ahl Laamara, R.,
\& Drissi, L. B. (2024). Finiteness of 3D higher spin gravity Landscape.
Classical and Quantum Gravity.

\bibitem{NG} Banks, T., \& Seiberg, N. (2011). Symmetries and strings in
field theory and gravity. Physical Review D---Particles, Fields,
Gravitation, and Cosmology, 83(8), 084019.

\bibitem{chark} Charkaoui, M., Sammani, R., Saidi, E. H., \& Laamara, R. A.
(2024). Asymptotic Weak Gravity Conjecture in M-theory on K 3 x K 3.
Progress of Theoretical and Experimental Physics, 2024(7), 073B08.\\
Charkaoui, M., Sammani, R., Saidi, E. H., \& Ahl Laamara, R. (2025). Minimal Weak Gravity Conjecture and gauge duality in M-theory on K3 × T2 [Accepted manuscript]. Physica Scripta. https://doi.org/10.1088/1402-4896/ae0ed5

\bibitem{1AA} Meer Ashwinkumar, Jacob M. Leedom, Masahito Yamazaki, Duality
Origami: Emergent Ensemble Symmetries in Holography and Swampland:
Phys.Lett.B 856 (2024) 138935, arXiv:2305.10224v2 [hep-th].

\bibitem{JT} Teitelboim, C. (1983). Gravitation and hamiltonian structure in
two spacetime dimensions. Physics Letters B, 126(1-2), 41-45.\newline
Jackiw, R. (1984). Liouville field theory: A two-dimensional model for
gravity?, in Quantum Theory Of Gravity, S. Christensen, ed., (Bristol), pp.
403--420, Adam Hilger.

\bibitem{1b} N. Afkhami-Jeddi, H. Cohn, T. Hartman and A. Tajdini, Free
partition functions and an averaged holographic duality , arxiv:2006.04839.

\bibitem{RPP} Ashwinkumar, M., Dodelson, M., Kidambi, A., Leedom, J. M., \&
Yamazaki, M. (2021). Chern-Simons invariants from ensemble averages. Journal
of High Energy Physics, 2021(8), 1-34.

\bibitem{JT1} Saad, P., Shenker, S. H., \& Stanford, D. (2019). JT gravity
as a matrix integral. arXiv preprint arXiv:1903.11115.

\bibitem{JT2} Cotler, J. S., Gur-Ari, G., Hanada, M., Polchinski, J., Saad,
P., Shenker, S. H., ... \& Tezuka, M. (2017). Black holes and random
matrices. Journal of High Energy Physics, 2017(5), 1-54.

\bibitem{JT3} Saad, P., Shenker, S. H., \& Stanford, D. (2018). A
semiclassical ramp in SYK and in gravity. arXiv preprint arXiv:1806.06840.

\bibitem{JT4} Saad, P. (2019). Late time correlation functions, baby
universes, and ETH in JT gravity. arXiv preprint arXiv:1910.10311.

\bibitem{red1} Mertens, T. G. (2018). The Schwarzian theory---origins.
Journal of High Energy Physics, 2018(5).

\bibitem{red2} J. Cotler and K. Jensen, A theory of reparameterizations for
AdS3 gravity, JHEP 02,079, 2019, [arXiv:1808.03263 [hep-th]].

\bibitem{AT} Achucarro, A., \& Townsend, P. K. (1986). A Chern-Simons action
for three-dimensional anti-de Sitter supergravity theories. Physics Letters
B, 180(1-2), 89-92.

\bibitem{W} Witten, E. (1988). 2+ 1 dimensional gravity as an exactly
soluble system. Nuclear Physics B, 311(1), 46-78.

\bibitem{W1} Witten, E. (2007). Three-dimensional gravity revisited. arXiv
preprint arXiv:0706.3359.

\bibitem{Narain} Narain, K. S. (1986). New heterotic string theories in
uncompactified dimensions \TEXTsymbol{<} 10, Phys.Lett. 169B, 41--46.

\bibitem{Narain2} Narain, K. S., Sarmadi, M. H., \& Witten, E. (1986). A
note on toroidal compactification of heterotic string theory.

\bibitem{Zchi} Zamolodchikov, A. (1986). Irreversibility of the Flux of the
Renormalization Group in a 2D Field Theory. JETP Lett. 43 730--732.

\bibitem{SW1} Siegel, C. L. (1951). Indefinite quadratische formen und
funktionentheorie I. Mathematische Annalen, 124(1), 17-54.

\bibitem{SW2} Weil, A. (1964). Sur certains groupes d'op\'{e}rateurs
unitaires. Acta math, 111(143-211), 14.

\bibitem{SW3} Weil, A. (1965). Sur la formule de Siegel dans la th\'{e}orie
des groupes classiques. Acta math, 113(1-87), 2.

\bibitem{1AB} Benjamin, N., Collier, S., Fitzpatrick, A.L. et al. Harmonic
analysis of 2d CFT partition functions. J. High Energ. Phys. 2021, 174
(2021). https://doi.org/10.1007/JHEP09(2021)174

\bibitem{SD2} Roelcke, W. (1966). Das Eigenwertproblem der automorphen
Formen in der hyperbolischen Ebene, I. Mathematische Annalen, 167(4),
292-337.

\bibitem{SD3} Roelcke, W. (1967). Das Eigenwertproblem der automorphen
Formen in der hyperbolischen Ebene. II. Mathematische Annalen, 168(1),
261-324.

\bibitem{Y} Datta, S., Duary, S., Kraus, P., Maity, P., \& Maloney, A.
(2022). Adding flavor to the Narain ensemble. Journal of High Energy
Physics, 2022(5), 1-29.

\bibitem{H} Maloney, A., \& Witten, E. (2010). Quantum gravity partition
functions in three dimensions. Journal of High Energy Physics, 2010(2), 1-58.

\bibitem{PT} P\'{e}rez, A., \& Troncoso, R. (2020). Gravitational dual of
averaged free CFT's over the Narain lattice. Journal of High Energy Physics,
2020(11), 1-12.

\bibitem{23} Dymarsky, A., Henriksson, J., $\&$ McPeak, B. (2025).
Holographic duality from Howe duality: Chern-Simons gravity as an ensemble
of code CFTs. arXiv preprint arXiv:2504.08724.

\bibitem{SS} Saidi, E. H., \& Sammani, R. (2025). Code CFTs and Topological Matter. arXiv preprint arXiv:2506.22088.

\bibitem{Soft1} H. Afshar, S. Detournay, D. Grumiller, W. Merbis, A. Perez,
D. Tempo et al., Soft Heisenberg hair on black holes in three dimensions,
Phys. Rev. D93 (2016) 101503, arXiv: 1603.04824 [hep-th]

\bibitem{Raja5} Sammani, R., Saidi, E. H., Laamara, R. A., \& Drissi, L. B. (2025). Landscape of Narain CFTs. Fortschritte der Physik, e70029.

\bibitem{Raja6} Sammani, R., Saidi, E. H., Laamara, R. A., \& Drissi, L. B. (2025). Fluctuating ensemble averages and the BTZ threshold. The European Physical Journal C, 85(4), 1-18.

\bibitem{Soft2} Daniel Grumiller, Alfredo Perez, Stefan Prohazka, David
Tempo, Ricardo Troncoso, Higher Spin Black Holes with Soft Hair, JHEP 10
(2016) 119, arXiv:1607.05360 [hep-th].

\bibitem{BH} J. D. Brown and M. Henneaux, \textquotedblleft Central Charges
in the Canonical Realization of Asymptotic Symmetries: An Example from
Three-Dimensional Gravity,\textquotedblright\ Commun. Math. Phys. 104, 207
(1986).

\bibitem{Emb1} Campoleoni, A., Fredenhagen, S., Pfenninger, S., and Theisen,
S. Asymptotic symmetries of three-dimensional gravity coupled to higher-spin
fields. Journal of High Energy Physics, 2010(11), 1-36. (2010).

\bibitem{Raja2} Sammani, R., Boujakhrout, Y., Saidi, E. H., Laamara, R. A.,
\& Drissi, L. B. (2023). Higher spin AdS 3 gravity and Tits-Satake diagrams.
Physical Review D, 108(10), 106019.

\bibitem{sci} Sammani, R., and Saidi, E. H. (2025). Higher spin swampland conjecture for massive AdS $ _ {3} $ gravity. SciPost Physics, 18(6), 173.

\bibitem{Raja3} Sammani, R., and Saidi, E.H. (2024). Black flowers and real
forms of higher spin symmetries. J. High Energ. Phys. 2024, 44. \\ Sammani, R., Saidi, E. H., \& Laamara, R. A. (2025). Black hole solutions of three dimensional E6-gravity. Journal of Mathematical Physics, 66(2).

\bibitem{Raja4} Y. Boujakhrout, R. Sammani, E. H Saidi, Topological 4D
gravity and gravitational defects, Physica Scripta (2024), Volume 99, Issue
11, id.115256, 14 pp.

\bibitem{MEER} Ashwinkumar, M., Kidambi, A., Leedom, J. M., \& Yamazaki, M.
(2023). Generalized Narain Theories Decoded: Discussions on Eisenstein
series, Characteristics, Orbifolds, Discriminants and Ensembles in any
Dimension. arXiv preprint arXiv:2311.00699.

\bibitem{PT2} Dymarsky, A., \& Shapere, A. (2021). Comments on the
holographic description of Narain theories. Journal of High Energy Physics,
2021(10), 1-26.

\bibitem{KN} Narain, K. (2024). Toroidal and Orbifold Compactifications. In
Handbook of Quantum Gravity (pp. 1-19). Singapore: Springer Nature Singapore.

\bibitem{vac} Porrati, M., \& Yu, C. (2019). Kac-Moody and Virasoro
characters from the perturbative Chern-Simons path integral. Journal of High
Energy Physics, 2019(5), 1-71.

\bibitem{TW} E. Witten,Three Lectures On Topological Phases Of Matter, La
Rivista del Nuovo Cimento, 39 (2016) 313-370, arXiv:1510.07698.

\bibitem{TDS} E.H Saidi, Gapped gravitinos, isospin 12 particles and N=2
partial breaking, Prog Theor Exp Phys (2019), arXiv:1812.04509v1 [hep-th].

\bibitem{CC} El Hassan Saidi, Complex D(2,1;$\zeta$) and spin chain
solutions from Chern-Simons theory, \qquad arXiv:2409.14862 [hep-th].

\bibitem{2A} A. Maloney and E. Witten, Quantum Gravity Partition Functions
In Three Dimensions," JHEP 02 (2010) 029, arXiv:0712.0155.

\bibitem{1a} Benjamin, N., Keller, C.A., Ooguri, H. et al. Narain to Narnia.
Commun. Math. Phys. 390, 425--470 (2022).
https://doi.org/10.1007/s00220-021-04211-x

\bibitem{2AA} Kawabata, K., Nishioka, T. \& Okuda, T. Narain CFTs from
quantum codes and their gauging. J. High Energ. Phys. 2024, 133 (2024),
arXiv:2308.01579v4 [hep-th]. https://doi.org/10.1007/JHEP05(2024)133

\bibitem{3AA} C\'{o}rdova, C., Ohmori, K. \& Rudelius, T. Generalized
symmetry breaking scales and weak gravity conjectures. J. High Energ. Phys.
2022, 154 (2022). https://doi.org/10.1007/JHEP11(2022)154

\bibitem{3AB} Tristan Daus, Arthur Hebecker, Sascha Leonhardt, John
March-Russell, Towards a Swampland Global Symmetry Conjecture using Weak
Gravity, Nuclear Physics B Volume 960, November 2020, 115167,
arXiv:2002.02456v3 [hep-th]

\bibitem{FUR} Yuma Furuta, On the classification of duality defects in c=2
compact boson CFTs with a discrete group orbifold, arXiv:2412.01319v1
[hep-th].
\end{thebibliography}
\end{document}